\documentclass[twocolumn,showpacs,preprintnumbers,amsmath,amssymb,epsf,pre,longbibliography,graphicx]{revtex4-1}

\usepackage{MnSymbol}
\usepackage{color}    
\usepackage{graphicx}
\usepackage{braket} 
\usepackage{dcolumn}
\usepackage{bm}
\usepackage{subfigure}
\usepackage{amssymb}
\usepackage{multirow}
\usepackage{natbib}
    \usepackage{amsmath}
\graphicspath{{plots/}}
\usepackage[english]{babel}
\usepackage{rotating}

\newcommand{\angstrom}{\text{\normalfont\AA}}

\renewcommand{\vec}[1]{\mathbf{#1}}

 \usepackage{hyperref}

\begin{document}

\title{Quantum effects in plasmas}

\author{M.~Bonitz$^{1,2}$, H.~K\"ahlert$^{1,2}$, D.~Krimans$^1$, C.~Makait$^1$, 
P.~Hamann$^{3,4}$, 
 J.~Vorberger$^3$, Zh.~Moldabekov$^{3}$, 
 S. X. Hu$^{5,6,7}$, V. V. Karasiev$^5$, D.~Kraus$^{4,3}$, H.~Kersten$^8$, J.-P. Joost$^1$, P.~Ludwig$^1$, and  T.~Dornheim$^{3,9}$
}
\affiliation{
 $^1$Institut f\"ur Theoretische Physik und Astrophysik, Christian-Albrechts-Universit\"at zu Kiel,
 Leibnizstra{\ss}e 15, 24098 Kiel, Germany}
\affiliation{
 $^2$KiNSIS, Christian-Albrechts-Universit\"at zu Kiel, 24098 Kiel, Germany}
\affiliation{$^3$Institute of Radiation Physics, Helmholtz-Zentrum Dresden-Rossendorf (HZDR), D-01328 Dresden, Germany}
\affiliation{$^4$Institut f\"ur Physik, Universit\"at Rostock, D-18057 Rostock, Germany}
 \affiliation{$^5$Laboratory for Laser Energetics, University of Rochester, Rochester, New York 14623, USA}
 \affiliation{$^6$Department of Physics and Astronomy, University of Rochester, Rochester, New York 14627, USA}
  \affiliation{$^7$Department of Mechanical Engineering, University of Rochester, Rochester, New York 14627, USA}
 \affiliation{$^8$Institut f\"ur Experimentelle und Angewandte Physik, Christian-Albrechts-Universit\"at zu Kiel,
 Leibnizstra{\ss}e 15, 24098 Kiel, Germany}
 \affiliation{$^9$Center for Advanced Systems Understanding (CASUS), Helmholtz-Zentrum Dresden-Rossendorf (HZDR), D-02826 G\"orlitz, Germany}

%---------------
\begin{abstract}
The year 2025 had been designated by UNESCO as the International Year of
Quantum Science and Technology. 125 years ago Max Planck’s discovery
of radiation quanta started the quantum era and 100 years ago quantum
mechanics was discovered by Schrödinger, Heisenberg, Bohr, Pauli, Dirac,
Born, Fermi and many others. By now, quantum mechanics is the theoretical
foundation of most fields of physics and chemistry, and it is the basis for modern nanotechnology.
How about plasma physics? How important are quantum effects in plasmas? 
In what experiments quantum effects are observed and where do they govern the behavior of plasmas? How can these effects be  treated theoretically and via computer simulations?
Starting with a brief historical overview we  discuss the broad parameter range that is characteristic for plasmas and outline where quantum effects are relevant. This is the case primarily for warm dense matter and inertial fusion plasmas. We provide an overview on the theoretical quantum methods that are available for these dense plasmas and how their respective advantages can be combined in order to achieve predictive capability. The key is a downfolding approach that is based on first principles simulations.
\end{abstract}

%\pacs{52.27.Lw, 52.20.-j, 52.40.Hf}
%52.25.Dg Plasma kinetic equations
%52.27.Aj Single-component, electron-positive-ion plasmas
%52.27.Gr non-ideal plasmas
%52.65.Yy Molecular dynamics methods
%Plasma simulation, 52.65.-y
\maketitle

 \section{Introduction}\label{s:intro}

Since its first appearance in 1900, quantum theory has completely changed our picture of the microcosmos. However, when German physicist Max Planck presented his concept of quanta of electromagnetic radiation on December 14 of that year \cite{planck_vdpg00_2} that allowed him to accurately reproduce recent measurements of the black-body radiation spectrum in the micrometer range, this was met with massive skepticism and opposition among physicists \cite{bonitz-in-kbmp1}.  Even Planck himself for many years remained hesitant in accepting quanta (photons) as intrinsic property of radiation. It took the work of many scientists over almost two decades, most importantly, Einstein (photoelectric effect) and Bohr (planetary model of the atom), to prove that this concept is indeed at the heart of the microcosmos. Much less known today is that also plasma physicists made important contributions to the success of quantum theory, in particular, with experiments in gas discharges \cite{meyer_1858, seeliger-12} that eventually led to the famous Franck-Hertz experiment \cite{franck-hertz-14} that proved the quantum nature of atoms, as we will recall in Sec.~\ref{ss:low-p}. 

With the discovery of the mathematical framework of quantum mechanics in 1925-1927 by Heisenberg, Schrödinger, Born, Bohr, Pauli, Fermi, Dirac and many others, quantum concepts quickly became the fundamentals of modern physics -- of the microworld of atoms, molecules, solids and liquids as well as of nuclear an elementary particle physics. But even at cosmic scales quantum physics is at the heart of many phenomena and processes, including the structure of planets and compact stars and the evolution of the early universe following the Big Bang, as imprinted in the Planckian nature of the cosmic microwave background radiation \cite{durrer_cmb_20}.

On the other hand, in research and education in plasma physics quantum effects have, until now, played a much less prominent role than in other fields.
The International year of quantum science and technology 2025 is the occasion to have a closer look at quantum effects in plasmas and on their growing importance for the future of the field.

In plasmas, quantum effects arise primarily in the behavior of the lightest particles. Quantum effects of the plasma electrons are relevant either at sufficiently high density and/or low temperature where the degeneracy parameter $\chi$ exceeds unity, cf. Eq.~\eqref{eq:chi-def}. This is the case in highly compressed plasmas and warm dense matter (see below). At the same time, even in low-pressure plasmas where electrons are non-degenerate, quantum effects occur. 
These plasmas often contain neutral particles, such as atoms or molecules, the properties of which are determined by quantum effects. 
Another example (even in fully ionized plasmas) are 
``close collisions" of electrons and ions occurring in a wide range of densities (e.g. laser absorption in coronal plasmas).  To compute transport and optical properties requires a quantum treatment even at high temperatures. For example, the Coulomb logarithm plasma physicists often use for the cross section of e-i collisions is defined as $L=\ln(b_{\textrm {max}}/b_{\textrm {min}})$, with $b_{\textrm {min}}$ being the smallest impact parameter that is usually defined as $b_{\textrm {min}}={\textrm {min}}(h/m_ev, Ze^2/mv^2)$,  invoking the de Broglie wavelength associated to the uncertainty principle [first term, where $h$ denotes Planck's constant, Eq.~\eqref{eq:h-def}]. A third example is ``classical'' high-temperature plasmas that form the basis for magnetic fusion. The central part -- the fusion reaction -- is governed by quantum effects, such as tunneling through the Coulomb barrier.

Let us discuss a bit more in detail the situation when the plasma electrons exhibit quantum behavior. This is the case in warm dense matter (WDM) -- a field on the border of plasma physics and condensed matter physics that attracted substantial interest in recent years, e.g.~\cite{graziani-book,Fortov2016, moldabekov_pre_18, dornheim_physrep_18}. Examples of WDM in astrophysics are plasma-like matter in brown and white dwarf stars \cite{saumon_the_role_1992, chabrier_quantum_1993,chabrier_cooling_2000}, giant planets, e.g. \cite{schlanges_cpp_95,bezkrovny_pre_4, vorberger_hydrogen-helium_2007, militzer_massive_2008, redmer_icarus_11,nettelmann_saturn_2013}  and the outer crust of neutron stars \cite{Haensel,daligault_electronion_2009}. Warm dense matter is also thought to exist in the interior of our Earth \cite{hausoel_natcom_17}.
In the laboratory, WDM is being routinely produced via laser or ion beam compression or with Z-pinches, see Ref.~\cite{falk_2018} for a recent review article.
Among the relevant facilities, we mention the National Ignition facility (NIF) at Lawrence Livermore National Laboratory \cite{moses_national_2009,hurricane_inertially_2016}, the Z-machine at Sandia National Laboratory \cite{matzen_pulsed-power-driven_2005,knudson_direct_2015}, the Omega laser at the University of Rochester \cite{nora_gigabar_2015},  the Linac Coherent Light Source (LCLS) in Stanford \cite{sperling_free-electron_2015,glenzer_matter_2016}, the European free electron  laser facilities FLASH and X-FEL in Hamburg, Germany \cite{zastrau_resolving_2014,tschentscher_photon_2017},  and the upcoming FAIR facility at GSI Darmstadt, Germany \cite{hoffmann_cpp_18,tahir_cpp19}.  

The most important technological application of quantum plasmas is inertial confinement fusion \cite{moses_national_2009,matzen_pulsed-power-driven_2005,hurricane_inertially_2016}. In recent experiments at the NIF major breakthroughs were achieved. For the first time the Lawson criterion for ignition was exceeded \cite{abu-icf_prl_22} and later target gain larger than unity was reported \cite{abu-icf_prl_24}.
In these experiments electronic quantum effects are expected to be crucial, primarily during the initial phase of the target implosion. 
Aside from dense plasmas, also many condensed matter systems, such as the electron gas in metals or electron-hole plasmas in semiconductors or quantum materials, exhibit WDM behavior -- if they are subject to strong excitation, e.g. by lasers or free electron lasers~\cite{Ernstorfer1033,PhysRevX.6.021003}, which are topics of high current interest, e.g.~\cite{caruso_jpmat_25}.\\

The goal of this paper is to present an overview on quantum effects and their role in current and future areas of plasma physics. Also, we give an overview on theory and simulation approaches that are of relevance to understand and, ultimately, predict the behavior of such plasmas.\\

 This paper is organized as follows: In Sec.~\ref{s:history} we give an overview on the most important  quantum effects and recall some historical details related to the initial discovery of quantum theory by Max Planck. In Sec.~\ref{s:q-effects} we give an overview on quantum effects in plasmas by analyzing a broad range of temperatures and densities and introducing the key dimensionless parameters. After this we analyze in some detail quantum effects in a large variety of different plasmas. In Sec.~\ref{s:methods} we concentrate on dense quantum plasmas and present an overview on some of the most important experimental techniques and theoretical approaches. To overcome the  limitations each of the methods possesses, we outline, in Sec.~\ref{s:predictive} a strategy how to combine various approaches in order to achieve simulations with predictive capability.

%------------
\section{Why should plasma physicists care about quantum effects?}\label{s:history}
Plasma is traditionally described as the state of matter that is reached after continuous heating: a solid turns into a liquid and a gas and, ultimately thermal energy creates a charged gas of electrons and positive atomic nuclei. At temperatures high enough that the electrons' kinetic energy exceeds the Coulomb binding energy there appears to be no room for quantum effects and no need to worry about complicated quantum experiments and theoretical methods. This expectation may seem to hold for high temperature plasmas, such as in magnetic fusion devices, but even there quantum effects are important. However, the scope of plasma physics has grown tremendously in recent decades and extends also to very dense plasmas the entire behavior of which is governed by quantum effects.

\subsection{Max Planck and the discovery of quantum effects}\label{ss:planck}
German physicist Max Planck has made important contributions to plasma physics. His early works were concerned with thermodynamics and transport in solutions. Well known are also contributions to kinetic theory, such as the Fokker-Planck equation. Also, he presented a solution to the divergency problem of the canonical partition function of Coulomb bound state which is important for the ionization equilibrium of plasmas (Saha equation). This results is known today as Planck-Larkin partition function, e.g., \cite{green-book}. But, of course, his main achievement is the discovery of the Planck constant, of the radiation law and the quantization hypothesis. We briefly discuss Planck's path to these achievements providing some details that are either missing or presented not entirely adequately in standard text books.

Max Planck was born in Kiel in 1858 where he spent the first 9 years of his life and where he later returned for his first professorship in theoretical physics (1885-1889). Planck studied physics in Munich and Berlin with the leading scientists of his time, including Helmholtz and Kirchhoff and encountered a (seemingly) mature theoretical apparatus. After the works of Newton (mechanics), Faraday, Maxwell, and Hertz (electrodynamics), Helmholtz and Claudius (thermodynamics) and many others, theoretical physics was viewed by many as near  ``completion'', with only ``a few dust grains'' left to explore \cite{planck_nat-wis_25}. This view of completeness was shattered just two decades later by Einstein and Planck himself with their discovery of Special Relativity in 1905 and elementary Quantum Theory in 1900, respectively. Here, we concentrate on the latter since it is at the heart of quantum effects in plasmas. Moreover, Planck's black body radiation law is an important diagnostic for plasmas.

When, in the context of the international quantum year 2025, physicists and historians returned to Planck's discovery and its perception, it turned out that many modern textbooks (not just of plasma physics) present a picture that is not entirely correct. The  story that is commonly told is that Planck originally presented a fit between two limiting cases -- the radiation laws of Wien and of Rayleigh and Jeans -- and that his motivation was to avoid the scenario of an ``ultraviolet catastrophe'' -- the divergence of radiation energy at short wavelengths. The problem with these stories is that the Rayleigh-Jeans law was published in 1905 \cite{rayleigh_nat_05}, i.e., five years after Planck's discovery, whereas the term ultraviolet catastrophe was introduced by P.~Ehrenfest only in 1911 \cite{ehrenfest_11}. Therefore, both did not play any role in Planck's discovery.

\subsubsection{The entropy of electromagnetic radiation}
\begin{figure}
    \centering
    \includegraphics[width=0.75\linewidth]{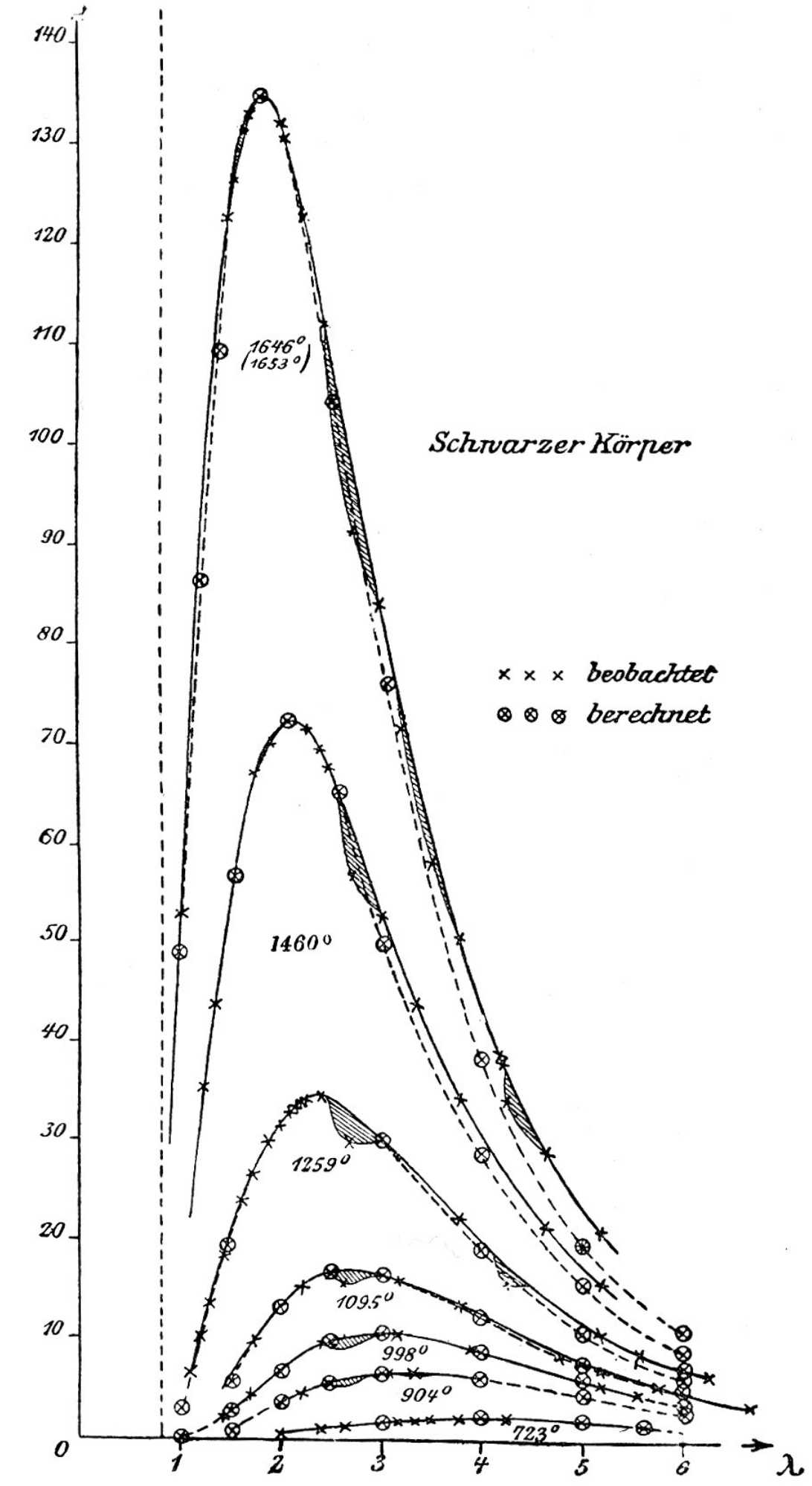}
    \caption{Radiation energy versus wavelength (in $\mu m$) for 7 temperatures. Upper curves (crosses): measurements by Lummer and Pringsheim, lower curves (dashed lines with circles): Wien's theory. From Ref.~\cite{lummer_vhdpg_99}}
    \label{fig:lummer-pringsheim}
\end{figure}

Planck started to work on black body radiation already in the mid 1890s. Then accurate experimental data due to Lummer, Pringsheim, Rubens and Kurlbaum at the Physikalisch-Technische Reichsanstalt  Berlin (PTRB) started to become available. The measurements were accurately reproduced by the radiation law derived by W. Wien \cite{wien_adp_96}. An example is shown in Fig.~\ref{fig:lummer-pringsheim} and confirms the excellent agreement with the measurements, in a broad range of wave lengths and temperatures \cite{lummer_vhdpg_99}. Planck was impressed by Wien's result and tried to substantiate it further by presenting an alternative derivation that was based on thermodynamics. He was convinced that black body radiation was emitted in a state of thermodynamic equilibrium and this state should, therefore, correspond to a maximum of the entropy of electromagnetic radiation, $S(U,V) \to $ max, where $U$ is the internal energy, and the volume $V$ is fixed and can be omitted. Correspondingly, he computed the entropy that corresponds to Wien's energy formula (\ref{eq:energy-wien}) that contains two constants $a$ and $b$, 
\begin{align} 
    U^{\rm W}_\nu(T) &= b \, e^{-a \nu/T}\,,\label{eq:energy-wien}\\
    S^{\rm W}_\nu(U) &=  - \frac{U}{a\nu}\left( \ln \frac{U}{b \nu} - 1\right)\,,\label{eq:entropy-wien}\\
    R^{\rm W}_\nu(U) &= \left( \frac{d^2S}{d U^2}\right)^{-1} = - a \nu U\,,
\end{align}
where the entropy (\ref{eq:entropy-wien}) readily follows from $\frac{dS}{dU} = \frac{1}{T}$,  where $T$ is the temperature and a fixed frequency $\nu$ is considered.
The second derivative governs the stability of the extremum of the entropy and is directly related to the heat capacity of the radiation, $R=-T^2 C_{VN}$.

\subsubsection{Two universal constants}\label{ss:2-constants}

From comparison of Wien's formula with the improved experiments of Lummer and Pringsheim, Planck identified the two unknown constants that have the dimension of action and of the entropy (energy divided by temperature), respectively,
\begin{align}
   b &\to  h = 6.55 \cdot 10^{-27}\mbox{erg s}\,, \label{eq:h-def}\\
   \frac{b}{a} &\to  k_B = 1.346 \cdot 10^{-16}\mbox{erg / grad}\,, \label{eq:kb-def}
\end{align}
which he called elementary quantum of action and Boltzmann's constant, respectively. Since Wien's radiation law contained no dependencies on the material but only on temperature, Planck was convinced that this law and the two constants would have universal validity and importance and would be valid ``...for all times and all, ... even extra-terrestrial, civilizations...'' \cite{planck_padw99}.  Additionally, he also proposed a system of natural units (comprising the universal constants $h$, the speed of light, and the gravitational constant that is commonly called ``Planck units'' and is actively used today in high energy physics). On the other hand,  in 2019 the international society of metrology decided to make Planck's constant $h$ one of the fundamentals of the new international system of units (thereby eliminating the problematic elementary mass standard). The value of the constant has now been fixed for all times to $h=6.626\,070\,15 \cdot 10^{-34}$Ws$^2$. 

The importance of Planck's constant $h$ [which he determined well before finding his radiation law (\ref{eq:energie-planck}) and energy quanta (\ref{eq:energy-quantum})] goes far beyond systems of units. It is evident today, from its appearance in the key formulas of quantum mechanics, atomic and molecular physics, quantum chemistry, nuclear physics as well as of quantum plasmas.

\subsubsection{Planck's first derivation of the radiation law}
\label{ss:planck1}

The accuracy of the measurements of Lummer and Pringsheim, as well as Rubens and Kurlbaum, at the PTRB in 1900, was so high that, aside from the overall good agreement, they also indicated serious deviations  of Wien's theory at large wavelengths (actually at large values of the product $\lambda T$) that are already visible in Fig.~\ref{fig:lummer-pringsheim}. When the accessible wavelengths exceeded 20 $\mu m$ and later 50 $\mu m$, it became obvious that Wien's formula could not be of universal validity, and a new radiation law had to be found.
The story of Planck's original derivation of the correct black body radiation law is known in detail. For example, one of Rubens' students, Georg Hettner, recalls 
\cite{hettner_rubens_22} that, on October 7 1900, Heinrich Rubens and his wife visited the Plancks at their home. There Rubens showed Planck new radiation measurement results for longer wavelengths where Wien's radiation law was clearly failing. But Rubens already had found in the literature a formula that seemed to be in agreement with the new measurements -- the recent result of Lord Rayleigh \cite{rayleigh_pm_00} which is essentially based on the equipartion assumption ($d$ is a constant proportional to $k_B$):
\begin{align}
    U_{\nu}^{\rm R}(T) & = d \cdot T \,,\\
    S_{\nu}^{\rm R}(U) & = d \cdot \ln U\,,\\
    R_{\nu}^{\rm R}(U) & = - \frac{U^2}{d}\,.
\end{align}
Planck was particularly intrigued by the simple analytical dependence of the specific heat ($R$) on the energy in both limiting cases -- linear and quadratic, respectively. Since he knew that, for large $U$, the quadratic dependence should dominate and, for small $U$, the linear one, he just added the two results (without any fitting parameters!) from which he immediately could derive his radiation law as well as the entropy of the radiation that would be valid in the entire range of frequencies:
\begin{align}
     R_{\nu}^{\rm P}(U) & \equiv R_{\nu}^{\rm W}(U) + R_{\nu}^{\rm R}(U) = -\left( a \nu U + \frac{U^2}{d}\right)\,,\label{eq:r-planck}\\
     U_{\nu}^{\rm P}(T) & = \frac{h\nu}{ e^{\frac{a\nu}{T}}-1} = 
     \frac{h\nu}{ e^{\frac{h\nu}{k_B T}}-1} \,,
     \label{eq:energie-planck}
     \\
    S^{\rm P}(U) &= k_B\left[ \left( \frac{U}{h\nu} +1 \right) \ln  \left( \frac{U}{h\nu} +1 \right) - \frac{U}{h\nu}\ln \frac{U}{h\nu}\right]\,.\label{eq:entropie-planck1}     
\end{align}
The result (\ref{eq:energie-planck}) is Planck's famous radiation law which he first reported on October 19 1900 at a meeting of the German Physical Society in Berlin \cite{planck_vdpg00_1}. The experimentalists quickly confirmed that the new formula is in excellent agreement with all their measurements, and this agreement persisted also with later measurements, at even larger wavelengths. The importance of Planck's formula \eqref{eq:energie-planck} can be hardly overrated: in fact Planck discovered Bose-Einstein statistics and the Bose distribution, a quarter century before the advent of quantum mechanics and quantum statistics.

\subsubsection{Planck's second derivation: quantization of the field energy}\label{ss:planck2}
Planck was well aware that the addition procedure (\ref{eq:r-planck}) could not be regarded a strict derivation, even though it apparently led to the correct result. After many attempts he finally found a satisfactory solution. Knowing the correct result (\ref{eq:entropie-planck1}) for the entropy, he used Boltzmann's connection between entropy and the number $Z_{\mu}$ of microstates (partition sum), 
\begin{align}
S =k_B \ln Z_{\mu}\,.
\label{eq:entropy-boltzmann}
\end{align}
In order to count states Planck required a discretization of the total energy $U_N$: he chose a subdivision into a large number $P$ of identical portions $\epsilon$. Further he introduced (a large number) $N$ of identical copies of oscillators (field modes of a fixed frequency $\nu$) that carry the same average energy $U$, leading to
\begin{align}
U_N= P \cdot \epsilon = N \cdot U   \,,
\label{eq:planck-ansatz-diskret}
\end{align}
where $\epsilon$ is still undefined.
He then computed the number of microstates as the number of possibilities to distribute $P$ energy elements across $N$ oscillators,
\begin{align}
        Z_\mu &= \frac{(P+N)!}{P! N!} \approx \frac{(P+N)^{P+N}}{P^P N^N}\,, \label{eq:kombi-mit-wdh}
\end{align}
where Stirling's formula was applied. 
This result immediately yields the microcanonical entropy, Eq.~(\ref{eq:entropy-boltzmann}), of the total system,
\begin{align}
    \frac{1}{k_B}S_N =& \frac{N}{k_B} S = \ln Z_\mu \approx 
    \nonumber\\
    & (P+N)\ln (P+N) - P \ln P - N \ln N\,.    
\end{align}
Finally, the entropy of a single oscillator with a fixed frequency $\nu$ becomes
\begin{align}
    S = k_B \left\{ \left( \frac{P}{N} + 1 \right)\ln \left( \frac{P}{N} + 1 \right) - \frac{P}{N}\ln \frac{P}{N}\right\}\,. \label{eq:planck-entropie2}
\end{align}
This agreed exactly with the already known entropy (\ref{eq:entropie-planck1}), if Planck identified $U/h\nu = P/N$, i.e., $k_B=h/a$ and the elementary energy portion equals 
\begin{align}
\epsilon = h\nu\,.    
\label{eq:energy-quantum}
\end{align}
 This is nothing but the photon energy that is today at the heart of quantum mechanics and quantum electrodynamics. 

This derivation and the result for the energy quanta Planck reported, for the first time, at a session of the German Physical Society in Berlin, on December 14 1900 \cite{planck_vdpg00_2, planck-nobelvortrag}, and this date is now commonly called ``birthday of quantum theory''.

\subsubsection{Discussion of Planck's result}
\begin{figure}[h]
    \centering
    \includegraphics[width=1.02\linewidth]{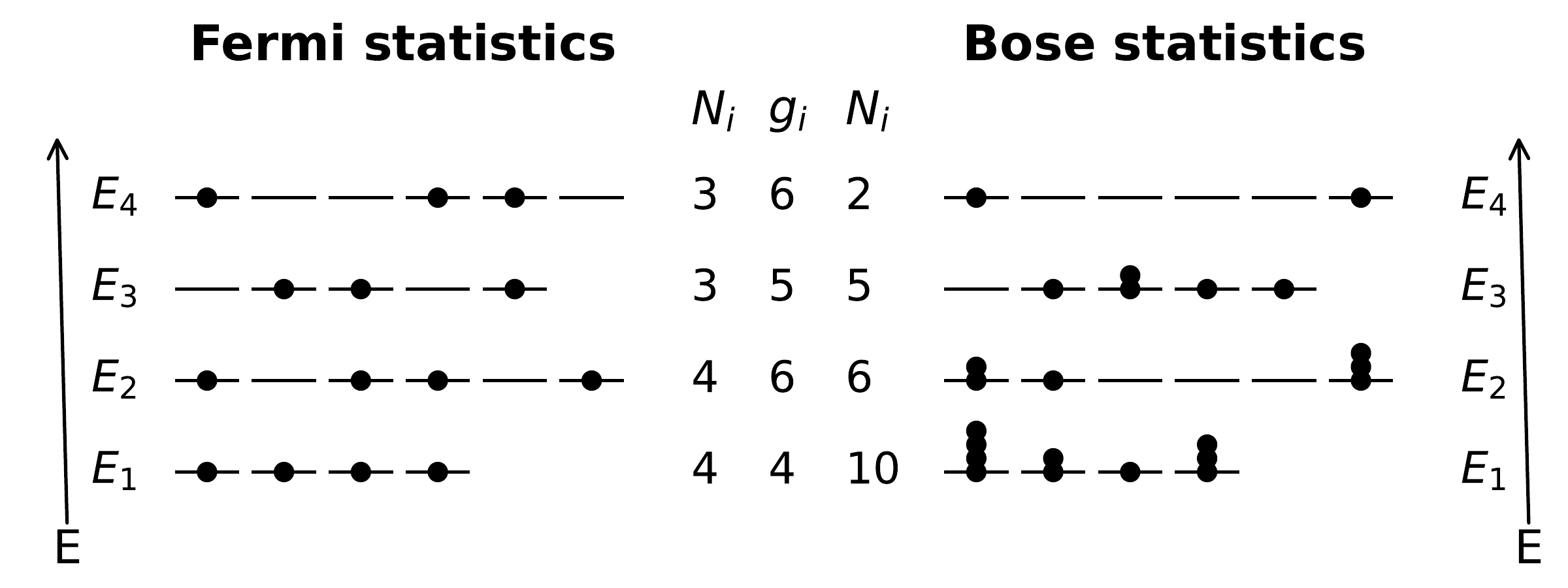}
    \caption{Illustration of Planck's second derivation by comparing Fermi (left) and Bose statistics (right). Each of the four energy levels $E_i$ hosts a total number $N_i$ of particles (in Planck's model $N_i \to P$ and particles [dots] correspond to energy units). Energy $E_i$ is $g_i$-fold degenerate (e.g. four horizontal bars for $E_1$). In the right part, each of the $g_i$ states can host between $0$ and $N_i$ particles, and the number of realizations is given by Eq.~(\ref{eq:kombi-mit-wdh}) which is characteristic for Bose statistics. In contrast, for fermions, due to the Pauli principle, each of the $g_i$ states can only host $0$ or $1$ particles which leads to a completely different partition sum (\ref{eq:z-fermi}) and entropy (\ref{eq:entropy-fermi}). Left and right figures show a possible configuration (microstate).}
    \label{fig:bose-fermi}
\end{figure}

Even though Planck was able to correctly rederive the entropy result (\ref{eq:entropie-planck1}) using his energy discretization procedure, when reading his paper \cite{planck_vdpg00_2}, the used model and Eq.~(\ref{eq:planck-ansatz-diskret}) remain foggy. In fact, in the frame of classical physics there is no such system with $N$ identical copies and the postulated stochastic process to distribute the energy portions. What Planck describes is nothing but the combinatorial stochastic process that is equivalent to Bose statistics. 
To recover the same results from modern quantum statistics, 
Planck's number $N$ of identical copies of an oscillator of energy $E_i$ (frequency $\nu_i$) has to be identified with the degeneracy $g_i$ of that energy, as is illustrated in Fig.~\ref{fig:bose-fermi}.

One easily verifies that the right illustration is a snapshot contributing to Planck's partition sum \eqref{eq:kombi-mit-wdh} and, thus, leads to the entropy \eqref{eq:planck-entropie2} where $P/N \to N_i/g_i$. Since this result is for a thermodynamic state of finite temperature, it has to be understood as the mean occupation of the level (or frequency $\nu$), $N_i/g_i \to n_i(T) $, which is a real number from the interval $[0,N]$. For completeness we also provide the result for fermions (left part of Fig.~\ref{fig:bose-fermi}) which is of direct relevance for electrons in quantum plasmas. The fermionic partition sum for level $E_i$ takes into account the Pauli principle and counts all possibilities to select $N_i$ different elements out of $g_i$
    \begin{align}
        Z_{\mu,i}^F &= \frac{g_i!}{N_i!(g_i-N_i)!}\,. \label{eq:z-fermi}
    \end{align}
Applying the same procedure (Stirling's formula and taking the logarithm) yields the entropy (per state) of an ideal Fermi gas at temperature $T$
    \begin{align}
        \frac{1}{g_i }\frac{S^F_{\mu,i}}{k_B} &= -\left( 1-\frac{N_i}{g_i} \right)\ln \left(1-\frac{N_i}{g_i}\right) - \frac{N_i}{g_i}\ln \frac{N_i}{g_i}\,,\label{eq:entropy-fermi}
    \end{align}
where again the mean occupation occurs, $N_i/g_i \to n_i(T)$ ,which now is restricted to $n_i \in [0,1]$.
In the limit $N_i/g_i \ll 1$, the first term in the bosonic and fermionic entropy, Eqs.~(\ref{eq:planck-entropie2}) and (\ref{eq:entropy-fermi}) vanishes and we recover the entropy of a classical ideal gas.

One may of course dispute whether the term ``birthday'' applies to quantum theory, since Planck's quantization idea was the result of a long process of scientific achievements to which many scientists made important contributions. Moreover, the quantization result was strictly rejected by most of the physicists of Planck's time since it was in striking conflict with Maxwell's celebrated theory of electromagnetism. 
Even Planck himself viewed his equipartition of the radiation energy rather a mathematical trick to compute the partition sum and the entropy \footnote{Some historians argue that Planck even questioned the equipartition, since he wrote in Ref.~\cite{planck_vdpg00_2} that if the ratio $P/N = U_N/\hbar\nu$  is not an integer, one should choose for $P$ the nearest integer. However, in the entropy formula (\ref{eq:planck-entropie2}) there is no need at all that this ratio is an integer. In fact, the theory assumes a finite temperature of radiation. A modern derivation of the entropy of an ideal Bose gas leads exactly to Planck's result when one replaces $P/N \to n_{\nu}(T)$, where $n_{\nu}$ is the mean occupation number of state (mode) $\nu$ which is a non-negative real number.}. For more than 10 years he would not see the justification that the energy quanta are an intrinsic physical property of radiation. And the Nobel committee as well was skeptical about Planck's result and decided to award the Nobel prize in 1911 to Wilhelm Wien ``for his discoveries regarding the laws governing the radiation of heat''. Even though it was known that Planck's formula reproduces the experiments far better, Wien's contributions to the theory, at that time, were viewed as the primary ones.

It took the work of many scientists to demonstrate the fruitfulness of the photon concept: most importantly Einstein's explanation of the photoelectric effect and Bohr's model of the hydrogen atom, involving the postulate of radiation-free electron orbits that are selected by quantizing the electron action in units of Planck's constant. Finally, in 1919 Max Planck was awarded the Nobel prize (for the year 1918) ``in recognition of the services he rendered to the advancement of Physics by his discovery of energy quanta''.
In his presentation speech the President of the Royal Swedish Academy of Sciences, Dr. A.G.~Ekstrand, emphasized:
``...Planck’s radiation theory is, in truth, the most significant lodestar for modern physical research, and it seems that it will be a long time before the treasures will be exhausted which have been unearthed as a result of Planck’s genius.'' The development of physics during the century following this award impressively confirmed these visionary words.

\subsection{Quantum effects}\label{ss:q-effects}
There exists a large variety of quantum effects that are relevant for plasmas. We start with those related to single particles and then consider quantum effects arising in an ensemble of many particles.
\subsubsection{Single-particle quantum effects}\label{ss:1-particle-qe}
\begin{figure}[h]
    \centering
    \includegraphics[width=0.95\linewidth]{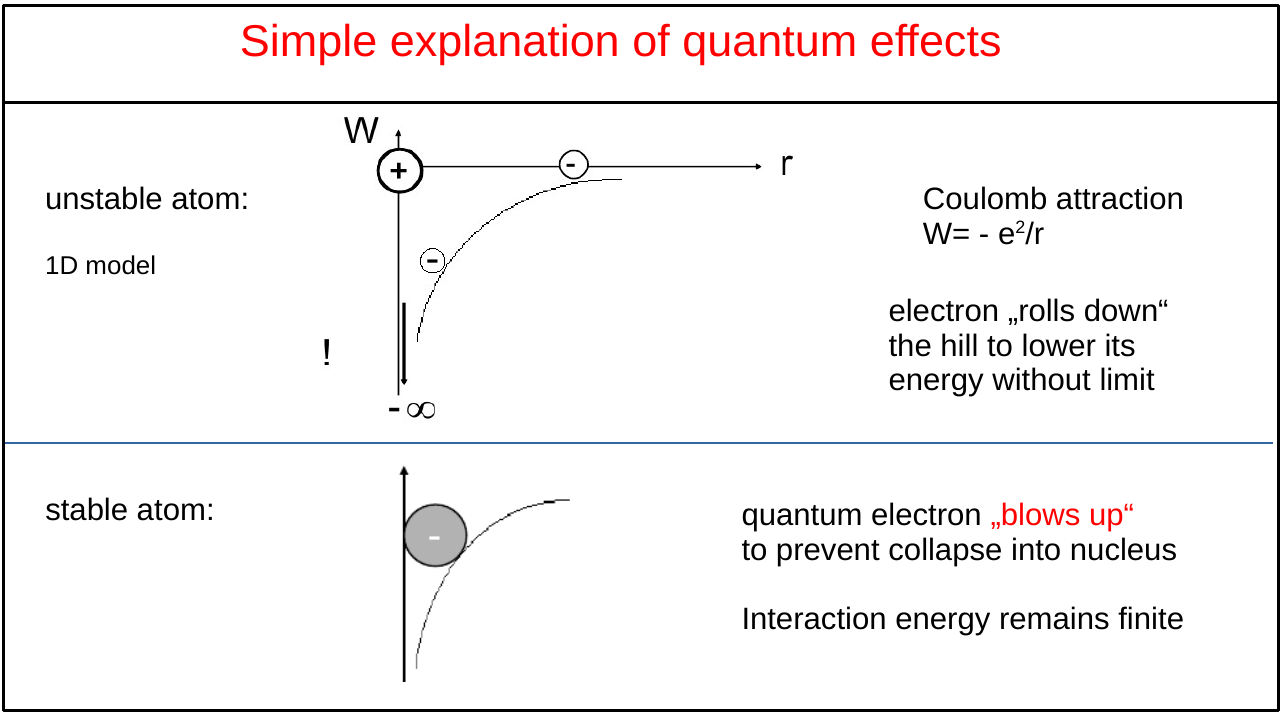}
    \caption{Illustration of the key quantum effect -- spatial delocalization -- for the example of an atom. A classical point particle (electron) would unavoidably collapse into the nucleus, as this lowers its energy $W$ (top figure). This is in contrast to the known stability of atoms. Nature provides a simple solution (bottom): during its approach of the nucleus the electron increases its size (grey circle), giving rise to a finite value of the interaction energy $W$. In quantum mechanics the finite extension is connected with a statistical interpretation.}
    \label{fig:q-effects}
\end{figure}

The first group of effects are those encountered by a single particle such as tunneling through a barrier or interference of the particle with itself, e.g., at a double slit, a grating, or crystal plane. The key property that is at the heart of all these effects is spatial delocalization, i.e., the finite spatial extension of a quantum particle, as opposed to the semiclassical notion of a point-like microparticle as is common in electrostatics  or plasma theory (point charge). However, a pointlike electron cannot form a stable atom as suggested by the ``planetary model'', due to radiation losses. Also Bohr's postulates of radiation-free special orbits do not present a satisfactory solution. The key problem in this classical picture is the assumption of point particles, and the problem immediately disappears if the electron is allowed to acquire a finite spatial extension which is illustrated in Fiq.~\ref{fig:q-effects}.

\begin{figure}[h]
    \centering
    \includegraphics[width=0.995\linewidth]{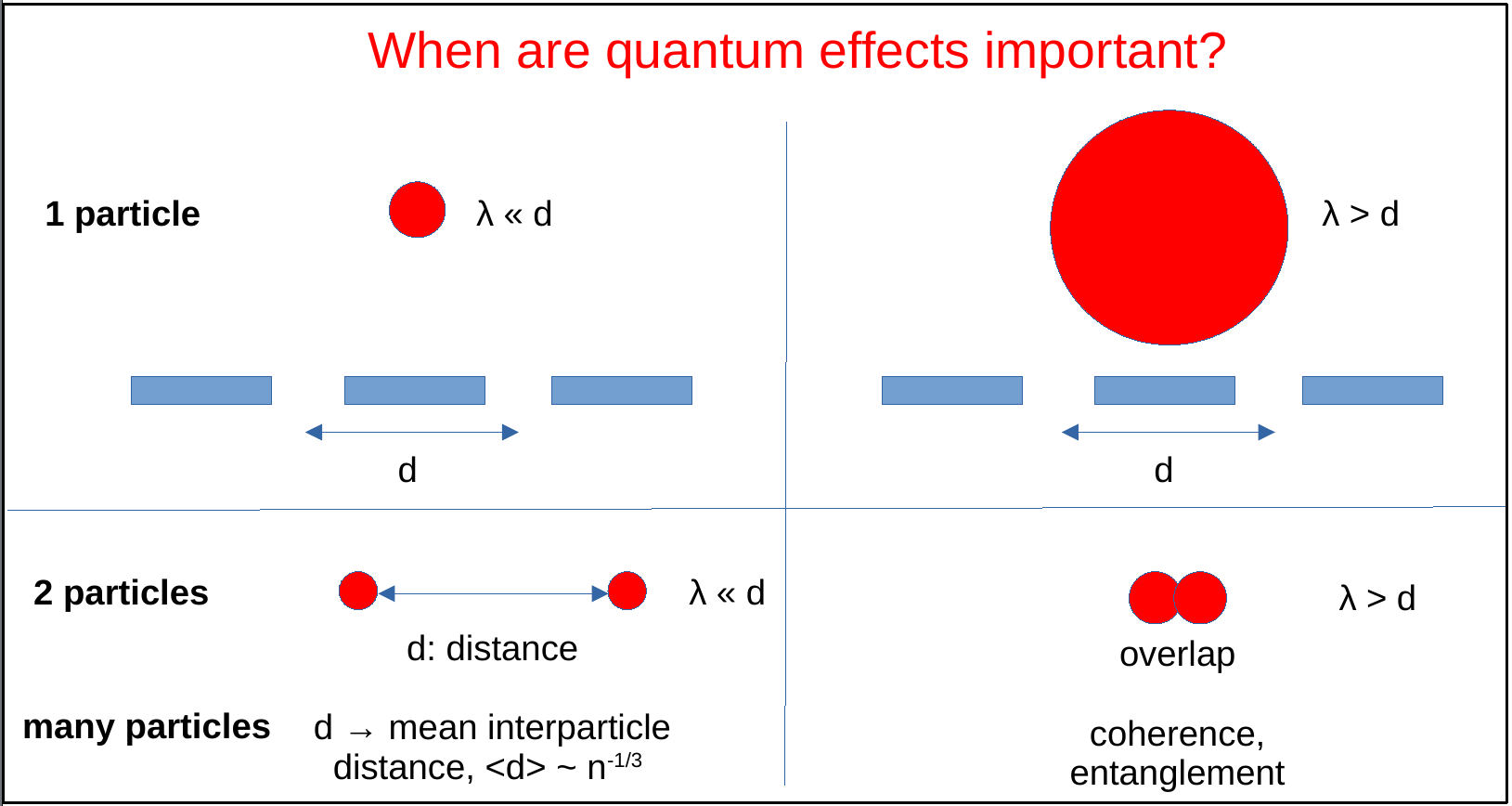}
    \caption{Illustration when quantum effects are relevant. Top: for one particle encountering an obstacle (a double slit) with extension $d$ quantum effects will be relevant when the quantum extension $\lambda$ exceeds $d$, as in the right picture giving rise to diffraction and interference. For two particles at a distance $d$, quantum effects will dominate if $\lambda$ exceeds $d$ resulting in coherence and entanglement. Finally, for many particles (bottom), $\lambda$ has to be compared to the statistical mean of the interparticle distances which scales with the density as $\langle d\rangle \sim n^{-1/3}$.}
    \label{fig:q-effects-importance}
\end{figure}

The existence of a finite extension $\lambda$ immediately explains other elementary quantum effects, such as diffraction and interference. Figure~\ref{fig:q-effects-importance} illustrates that the appearance of these effects depends on the ratio of the particle size, $\lambda$, to the relevant geometrical dimension, $d$, such as the distance of two slits (top part). Interference will be observed if $\lambda \gtrsim d$ (right part). Similarly, a finite spatial extension allows a quantum particle to penetrate into a potential barrier (tunneling). Of course, it is a separate question to compute or measure the extension $\lambda$. In many cases, such as for non-interacting particles in thermodynamic equilibrium, reasonable approximations for the mean extension can be derived leading to the de Broglie wavelength (\ref{eq:lambda-def}) and to a dimensionless degeneracy parameter (\ref{eq:chi-def}) characterizing the importance of quantum effects.

So far we have used qualitative arguments to explain important single-particle effects in terms of a finite size $\lambda$. Quantum mechanics, such as the Schrödinger equation
\begin{align}
i\hbar \frac{\partial}{\partial t} \psi(\textbf{r},t)
    = -\frac{\hbar^2}{2m}\nabla^2 \psi(\textbf{r},t) + W(\textbf{r},t)\psi(\textbf{r},t)
    \label{eq:schroedinger}\,,
\end{align}
yields not only the size but also the precise shape of a quantum particle expressed in terms of the probability density, $|\psi(\textbf{r},t)|^2$, and its time evolution. Returning to the example of an electron in the field of a positive nucleus, cf. Fig.~\ref{fig:q-effects}, the collapse into the attractive Coulomb potential W is prevented by the first term on the right: the quantum kinetic energy depends on the shape of the wave function (essentially, the curvature of the probability density). Strong spatial localization would give rise to an increase of curvature and of kinetic energy and is energetically not advantageous. The actual shape of $\psi(\textbf{r},t)$ is a compromise between the expanding and contracting trends provided, respectively, by the first and second terms on the right. 

\subsubsection{Many-particle quantum effects}\label{ss:many-particle-qe}
\begin{figure}[h]
    \centering
    \includegraphics[width=0.995\linewidth]{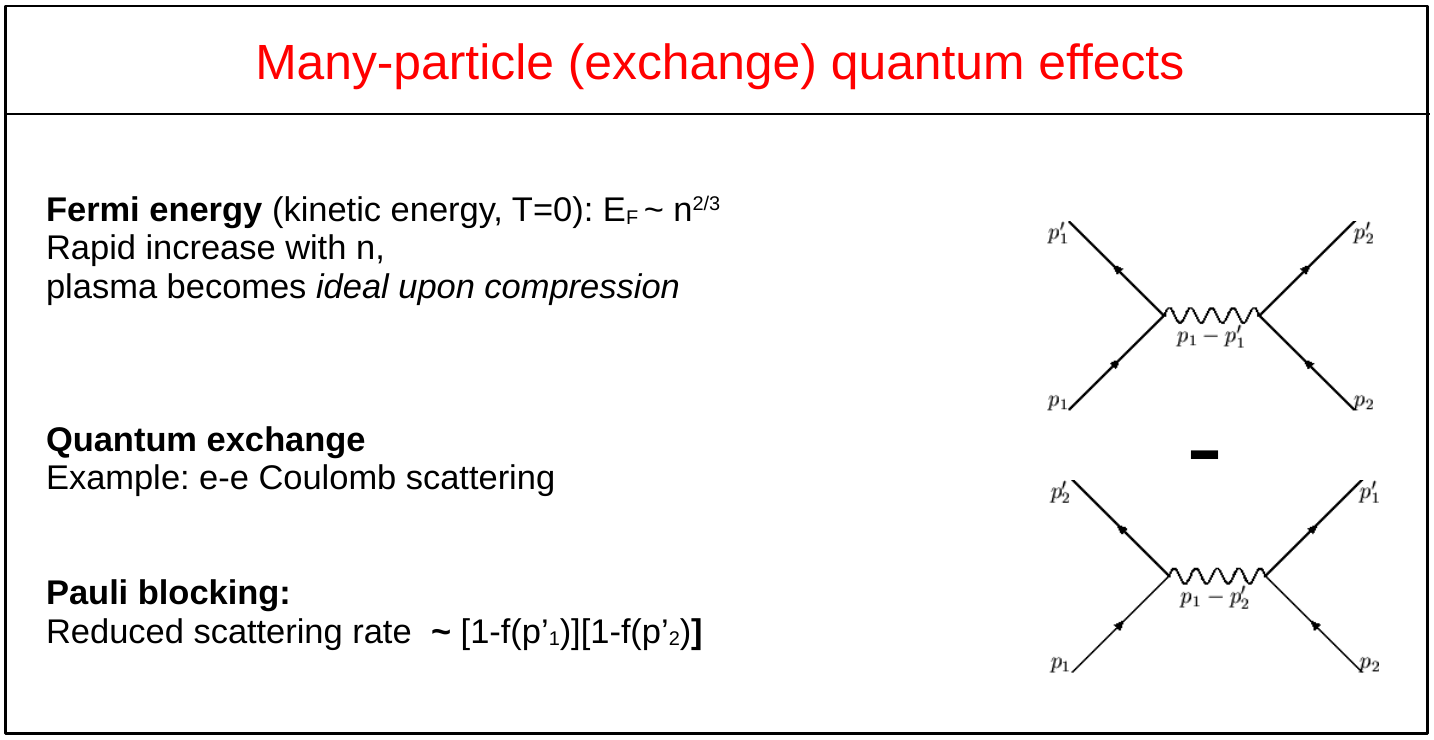}
    \caption{Illustration of N-particle quantum effects that arise from spin statistics of fermions. In the top left part we indicate the strong density dependence of the Fermi energy that leads to a decrease nonideality effects with density, cf. Sec.~\ref{ss:qp_parameters}. The top right graphic shows a scattering process of two electrons entering from below with momenta $p_1$ and $p_2$ and exiting with $p'_1$ and $p'_2$. Quantum exchange gives rise to an additional (negative) scattering contribution where $p'_1$ and $p'_2$ are exchanged. Finally, the probability of a scattering process (scattering rate) is reduced by Pauli blocking.}
    \label{fig:q-effects-statistics}
\end{figure}

Plasmas usually contain a macroscopic number of particles what gives rise to additional quantum effects. Since in almost all cases of current interest in plasma physics, quantum effects of  electrons (fermions) are important, we will focus on particles with half integer spin below. We already encountered Fermi statistics in Fig.~\ref{fig:bose-fermi} and noted the dramatic differences to the behavior of bosons, such as photons. Since fermions are characterized by an anti-symmetric N-particle wave function each single-particle state can be occupied by not more than 1 particle. This is the Pauli principle and, at zero temperature, the mean occupation of an orbital with energy $E_i$ is given by the step function $n^{\rm F}(E_i)=\Theta(E_F-E_i)$. The highest occupied orbital has the energy $E_F$ -- the Fermi energy, Eq.~(\ref{eq:ef-def}). Since additional particles can only be added into higher energy states, this energy increases rapidly with density, 
$E_F \sim n^{2/3}$, cf. Fig.~\ref{fig:q-effects-statistics}. This scaling has dramatic consequences for the density and temperature dependence of the coupling strength of quantum plasmas that we discuss in Sec.~\ref{ss:qp_parameters}.

Another important example is the effect of Fermi statistics on electron-electron scattering, which is crucial for transport properties and the thermalization of a quantum plasma. A two-particle scattering process is sketched in the top part of Fig.~\ref{fig:q-effects-statistics}: two electrons enter the scattering event with momenta $p_1$ and $p_2$, respectively and exit with $p'_1$ and $p'_2$. However, since the electrons are indistinguishable, there appears another scattering contributions (which is negative for fermions) where the electrons in the exit channel are exchanged, cf. sketch in the bottom of Fig.~\ref{fig:q-effects-statistics}. Finally, the probability of the scattering event is also directly affected by additional electrons: if the final states are already occupied by other electrons with probabilities $f(p'_1)$ and $f(p'_2)$, the scattering rate is reduced by a factor $[1 - f(p'_1)] [1- f(p'_2)]$. In particular, in the ground state ($T=0$), $f(p) \to n^{\rm F}$, and scattering into states below the Fermi energy is completely suppressed, for more details, see Ref.~\cite{bonitz_qkt}.

\section{Quantum effects in plasmas}\label{s:q-effects}
After the qualitative discussion of important quantum effects in Sec.~\ref{s:q-effects}, we now turn to a quantitative characterization of these effects in Sec.~\ref{ss:qp_parameters}. After this we give a brief overview on different areas of plasma physics and the role of quantum effects for them. We start with low-pressure plasmas, in Sec.~\ref{ss:low-p}, followed by high-density plasmas in Sec.~\ref{ss:high-p}. We conclude by briefly discussing quantum effects of the heavy plasma particles, in Sec.~\ref{ss:heavy-particles}, and unconventional quantum plasmas such as condensed matter systems and the quark-gluon plasma, in Sec.~\ref{ss:unconventional}.
  
  \subsection{Parameters of quantum plasmas}\label{ss:qp_parameters}
\begin{figure}[t]
\includegraphics[width=0.48\textwidth]{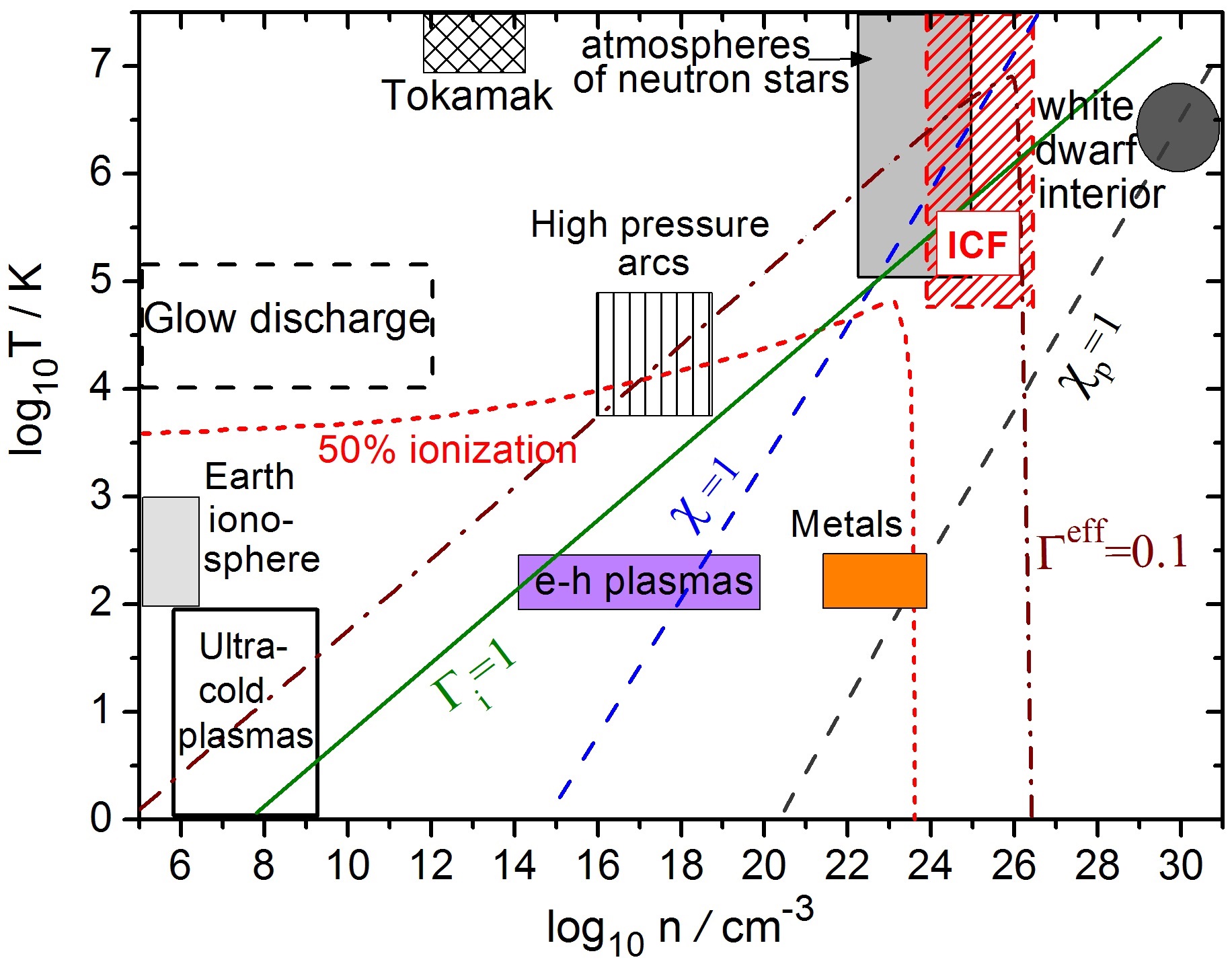}
\caption{Density-temperature plane with examples of plasmas and characteristic plasma parameters. ICF  denotes inertial confinement fusion. Metals (semiconductors) refers to the electron gas in metals (electron-hole plasma in semiconductors). Weak electronic coupling is found outside the line $\Gamma^{\rm eff}=0.1$, cf. Eq.~(\ref{eq:gamma-eff}). Electronic (ionic) \textit{quantum effects} are observed to the right of the line  $\chi = 1$ ($\chi_{\rm p} = 1$), cf. Eq.~\eqref{eq:chi-def}. The coupling strength of quantum electrons increases with $r_s$ (with decreasing density). Atomic ionization due to thermal effects (due to pressure ionization) is dominant above (to the right of) the red line, $\alpha^{\rm ion}=0.5$,  for the case of an equilibrium hydrogen plasma~\cite{Sheffield}). The values of $\chi_p$ and $r_s$ refer to the case of hydrogen. Figure modified from Ref.~\cite{bonitz_pop_19}.
}
\label{fig:0}
\end{figure}

Let us recall the basic parameters of dense quantum plasmas and warm dense matter \cite{bonitz_qkt,bonitz_pop_19,ott_epjd18}: 
the first are the electron degeneracy parameters 
\begin{align}
\Theta & = \frac{k_B T}{E_{F}}  \label{eq:theta-def}  \,,\\
\chi & = n\Lambda^3,  \label{eq:chi-def}
\end{align}
  that contain characteristic length and energy scales -- the thermal de Broglie wave length, and the Fermi energy of electrons (in 3D), respectively: 
\begin{align}
 \Lambda &=\frac{h}{\sqrt{2\pi m k_B T}}\,, \label{eq:lambda-def}\\
    E_{F} &= \frac{\hbar^2}{2m}(3\pi^2 n)^{2/3} = k_B T_F\,, 
    \label{eq:ef-def}
\end{align}
where $n$ and $T$ are the electron density and temperature.
Quantum degeneracy effects of the ions are important only in plasmas that are very strongly compressed, cf. Sec.~\ref{ss:heavy-particles}. The reason is that the degeneracy parameter of the ions
is a factor $(m T/m_iT_i)^{3/2}$ smaller than the one of the electrons.
The second important parameter in dense plasmas is the classical coupling parameter of ions 
\begin{align}\label{eqn:gammai}
\Gamma_i= \frac{Q_i^2}{a_ik_BT_i}\,,    
\end{align}
 where $Q_i$ is the ion charge, and $a_i$ is the mean inter-ionic distance.
Further, the quantum coupling parameter (Brueckner parameter) of electrons in the low-temperature limit is,
    \begin{equation}
    r_s = \frac{a}{a_B}, \qquad a_B = \frac{\hbar^2\epsilon_b}{m_r Q_i e},  
    \label{eq:rs-def}
    \end{equation}
    where $a=(4/3\pi n)^{-1/3}$ denotes the mean distance between two electrons,  $a_B$ is the Bohr radius, and $m_r=m m_i/(m+m_i)$ and $\epsilon_b$ are the reduced mass and background dielectric constant, respectively. For plasmas, $m_r \approx m$, $\epsilon_b=1$, and $a_B=0.529\angstrom$. Alternatively, the coupling strength in the degenerate limit, is defined by
    \begin{align}
        \Gamma^2_q = \frac{(\hbar \omega_{pl})^2}{E^2_F} = r_s \cdot \frac{16}{9\pi}\left(\frac{12}{\pi} \right)^{1/3} \approx 0.88 \cdot r_s \,. 
    \end{align}
Sometimes it is useful to introduce an effective coupling parameter that interpolates between the classical and strongly degenerate limits,
\begin{align}
    \Gamma^{\rm eff} = \frac{e^2/a}{\left[(k_BT)^2+E_F^2\right]^{1/2} } = \frac{e^2}{ak_BT}\frac{1}{\left( 1 + \Theta^{-2} \right)^{1/2}}\,.
    \label{eq:gamma-eff}
\end{align}
A rough estimate for the boundary between ideal and nonideal plasmas is the line $\Gamma^{\rm eff} =0.1$ that has been included in Fig.~\ref{fig:0}.
Finally, the degree of ionization of the plasma -- the ratio of the density of free electrons, $n$, to the total (free plus bound) electron density -- 
\begin{align}
\alpha^{\rm ion} = \frac{n}{n_{tot}}\,,
\label{eq:alpha-ion}
\end{align}
determines how relevant plasma properties are compared to neutral gas or fluid effects.

The parameters $\chi, \chi_i$ and $\Gamma_i$ are shown in Fig.~\ref{fig:0} where we indicate where these parameters equal one.   
Note that the classical coupling parameter, $\Gamma_i$, increases with density whereas the quantum coupling parameters, $r_s$ and $\Gamma_q$, decrease with $n$. We underline that the parameters $a_B, E_{F}, \Lambda, \theta$ contain the density of free electrons and $\Gamma_i, \chi_i$ the density of free ions. This means the lines of constant $\Gamma_i, \chi, \chi_i, r_s$ shown in Fig.~\ref{fig:0} refer to the free electron (ion) density. 

In cases when the plasma is only partially ionized the free electron density has to be replaced by $n\to \alpha^{\rm ion}\times n$. 
The degree of ionization decreases when the temperature is lowered, according to the Saha equation, $\alpha^{\rm ion} \sim e^{-|E_b|/k_BT}$, where $E_b$ denotes the binding energy of the atom, and in Fig.~\ref{fig:0} we indicate the line where a  hydrogen plasma has a degree of ionization of $0.5$. Qualitatively, a quantum plasma is found to the right of this line. Figure~\ref{fig:overview} shows a zoom into the warm dense matter range and also contains lines of constant $r_s$- and $\Theta$-values. 

\subsection{Quantum effects in low-temperature and low-pressure plasmas}\label{ss:low-p}
Let us start the overview with low-pressure plasmas where not only the heavy particles are classical, but also the degeneracy parameter of the electrons, Eq.~(\ref{eq:chi-def}) is much smaller than one. Then, electronic quantum effects, as discussed in Sec.~\ref{s:q-effects}, are irrelevant. However, even for such plasmas quantum effects may be of high relevance, as we will briefly discuss below.

\subsubsection{Plasmas containing atoms and molecules}\label{ss:atoms-molecules}
We first consider low-temperature plasmas with $k_BT \ll |E_b|$, i.e., most of the electrons are bound in atoms and molecules, and the degree of ionization is typically low, $\alpha^{\rm ion} \lesssim 0.1$. These plasmas are of high importance for many technological applications ranging from atomic layer deposition, lithography, lighting and treatment of surfaces and biological tissues. Even applications in medicine (``plasma medicine'') have been convincingly demonstrated, see Refs.~\cite{Kuchenbecker_2009,graves_2009_molecular,Kong_2009} and references therein.

While direct measurements of plasma density and temperature are sometimes difficult in these plasmas, accurate diagnostics are possible using absorption and emission spectroscopy. They are completely relying on the quantum properties of atoms and molecules and require detailed knowledge of their ground and excited states. 
Another example of quantum effects are radiation and luminescence of the plasma. There is extensive work on plasma spectroscopy, e.g. Refs.~\cite{Fantz_2006,hutchinson_cup_02}. Spectral methods also allow one to diagnose the electric field strength by observing the modification of spectral lines (Stark effect). Very strong fields may lead to ionization of atoms and molecules via tunnel or field ionization [see also Sec.~\ref{ss:relativistic}] and, thus, to a change of the degree of ionization. Similar effects are observed at high density, due to many-particle effects which will be discussed in the context of the Saha equation in Sec.~\ref{ss:fpimc-saha} 

\begin{figure}
    \centering
    \includegraphics[width=0.65\linewidth]{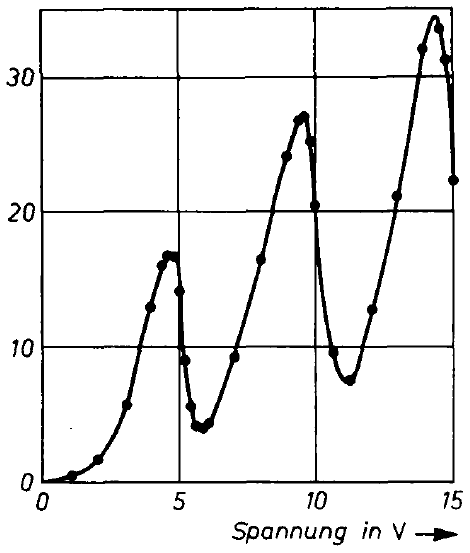}
    \caption{Current-Voltage characteristic in the original Franck-Hertz experiment. Measurements were made with a gas tube filled with mercury vapor. From Ref.~\cite{franck-hertz-14}.}
    \label{fig:franck-hertz-ui}
\end{figure}

A key quantum effect in non-thermal low pressure plasmas is electron impact ionization or excitation of atoms. 
The pioneering experiment was conducted 1914 by James Franck and Gustav Hertz and provided one of the first direct experimental proofs for the existence of quantized energy levels in atoms and earned them the Nobel prize in 1925. Their experiment was based on bombarding atoms by electrons and detecting the kinetic energy loss of the scattered electrons. 
In particular, electrons were accelerated through an evacuated tube filled with mercury vapor. As the accelerating voltage increased, the measured current also increased, until it suddenly dropped at certain voltages (in this case $4.9$ eV and multiples thereof), see Fig.~\ref{fig:franck-hertz-ui}. These drops occur because electrons collide with mercury atoms and lose kinetic energy in an inelastic collision that excites an atom. Franck and Hertz originally thought that electrons would ionize the mercury atom but it soon was shown by Davis and Goucher \cite{davis_pr_1917} that this occurs only at $11.4$ eV. Instead of ionizing the atom, the inelastic electron collision leads to an excitation between discrete atomic levels (from the $6^1 S_0$ state to the $6^3P_1$ state), in qualitative agreement with Bohr's atom model. Obviously, one would expect that upon de-excitation the atom should emit radiation in the UV range ($\lambda \approx 253$ nm) which was observed a few years later.

\begin{figure}
    \centering
    \includegraphics[width=0.3\linewidth]{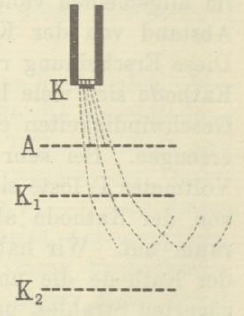}
    \includegraphics[width=0.65\linewidth]{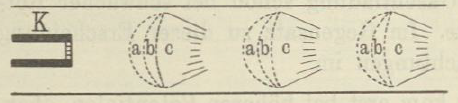}
    \caption{\textbf{Left}: Electrons are emitted from cathode K and are accelerated towards anode A and decelerated again by negative grids K$_1$ and K$_2$. As a result electrons even turn back (dashed lines). Due to the spread in kinetic energy different atomic transitions occur, accompanied by emission of different colors. \textbf{Right}: Glowing gas pattern (three spatially separated curved striations). Figure from Ref.~\cite{seeliger-12}.}
    \label{fig:seeliger}
\end{figure}

A strong simplification in the common explanation of Fig.~\ref{fig:franck-hertz-ui} is that the electron density in the gas tube is uniform and all regions contribute equally to the measured current. In contrast, a 
frequently observed feature in gas discharges is a spatial modulation of the atomic emission. For example, already in 1912 Gehrcke and Seeliger observed a non-monotonic current-voltage characteristic that is accompanied by a spatial modulation of the emission \cite{seeliger-12}. Figure~\ref{fig:seeliger} is from their paper and sketches the arrangement of the electrodes and grids (left) and the formation of several slightly curved localized emission areas (striations, right).

\begin{figure}
    \centering
    \includegraphics[width=0.995\linewidth]{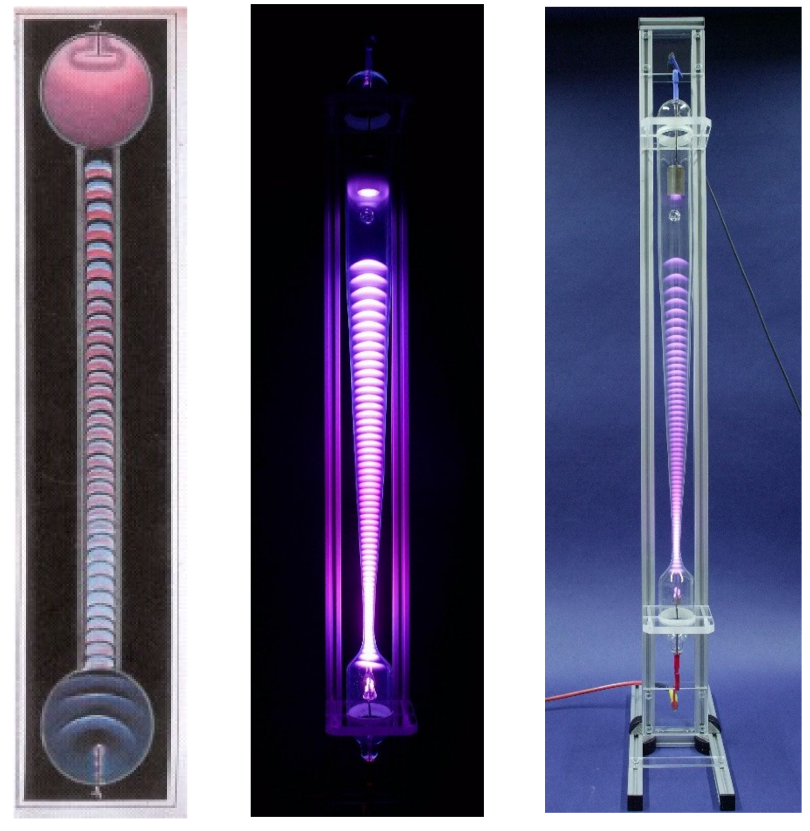}
    \caption{Standing striations in a gas discharge. Left: historic drawing from Ref.~\cite{meyer_1858}. Center and right: photos from a low-pressure hydrogen glow discharge. The cathode is at the top and the anode at the bottom of the glass tube. Experiment by the group of H. Kersten, photo by U. Haeder.}
    \label{fig:striations}
\end{figure}

Striations are fascinating emission phenomena in low-pressure plasmas, for a colorful view, see Fig.~\ref{fig:striations}.
These structures arise due to non-uniform ionization and excitation processes, as a result of electron energy changes and space charge effects. The electrons are accelerated in the electric field on their way between cathode and anode. With the kinetic energy they gained in the field they can excite or ionize gas atoms by where they lose energy and have to be accelerated again. This mechanism creates repeating spatial zones, e.g. high excitation results in bright regions and low excitation in dark regions. Due to the different mobility of electrons and ions space charge regions of excess of ions or electrons, respectively, are formed, generating local electric field variations, that reinforce the pattern. The striations can even  move through the plasma (e.g. ionization waves) or give rise to plasma oscillations \cite{kewitz_10}. 

It is interesting to point out that striations in gas discharges have been documented for as long as almost 200 years. Apparently, the first observation of these interesting phenomena were mentioned in 1843 by M. Abria  \cite{abria_1843}.  A particularly prominent example is the interesting study by Th. Meyer (an assistant of J. Plücker in Bonn) who published in 1858 a booklet entitled ``Observations about the layered electrical light and about the strange influence of a magnet on them'' \cite{meyer_1858} that contains detailed colorful drawings of the observations, cf. left part of Fig.~\ref{fig:striations} that are strikingly similar to those measured in recent years. Indeed, the center and right panels of Fig.~\ref{fig:striations} shows experimental observations of striations in a hydrogen discharge performed at Kiel university by H. Kersten and co-workers.

\begin{figure}
    \centering
    \includegraphics[width=0.995\linewidth]{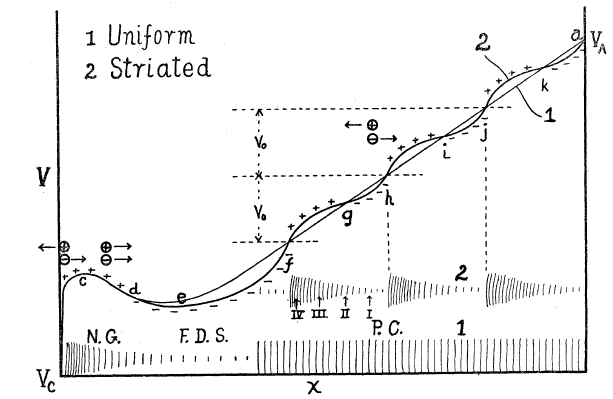}
    \caption{Electric potential distribution along the discharge tube axis.
    Line 2: Theoretical potential distribution causing striations, compared to Line 1: a uniformly increasing potential (in the absence of striations). The exciting inelastic electron collisions occur at points f, h and j. The arrows indicate the direction of average drift of ions in various parts of the discharge and the roman numbers indicate the structures of strata. (N.G. – negative glow, F.D.S. – Faraday dark space, P.C. – positive column). From Ref.~\cite{compton_pr_24}.}
    \label{fig:compton-striations}
\end{figure}

After many accurate observations and qualitative descriptions of striated discharges, in the 20th century also many papers have been published that were devoted to a theoretical explanation and (quantitative) modeling of the striations. An early explanation was given by K.T. Compton \textit{et al.} \cite{compton_pr_24}, who discussed the theoretical potential distribution in the case of a long discharge tube and the effect of local inelastic electron impacts and the average directed drift of ions in various parts of the tube. Figure~\ref{fig:compton-striations} illustrates their explanation.
For a review on the Franck-Hertz experiment on the occasion of its 100th anniversary and further references, see Ref.~\cite{robson_2014}.

To summarize this section, low pressure non-thermal plasmas have played a key role for the understanding of the structure of atoms (Franck-Hertz experiment). Moreover, these plasmas host beautiful discharge patterns but, at the same time, are also striking visual manifestations of quantum effects in the structure of atoms and molecules, as well as electron-neutral collisions.

\subsubsection{Interface of plasmas and solids}\label{ss:psi}
\begin{figure*}
    \centering
\includegraphics[width=0.85\linewidth]{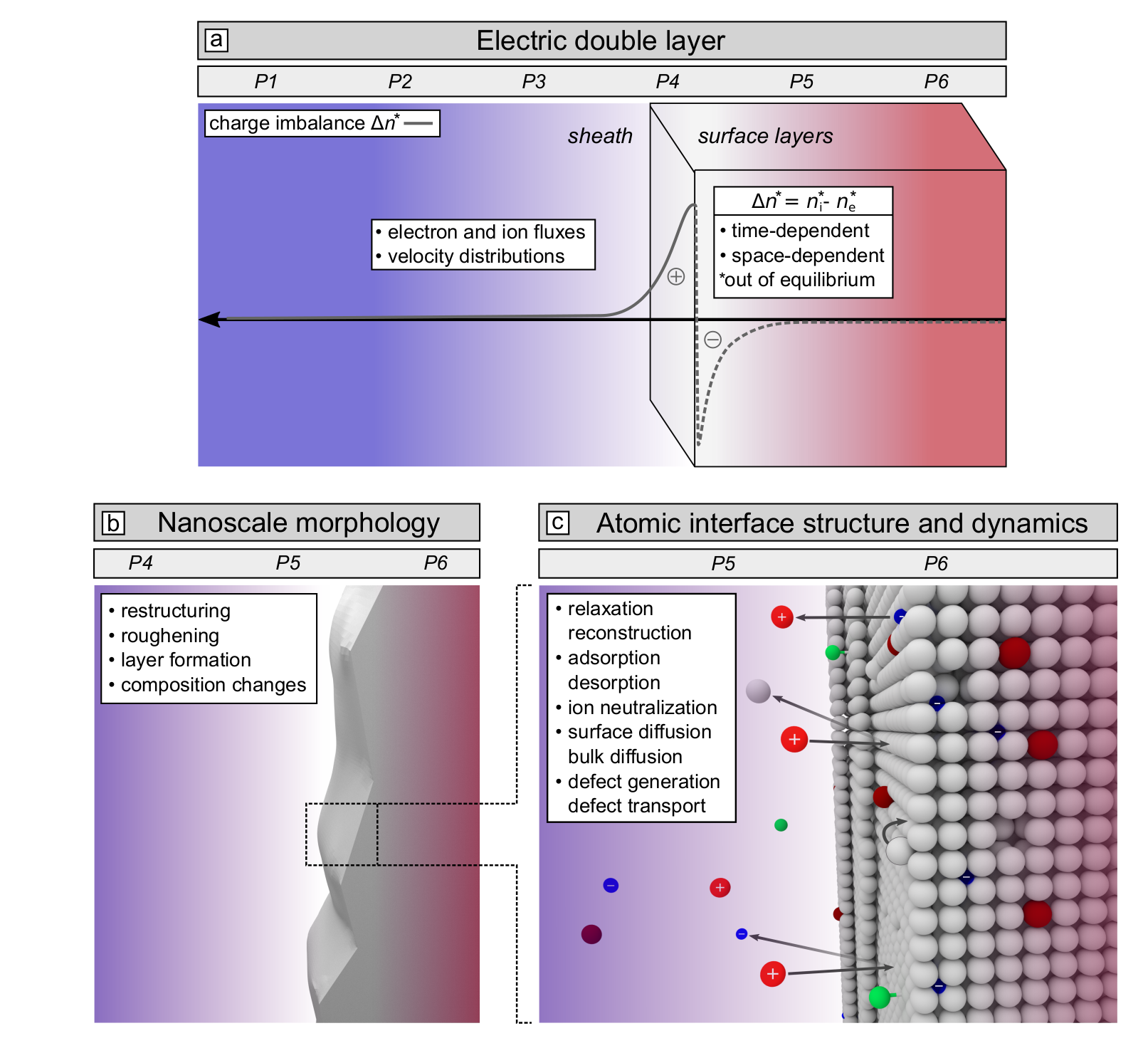}
    \caption{Sketch of quantum effects at the interface between a low-temperature plasma and a solid.
    The physical processes at the plasma-solid interface are shown --from the largest to the smallest length scale. \textbf{Top}: The electric double layer (on the scale of the Debye length, on the plasma side, and a few nanometers, in the solid) resulting from electron depletion in the plasma sheath is characterized by the local difference of the nonequilibrium ion and electron densities and is accompanied by electron accumulation in the solid which is influenced by the processes in figure parts b) and c). \textbf{Bottom left}: on the scale of the surface roughness (typically nanometers) the surface exhibits local variations of the morphology and chemical composition. \textbf{Bottom right}: atomic scale modification of the surface and the plasma sheath caused by individual particle impacts, charge transfer, chemical reactions etc. 
  The relevant processes are indicated inside the figure parts.
    Figure taken from Ref.~\cite{Bonitz_fcse_19} with the permission of the authors.}
    \label{fig:plasma-interface}
\end{figure*}
Plasmas in the laboratory are bounded typically by condensed matter systems -- vessel walls, electrodes etc. Solids are governed by quantum effects, so these effects naturally influence the plasma, primarily in the vicinity of the surface. Therefore, understanding
plasma-surface interaction is important for predicting the plasma behavior \cite{graves_2009_molecular,neyts2017molecular,Bonitz_fcse_19}. 
There is a large variety of such processes that are sketched in  Fig.~\ref{fig:plasma-interface}. 
On large scales (top panel), the surface influences the plasma sheath which is characterized by an access of positive charge. The missing electrons partly accumulate in the top layers of the solid forming a double layer \cite{heinisch_2011_physisorption,heinisch_2012_electron,bronold_2015_absorption}. Zooming into the surface (bottom left) the top layers of the solid exhibit distortions due to the impact of plasma particles. This changes the sticking behavior of electrons and ions to the surface which, in turn, affects the plasma sheath. To properly understand this interaction between plasma and surface requires to zoom in further (bottom right panel) and to resolve the atomic processes that are initiated by the impact or electrons or ions onto the surface. This involves a microscopic treatment of the neutralization of ions \cite{marbach_2012_resonant,balzer_cpp_21} and of the energy loss of projectiles in the surface \cite{balzer_prb16}. A question of recent interest is the interaction of plasma particles with quantum materials, such as monolayers of graphene or transition metal dichalcogenides. Here correlation effects in the target have a profound influence on the interaction with the plasma particles \cite{balzer_prl_18,borkowski_pss_22, lovato_pssb_25}. Correlations within the surface also have a drastic effect on
secondary electron emission \cite{niggas_prl_22} and, thus, act back into the plasma.

An important problem to be solved in simulations is the large difference of characteristic time scales in the plasma and the solid, respectively. Moreover, typical plasma applications require to simulate large times of device operation cycles. For this reason a variety of concepts have been developed where quantum simulations are accelerated or extended towards macroscopic times, e.g.
\cite{Bal2015,filinov_psst18_1,filinov_psst18_2,bonitz_psst18,schluenzen_cpp_18}.
An overview on experiments and simulations investigating the plasma-solid interface has been given in Ref.~\cite{Bonitz_fcse_19}.

\subsubsection{High temperatures. Magnetic Fusion}
Magnetic fusion research in the laboratory, using the tokamak or stellarator concepts, has seen dramatic progress in recent years, and steadily approaches ignition and gain. An overview on the achieved values for the triple product (Lawson criterion) is presented in Fig.~\ref{fig:triple_product}.
In magnetic fusion plasmas, electron densities are comparatively low and temperatures high. As a result, the plasma is fully ionized with the electrons being non-degenerate, i.e., $\chi\ll 1$ and $\Theta \gg 1$, see Fig.~\ref{fig:0}. Nevertheless, quantum effects play a critical role for high-energy collision processes, in particular,
the fusion cross sections. These cross sections are sensitive to quantum effects and depend on the tunneling probability of the deuterium (tritium) nuclei through the repulsive Coulomb barrier. 
Additional quantum effects arise from 
electronic screening of the Coulomb repulsion but are weak in the low density regime. Further, it has been suggested that, at high densities, quantum effects give rise to 
non-exponential ion momentum distribution functions \cite{starostin_physica02, savchenko_pop01, starostin_jetp17} which would have the potential to significantly enhance the fusion probability. Indeed, quantum momentum distributions in the presence of Coulomb interactions have a large momentum asymptotic of $\mathcal{O}(p^{-8})$~\cite{Hofmann_PRB_2013}, instead of an exponential tail, as in the case of Maxwellian ions or fermions. This asymptotic could be confirmed in fermionic PIMC simulations \cite{hunger_pre_21}, but the occupation of the tail turned out be very low.

Besides laboratory plasmas, fusion reactions are crucial in stars, and there is a broad variety of reactions involving heavier elements than hydrogen. A theoretical analysis of the role of tunneling effects has been given in Ref.~\cite{balantekin_rmp_98}. Here, additional quantum effects play an important role, such as tunneling processes involving excited states of the nuclei, e.g. 
\cite{sun_prc_25}.

\subsubsection{Plasmas interacting with strong lasers. QED effects}\label{ss:relativistic} 
In this section we briefly outline different types of quantum effects in plasmas that are not primarily related to the quantum properties of the plasma particles such as atoms and electrons. In contrast, quantum effects may also be introduced into the plasma by the action of an intense electromagnetic field. Consider first the situation of lasers of moderate intensity and/or high frequency. In this regime, collision effects are important, but they will be strongly modified by the action of the laser field, giving rise to 
nonlinear conductivity and the excitation of high odd harmonics of the field \cite{silin_jetp_65,decker_pop_94}. A kinetic theory of laser matter interaction was developed by 
Kremp \textit{et al.} \cite{kremp_99_pre,bornath_lpb_00} and revealed interesting electron-ion scattering effects leading to collisional plasma heating via inverse bremsstrahlung and high harmonics. Numerical  solutions of the quantum kinetic equations revealed that, in nonequilibrium situations, also the generation of even harmonics is possible \cite{haberland_01_pre}. Also, the effect of the field on
dynamical screening and  plasmons has been analyzed \cite{bonitz_99_cpp}.

The situation changes fundamentally upon further intensity increase: then correlation effects become less important, and the dynamics is well described by mean field models (Vlasov equation) in the presence of the field. In the case 
of ultra-intense laser pulses electrons gain relativistic velocities, giving rise to pair creation effects. Furthermore, vacuum fluctuations and quantum electrodynamics (QED) effects, such as the Schwinger effect (vaccum pair creation in an ultrastrong field) \cite{Sauter1931, schwinger_pr_51} become relevant, for a kinetic theory analysis see Refs.~\cite{smolyanski_cpp09, smolyanski_cpp13}.
With the availability of new multi-petawatt laser facilities these effects are now becoming accessible for direct experimental observation, e.g. \cite{Qian_njp_26}.  This is also of direct relevance to astrophysical environments, such as pulsar magnetospheres. Among the important questions are the radiation of hard x-rays and even gamma rays by an ultra-relativistic particle and the back-reaction of this radiation on its trajectory (``radiation reaction'') \cite{keitel_prl_10, Ridgers_jpp_17, Qian_njp_26}. Recent studies indicate the importance of quantum effects on radiation reaction in ultra-strong electromagnetic fields \cite{Los2026}. For more details on recent developments in this field we refer to Ref.~\cite{Qian_njp_26}.

\begin{figure}[h]
\includegraphics[width=0.48\textwidth]{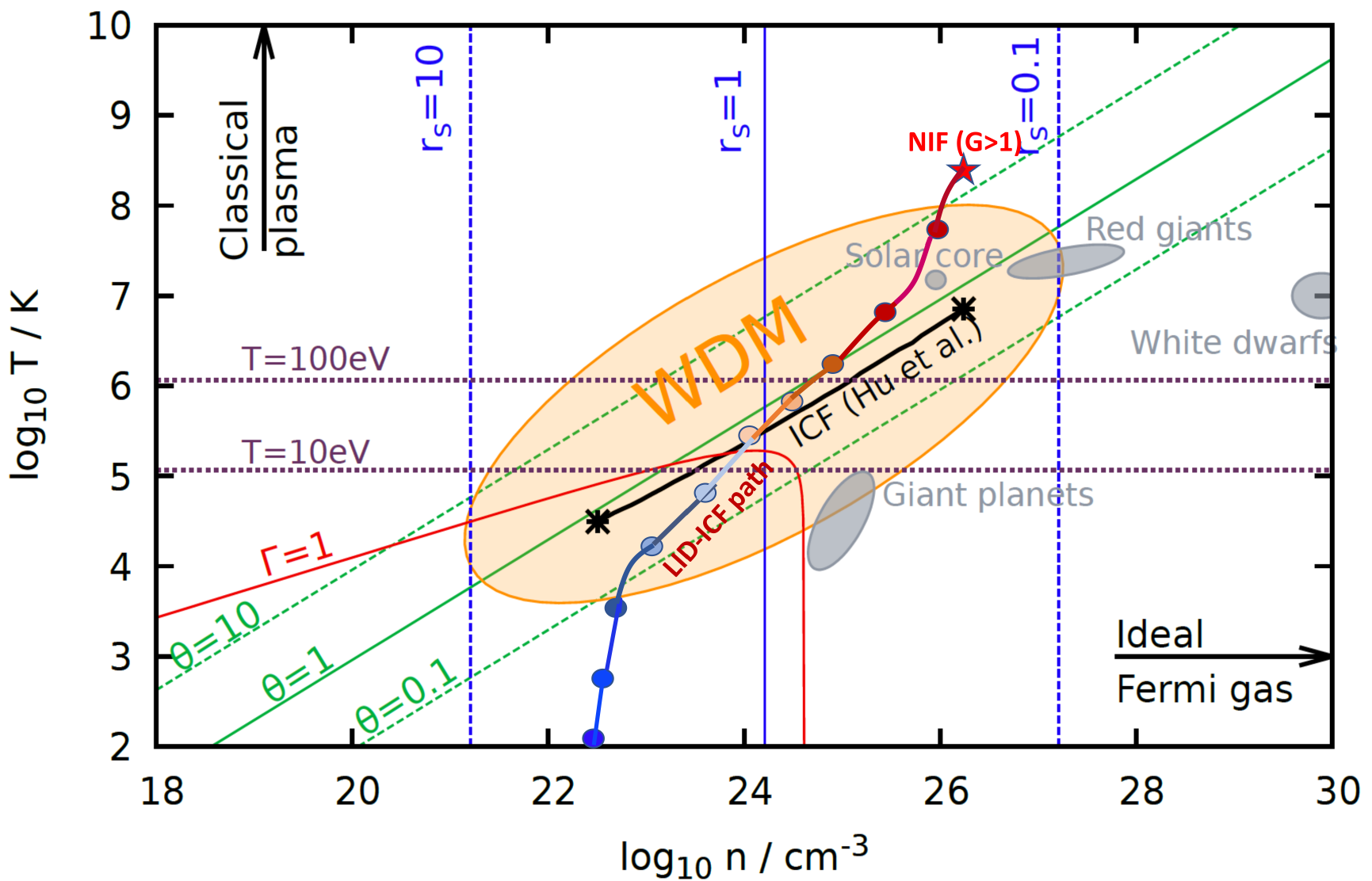}
\caption{Density-temperature plane around WDM parameters with a few relevant examples.
Electronic  \textit{quantum effects} are observed for  $\Theta \lesssim 1$, cf. Eq.~\eqref{eq:theta-def}. The coupling strength of quantum electrons increases with $r_s$ [with decreasing density, Eq.~\eqref{eq:rs-def}]. Note that the values of $\Theta$ and $r_s$ refer to jellium (electrons in fully ionized hydrogen). The density-temperature path of DT-shell for laser-indirect-drive (LID) ICF is added to illustrate that the in-flight shell conditions are mostly in  the warm-dense matter regime before reaching ignition with fusion gain larger than one on the National Ignition Facility; the colors demonstrate the temperature increase; for more details, see Sec.~\ref{ss:icf-benchmarks}.  Adapted from Ref.~\cite{bonitz_pop_24}.
}
\label{fig:overview}
\end{figure}

\subsection{Quantum effects in high density plasmas}\label{ss:high-p}
High density plasmas, where the electrons are quantum degenerate, are presently the primary candidates for quantum effects. An overview is given in Fig.~\ref{fig:overview} where a broad range of electron densities and temperature is shown. 
Quantum plasmas are located below the line $\Theta=1$ and comprise many astrophysical objects, such as giant planets and white dwarf stars, but also neutron stars. On the other hand, quantum plasmas are now routinely created in the laboratory. One purpose of these experiments is to better understand astrophysical objects, see Sec.~\ref{ss:lab-experiments}, but the main goal currently is to realize inertial confinement fusion (ICF), as will be discussed in Sec.~\ref{ss:icf-overview}.

\subsubsection{Warm dense matter. Bound states and different phases}\label{ss:wem}

\begin{figure}[h]
    \centering    \includegraphics[width=0.995\linewidth]{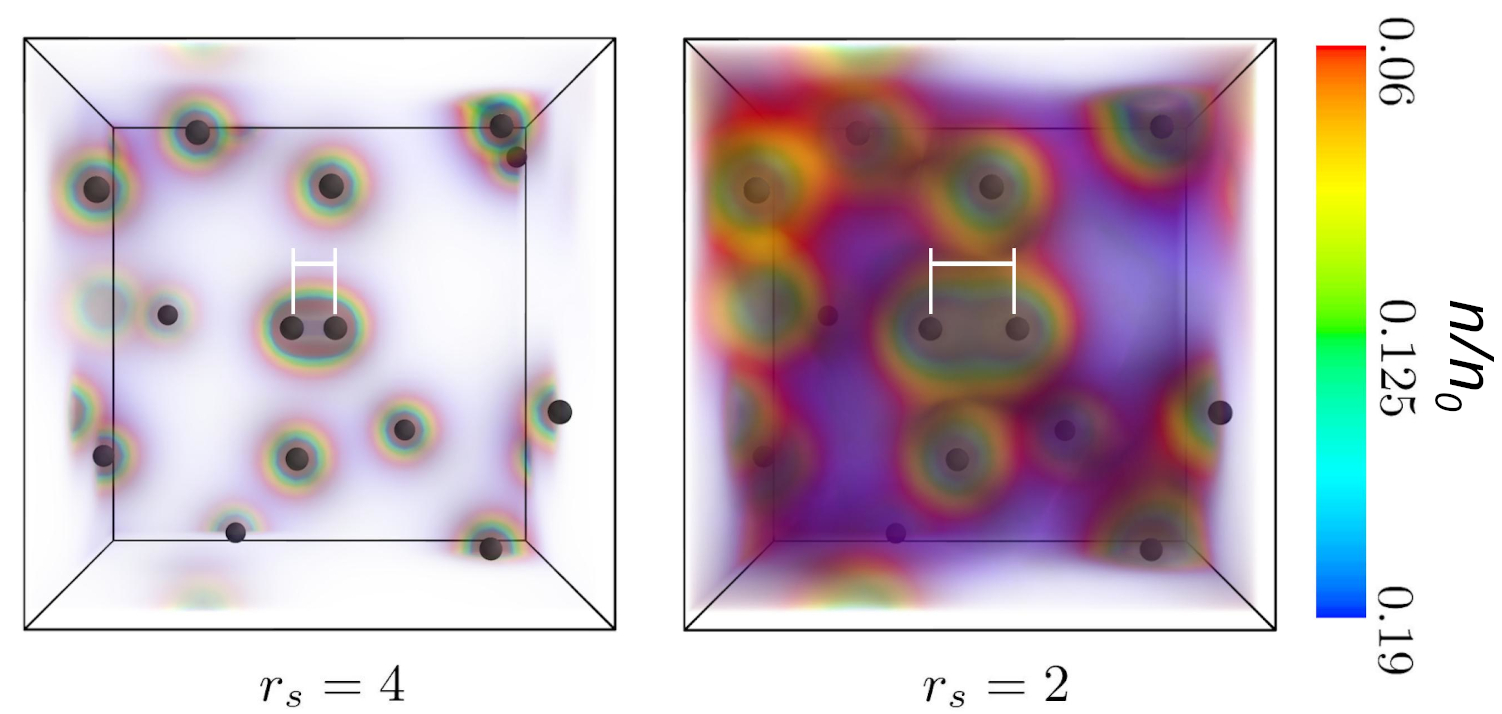}
    \caption{First principles fermionic PIMC results show the electron density in warm dense hydrogen for two snapshots with 14 protons (black dots), for $r_s=4$ (left) and $r_s=2$ (right), at the Fermi temperature, $\Theta = 1$. The electron probability density is indicated by the colors. In the central region of the simulation cell, two protons are positioned $d=0.74\,$\AA$\,$ apart (indicated by white bars), and this molecular configuration remains unchanged in both cases, even though the box extension in the right figure is reduced by a factor $2$. 
    The density is normalized to the mean density $n_0$. The results illustrate that as we enter the WDM regime at $r_s=2$, the distinction between bound and free electrons becomes increasingly ambiguous. Source: Moldabekov \textit{et al.} \cite{MOLDABEKOV2025104144}. Adapted from Ref.~\cite{Moldabekov_JCTC_2024}.}
    \label{fig:bound-states-pimc}
\end{figure}

Figure \ref{fig:bound-states-pimc} illustrates how the quantum plasma configuration changes inside the WDM range when moving approximately along the black line in Fig.~\ref{fig:overview}, from $r_s=4$ (left panel) to $r_s=2$ (right panel), at $\Theta=1$. The figure shows two snapshots from the simulation box of a fermionic PIMC simulation of hydrogen [for simulations details, see Sec.~\ref{ss:fpimc-progress}]. In this parameter range the physical properties of the quantum plasma change drastically. In the left plot the plasma contains predominantly bound states: hydrogen atoms and molecules; one molecule is depicted in the center with 2 protons being approximately $1.4 a_B$ apart. In the right panel, the density is increased by a factor 8 and the temperature is increased by a factor $4$ [cf. the definitions \eqref{eq:rs-def} and \eqref{eq:theta-def}]. Correspondingly, the electron clouds of individual atoms start to overlap significantly, leading to a breakup of bound states and an increase of the degree of ionization $\alpha^{\rm ion}$, even though the concept of the degree of ionization itself has become ambiguous. Physically, electrons are still weakly attached to protons but there is a high electron density in between the protons.

\begin{figure}[t]
    \centering
\includegraphics[width=0.995\linewidth]{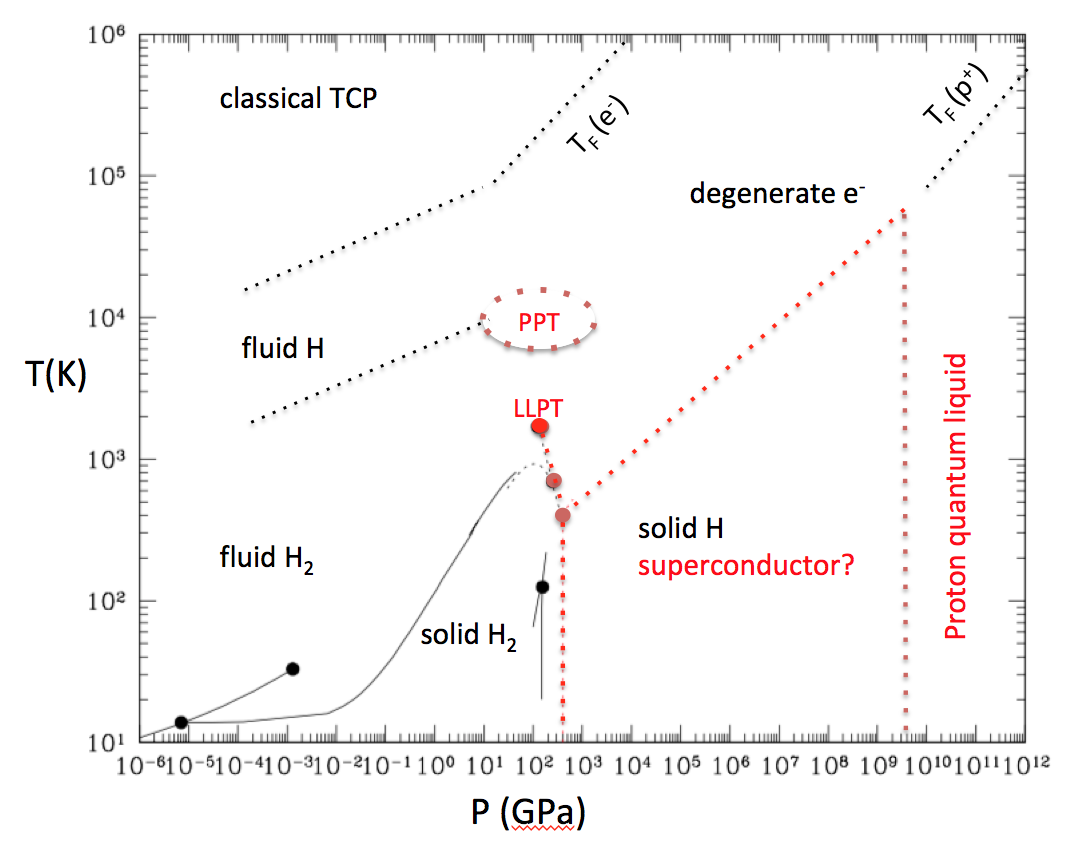}
    \caption{Hydrogen phase diagram in the quantum plasma range. Solid black lines show the boundaries between the gas, liquid, and solid phases as measured in static
experiments. The solid circles show the location of critical or triple points (black: observed, red: predicted). The black dashed lines are crossovers between the classical behavior of electrons and protons at high temperatures following from the condition $\Theta=1$, cf. Eq.~\eqref{eq:theta-def}. $T_{\rm F}(e^-)$ and $T_{\rm F}(p^+)$ are the Fermi temperatures of electrons and protons, respectively, cf. Eq.~\eqref{eq:ef-def}. PPT indicates the critical region of the hypothetical Plasma Phase Transition and LLPT denotes the critical point of the liquid-liquid transition. Figure taken from Ref.~\cite{bonitz_pop_24} with the permission of the authors.
}
    \label{fig:h-phase-diagram}
\end{figure}

This effect of thermal and pressure ionization of atoms is associated with a reduction of the ionization potential (ionization potential depression), which will be discussed in more detail in Sec.~\ref{ss:fpimc-saha}. At the same time, in a quantum simulation in the ``physical picture'', such as FPIMC [and also density functional theory], the distinction between free and bound electrons gradually vanishes. These two snapshots indicate the high complexity of the physical states in the warm dense matter region. In fact, when considering a much larger density-temperature range beyond WDM, as is done in Fig.~\ref{fig:h-phase-diagram}, there is large variety of different phases in hydrogen that was discussed in detail in Ref.~\cite{bonitz_pop_24}.

\subsubsection{Inertial confinement fusion (ICF)}\label{ss:icf-overview}

\begin{figure}[h]
    \centering
    \includegraphics[width=0.5\textwidth]{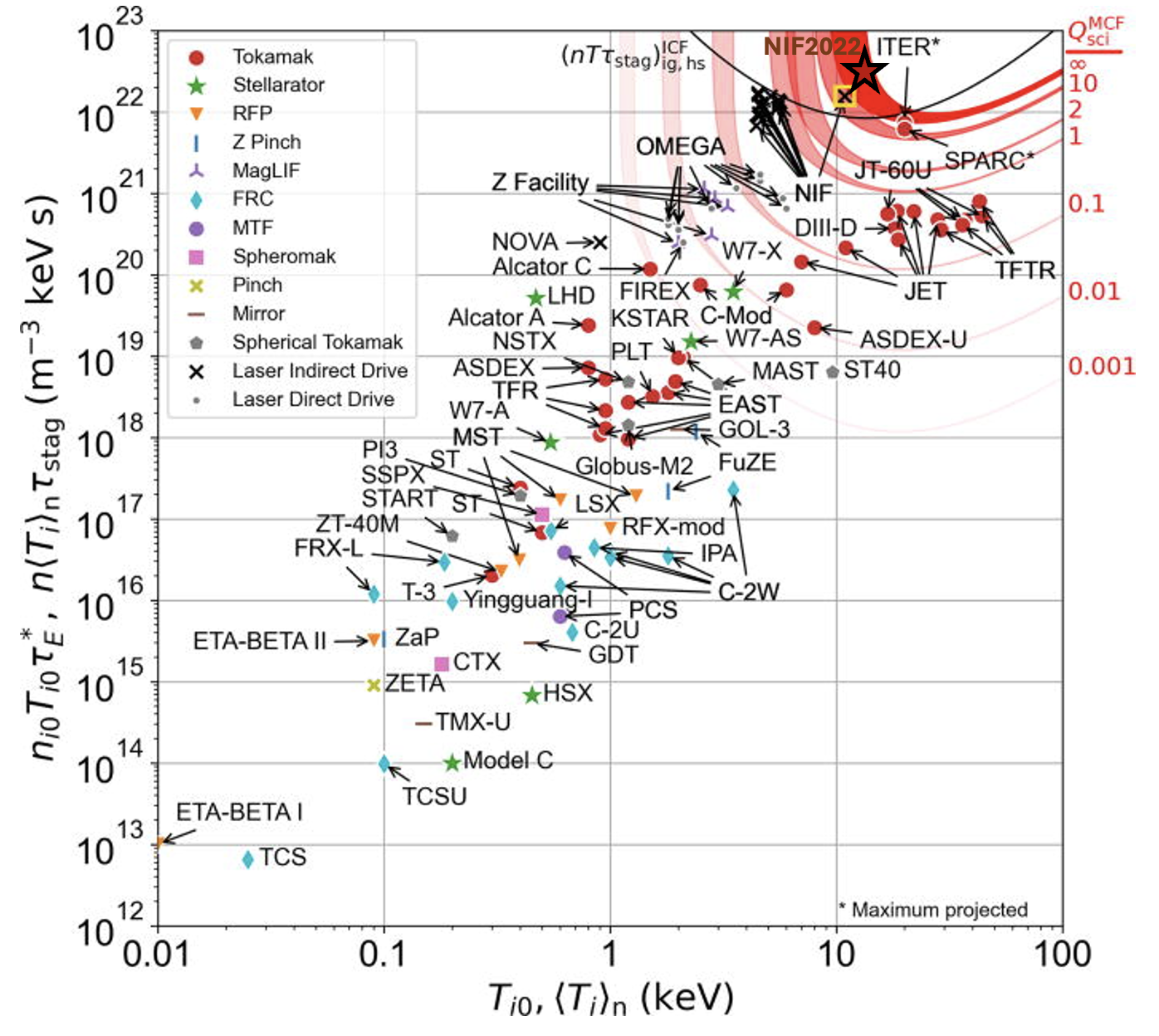}
    \caption{Evolution of the fusion triple product of number density $n$, ion temperature $T_i$ and energy confinement time $\tau_E$ or stagnation time $\tau_{\textnormal{stag}}$ for magnetic confinement and inertial confinement fusion experiments, respectively. The parameter $Q$ denotes the ratio of power generated by fusion and the input power by external energy sources. For an ignited fusion plasma, $Q$ reaches infinity as the heating to sustain fusion is provided by the fusion reactions. At NIF, the Lawson criterion for fusion ignition ($Q=\infty$) was fulfilled for the first time in August 2021 (shot N210808) and the target gain (fusion energy compared to the laser energy used to irradiate the target) exceeded unity in December 2022 (shot N221204) with 3.15\,MJ of fusion energy achieved with 2.05\,MJ of drive laser energy~\cite{abu-icf_prl_24}. Since then, ignition has been repeated several times with a current record fusion energy of 8.6\,MJ~\cite{record_shot}. The milestone of fusion ignition is yet to be achieved for magnetic confinement fusion.}
    \label{fig:triple_product} 
\end{figure}

The idea to create burning fusion plasmas by spherical implosion driven by pulsed high-energy lasers~\cite{nuckols1972} emerged only one decade after Maiman’s demonstration of the first laser. For this approach, the generation of a fully degenerate deuterium-tritium plasma at around 10,000 times solid density ($\sim$1000\,g/cm$^3$) is required as the fuel. However, due to the technical and physical challenges involved, about five decades elapsed before the Lawson criterion was surpassed~\cite{zylstra2022,abu-icf_prl_22} and a burning fusion plasma with net energy gain relative to the invested drive laser energy was achieved at the National Ignition Facility of Lawrence Livermore National Laboratory~\cite{abu-icf_prl_24}. Nonetheless, the approach of laser-driven inertial fusion remains the first -- and thus far the only –- approach to attain this critical milestone on the path toward fusion energy~\cite{Wurzel2025}, see Fig.~\ref{fig:triple_product}. For a recent review on the experimental developments and achievements, see Ref.~\cite{Edwards_rmmp_25}.

This success was achieved by an indirect drive approach where the high-energy lasers heat a hohlraum to radiation temperatures exceeding 300\,eV. The smooth radiation field peaking in the soft X-ray regime then compresses a spherical capsule containing the D-T fuel. Once the imploding DT shell stagnates, a hot spot is formed so that ignition and fusion burn is initiated thereby giving net energy gain. The current record yield of 8.6\,MJ~\cite{record_shot} presents a target gain of $\sim$4.1, compared to the input laser energy of 2.08\,MJ, and a fuel gain of $\sim$400, since only about 20\,kJ of the input laser energy reach the D-T fuel.

However, the yield still falls short of the electrical energy powering the lasers (300\,MJ), and an ICF power plant will most probably require more energy-efficient ignition concepts. Numerous proposals for such concepts, for instance direct laser irradiation of the capsule (“direct drive”) \cite{craxton2015direct}, shock ignition \cite{Theobald_PoP_2012}, fast ignition \cite{Tabak_PoP_1994} or pulsed power drivers \cite{Slutz_PRL_2012}, have been proposed. However, these approaches are, so far, mostly based on simulations scaling experiments at lower energies to the ignition regime. Due to the complex physics involved, including electronic quantum effects and ion correlations [cf. the fusion path sketched in Fig.~\ref{fig:overview}], substantial experimental validation programs are required before a credible design of a power plant can be identified. 

Following the success of NIF’s fusion ignition in the laboratory, private investments supporting various concepts of fusion, not limited to inertial fusion, have emerged. Numerous start-up companies have appeared around the globe and several of these have already started to build demonstration facilities for ignition concepts which could scale to a power plant. In several countries, the startup companies are strongly supported by the government in public-private partnerships, such as the DOE-funded fusion energy hubs STAREFIRE, RISE and COLoR in the US or the Fusion 2040 program in Germany. 

\subsubsection{Laboratory experiments and applications to astrophysics}\label{ss:lab-experiments}
Laboratory research on dense plasmas has mostly been driven by inertial confinement fusion research for several decades. However, the large-scale experimental infrastructures for ICF research were at some point also made available for studies with different fundamental science motivation. In the 1990s, the NOVA laser at LLNL started to allow experiments to study dense astrophysical plasmas and basic quantum phenomena in dense plasmas in the laboratory, giving birth to the broader field of high energy density physics that was no longer limited to ICF~\cite{Rosen1996,Drake2018}. Prominent examples from these campaigns are the debate on the compressibility due to dissociation and ionization of deuterium shock-compressed to pressures around 1\,Mbar~\cite{Collins1998,Nellis2002} or the discovery of laser-accelerated MeV ions beams~\cite{Snavely2000}. At the same time, numerous smaller laser facilities were built allowing the creation and characterization of dense plasmas, mostly aiming for equation of state and transport properties. Examples include the benchmark of equation of state in the multi-Mbar regime by shock Hugoniot measurements, (XRTS OMEGA, JLF, VULCAN). Finally, the National Ignition Facility devotes $\sim$10\% of its shots to basic ``Discovery Science''~\cite{Remington2021} with outstanding results, such as shock Hugoniot measurements, to access quantum effects in the compressibility at pressures approaching 1\,Gbar~\cite{Doeppner2018, Kritcher2020}, the insulator-metal transition in ramp-compressed deuterium~\cite{celliers2018insulator}, the onset of pressure ionization in highly compressed dense plasmas~\cite{Doeppner2023} and the free-free opacity of degenerate hydrogen plasmas at conditions comparable to the interiors of M-dwarfs~\cite{Luetgert2022,Schumacher2026}. 

On the other hand, adding laser facilities to heavy ion accelerators has allowed one to investigate the stopping power of heavy ions in laser-generated plasmas governed by electron-ion collisions~\cite{Frank2013,Cayzac2017}. At the same time, intense heavy ion beams can also be used to generate dense plasmas employing volumetric stopping and thus isochoric heating, which enables studies on longer timescales and larger sample volumes in comparison to typical laser experiments~\cite{Hesselbach2024,Luetgert2024}. However, accelerator-based sources have so far only reached temperatures of a few thousand kelvins, and more intense pulses are required to establish access to the core WDM regime~\cite{Schoenberg2020}. Such facilities are currently under construction and corresponding experiments are planned at the FAIR and HIAF facilities.

Next to lasers and particle accelerators, pulsed power devices, the largest one being the Sandia Z machine, have contributed substantially to the study of quantum effects in dense plasmas, in particular equation of state~\cite{Knudson2001,knudson_direct_2015} and radiative opacities~\cite{Bailey2015}.

The advent of X-ray free electron lasers such as LCLS (since 2009), SACLA or European XFEL provides a revolutionary diagnostic tool to study quantum effects in dense plasmas~\cite{Kraus2026}. Nanosecond high-energy lasers built at such facilities allow the creation of warm dense matter states comparable to planetary interiors via shock compression and highly precise characterization via the XFEL pulses mostly employing spectrally and angularly resolved X-ray scattering methods~\cite{Fletcher2015, Kraus2017, Frydrych_NatComm_2020, Kraus2025, Andriambariarijaona2026}. Moreover, femtosecond high intensity lasers have been established at XFELs to produce even more extreme energy densities, accessing and characterize complex non-equilibrium conditions~\cite{Fletcher2022,White2025,Schoenwalder2026} and stellar interior conditions~\cite{LasoGarcia2024}. Due to these successes and further prospects for inertial fusion energy research, upgrades to the drive laser capabilities at XFEL facilities from the 100-J-level to kJ energies per pulse are foreseen in the near future. This will enable experiments exploring even higher energy densities at these facilities. Moreover, at lower energy densities, substantially increased drive laser energies will allow for higher precision due to larger sample volumes and thus a reduction of gradients in space and time.

\subsection{Quantum effects of heavy particles}\label{ss:heavy-particles}
Due to their large mass, nuclei or ions are usually expected to behave classically, because the degeneracy parameter \eqref{eq:chi-def} scales with the particle mass to the power $-3/2$. However, for an accurate computation of the equation of state of hydrogen and helium in the warm dense matter region [cf. Fig.~\ref{fig:overview}], it is important to take into account quantum effects of protons or $\alpha$-particles, respectively. 
For the case of hydrogen the lines $\Theta_e=1$ and $\Theta_p=1$ are plotted in Fig.~\ref{fig:h-phase-diagram}. In that figure, the area ``solid H'' denotes a quantum plasma where the electrons are strongly degenerate and delocalized whereas the protons are point-like or weakly delocalized. At the same time their classical coupling parameter \eqref{eqn:gammai} exceeds $170$ and the quantum coupling parameter  \eqref{eq:rs-def}  $100$, which is sufficient to form a crystal lattice. The critical parameters of this Coulomb crystal were investigated in Refs.~\cite{bonitz_prl_5} and \cite{bonitz_jpa_06}.
Such ion crystals  in a two-component plasma are expected to exist, e.g., in white dwarf stars or neutron star crusts (formed primarily from carbon and oxygen nuclei). An interesting prediction [Ashcroft, Abrikosov] that still awaits confirmation is that this crystal may be superconducting, cf. Fig.~\ref{fig:h-phase-diagram}.
This ion crystal melts via heating or by compression. In the latter case $r_{sp}$ falls below the melting point and protons form a quantum liquid. Similar effects have been predicted for electron-hole plasmas with sufficiently flat valence bands (large effective mass of the holes) \cite{bonitz_prl_5, bonitz_jpa_06}.
Interestingly, state-of-the-art direct PIMC simulations are capable to even resolve the small nuclear quantum effects in hydrogen at $\Theta_e=1$ on sufficiently small length scales~\cite{Dornheim_MRE_2024}.

\subsection{Unconventional quantum plasmas}\label{ss:unconventional}
Aside from ``conventional'' electron-ion plasmas similar systems of charged particles exist in a variety of other fields. The primary examples are plasmas in condensed matter systems or at ultra-high energy density which we briefly discuss in the following.

\subsubsection{The electron gas in metals. Jellium}\label{ss:jellium}
Already in the early days of quantum theory of solids it became evident that the delocalized conduction electrons in metals exhibit many similarities with a plasma. The mathematical model that correctly takes into account electroneutrality is the uniform electron gas (UEG, ``jellium''). Indeed, there has been an active exchange of ideas between the plasma physics and the theory of metals, regarding the description of thermodynamic and dynamic properties of the UEG, including plasma oscillations (plasmons) and the response to electric and magnetic fields, for an overview see, e.g. Ref. \cite{giuliani2005quantum}. The properties of the UEG and their analytical parametrizations played a fundamental role for the success of density functional theory (DFT, local density approximation, LDA) and are crucial for the DFT simulation of materials, for details see Sec.~\ref{ss:dft-progress}. With the recent progress in dense quantum plasmas and warm dense matter, DFT methods from condensed matter physics were introduced here as well where they turned out to be successful and highly efficient, e.g., \cite{graziani-book}. However, the crucial relevance of finite temperature effects in WDM, in turn, required extensions of the jellium model and the LDA. These important developments were recently made in the field of WDM and led to the theory of the warm dense UEG, e.g.~\cite{Brown_PRB_2013,KARASIEV20143240,schoof_prl15}, for an overview, see Ref.~\cite{dornheim_physrep_18} and Sec.~\ref{ss:ueg-benchmarks}. Simultaneously, improved finite temperature functionals for DFT, beyond LDA, were developed, cf. Sec.~\ref{ss:fpimc-dft}. 

\subsubsection{The electron-hole plasma in semiconductors}\label{ss:e-h-plasma}
\begin{figure}[t]
    \centering
\includegraphics[width=0.98\linewidth]{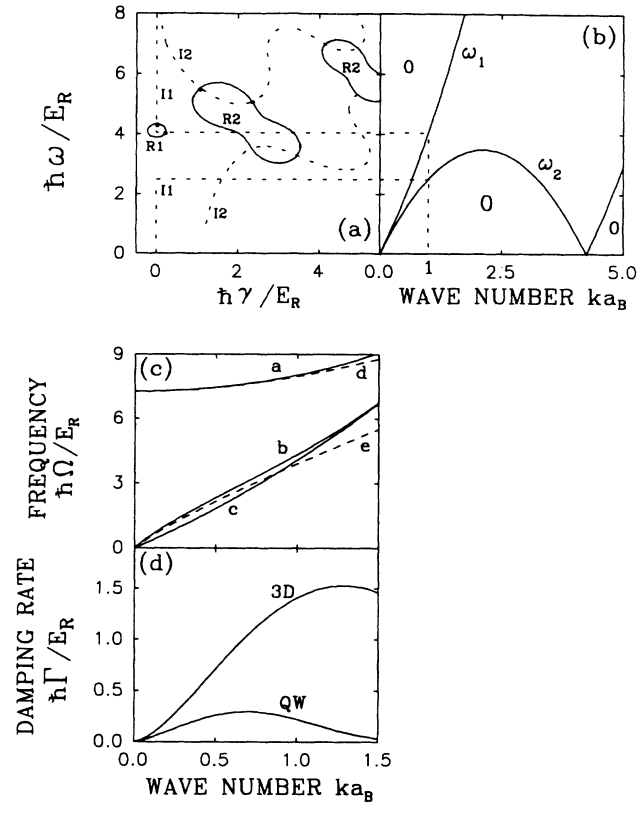}
    \caption{Collective excitations of the electrons in an e-h-plasma in equilibrium. The parameters are for GaAs ($\epsilon_b=12.7$, $a_B=135$ \AA, $E_R=4.2$ meV). (a) Zeroes of the analytical continuation of the RPA dielectric function (DF), Eq.~\eqref{eq:rpa-df}, for $k=1/a_B$. Plasmons (thick dots) are crossings of the (full) lines Re$\,\epsilon=0$ [R1 is for $T=0 $K and R2 for $T=100$ K] and (dashed) lines [I1 is for $T=0 $K and I2 for $T=100$ K] Im$\,\epsilon=0$. (b) Pair continuum and undamped (0) regions, $\omega_{1,2}= (2k_F\pm q)q/2m$; between the lines $\omega_1$ and $\omega_2$ plasmons are strongly Landau damped due to the excitation of electron hole pairs. (c) Plasmon dispersion: for a 3D plasma (a) and a quantum wire (QW, curve b). (d) Plasmon damping - the optical plasmon [lines a, b in (c)] is undamped at $T=0$ K. The curves show the second complex zero of the DF that is located inside the pair continuum [panel (b)] and is strongly damped. For details see text. 
    Figure taken from Ref.~\cite{bonitz-etal.94pre} with the permission of the authors.}
    \label{fig:e-h-dispersion}
\end{figure}

With the development of semiconductor theory the similarity with two-component electron-ion plasmas seemed to be striking. Consequently, many plasma physics concepts, including kinetic and hydrodynamic models, were  quickly applied to transport phenomena as well as plasma oscillations and instabilities in semiconductors.
An example is shown in Fig.~\ref{fig:e-h-dispersion}. There we plot the electron plasmon obtained from the complex zeroes of the RPA dielectric function in the plane of complex frequencies $z=\omega - i\gamma$
\begin{align}\label{eq:df}
    \varepsilon( \vec q, z) &= 1 - \sum_a V_{aa}(q)\Pi_a( \vec q, z)\,,\\
    \Pi_a(\vec q,z) &= (2s+1)
\int
{
\frac{d \vec p}{(2\pi\hbar)^3}
\frac{
 f_a\left(E^{(a)}_{\vec p}\right) - f_a\left(E^{(a)}_{\vec p+\hbar\vec q}\right)}
{\hbar z+E^{(a)}_{\vec p}-E^{(a)}_{\vec p+\hbar\vec q}
}
}\,,
    \label{eq:rpa-df}
\end{align}
where $V_{aa}(q)$ is the Fourier transform of the Coulomb potential and $\Pi_a$ is the retarded longitudinal polarization function that was derived independently by Klimontovich and Silin \cite{klimontovich-etal.52a,klimontovich-etal.52b} and Lindhard \cite{lindhard54}. The summation in Eq.~\eqref{eq:df} is over the band structure, where $E^{(a)}(\textbf{p})$ denotes the momentum dispersion of band $a$ and $f_a$ its occupation. 

The complex zeroes $(\omega;\gamma)$ of the dielectric function follow from solving simultaneously  Re $\epsilon(\omega,\gamma)=$ Im $\epsilon(\omega,\gamma) \equiv 0$ \cite{bonitz_qkt}, which is illustrated in Fig.~\ref{fig:e-h-dispersion} (a) by the crossing of the full (real parts) and dashed (imaginary parts) lines. Repeating this for different wave numbers yields the plasmon dispersion, $\omega(k)$ and damping, $\gamma(k)$, which generalized the common weak damping approximation that assumes $|\gamma| \ll \omega$, see also Refs.~\cite{bonitz_qkt, hamann_cpp_20}. The results are plotted in panels (c) and (d) of Fig.~\ref{fig:e-h-dispersion} for two cases: a 3D quantum plasma (line a) and a quasi-onedimensional quantum wire (QW). In the former case, the dispersion starts at the plasma frequency, $\omega(0)=\omega_{\rm pl}$ whereas in the 1D case the dispersion starts at $\omega(0)=0$, cf. line b. Note that, at $T=0$ K, there are two complex zeroes of $\epsilon$ shown by the dots at the ellipse R1 in panel (a). One is located at $\gamma=0$ and corresponds to the optical plasmon. The other one is at a lower frequency and a finite damping which is plotted in panel (d).
This damping behavior can also be understood from panel (b) where we plot the pair continuum of particle-hole excitations. It is obtained from analyzing the imaginary part of $\epsilon$ which is non-zero in between the lines $\omega_1$ and $\omega_2$ given by $\omega_{1,2}= (2k_F\pm q)q/2m$. Inside the pair continuum the imaginary part of $\epsilon$ is finite and no collective excitations are possible due to strong collisionless (Landau) damping. 

The pair continuum governs the behavior of collective modes in condensed matter systems at low temperature and differs substantially from the damping of plasma oscillations in electron-ion plasmas. There are more important differences which have their origin in quantum effects. In fact, the well-known Vlasov dielectric function follows as the classical limit of the polarization \eqref{eq:rpa-df}, i.e. $\hbar q \to 0$ and, further, assuming a parabolic dispersion, $E^{(a)}_{\vec p} \to \frac{p^2}{2 m_a}$, giving rise to 
\begin{align}
    \Pi^{\rm vlasov}_a(\vec q,z) &= (2s+1)
\int
{
\frac{d \vec p}{(2\pi\hbar)^3}
\frac{\frac{\partial f_a}{\partial \textbf{p}}\cdot \textbf{q}
}
{\textbf{q}\cdot \textbf{p}/m_a - z
}
}\,.
    \label{eq:vlasov-df}
\end{align}
In the classical limit, the plasmon dispersion $\omega(q)$ significantly differs from the quantum result, except for the long-wavelength limit.
The Vlasov dielectric function and various hydrodynamic models have frequently been applied to plasma oscillations and instabilities in electron-hole plasmas in semiconductors by parts of the plasma physics community. However, the corresponding predictions  have not been reproduced by experiments. The reason is that the parabolic assumption for $E^{(a)}_{\vec p}$ is a very crude approximation. More importantly, due to the existence of a band gap, under normal conditions, there exist no electrons in the upper (conduction) band and no holes in the lower (valence) band. Electrons and holes are only generated via external excitation, e.g., by a laser pulse, giving rise to nonequilibrium distribution function $f_a(\textbf{p},t)$ which thermalize via collisions between charge carriers and with the lattice (phonons). Moreover, the electron-hole plasma is not stable because electrons in the upper band typically within a few picoseconds return to the lower band, and electron and hole populations vanish. Thus, for a reliable description of collective modes in semiconductors, the band structure has to be properly taken into account which is commonly done by using density functional theory calculations, cf. Sec.~\ref{ss:dft-progress}. Moreover, in many condensed matter plasmas, correlation effects play an important role which requires to go beyond the simple mean field approximation \eqref{eq:rpa-df}, e.g. by including local field corrections. This situation is very similar to dense quantum plasmas, and a brief discussion can be found in Sec.~\ref{ss:ueg-dynamics}.

In conclusion we note that signatures of laser excited nonequilibrium plasmons have been observed in electron-hole plasmas in dedicated pump-probe experiments, e.g., \cite{lampin_prb_99,bonitz_prb_0} where they led to an accelerated thermalization of the plasma. At the same time, upon strong excitation, when the plasma is nearly isotropic, no plasma instabilities are possible but only an undamping of the nonequilibrium modes, as was shown in Refs. \cite{bonitz94pp,bonitz95pp}.

\subsubsection{Quark-gluon plasma}\label{ss:qcd}
The quark-gluon plasma (QGP) is an assembly of quarks, antiquarks and gluons that resemble, to some extent, an electron-ion quantum plasma. It is, therefore, of interest to briefly discuss the existing analogies and  differences here. Quarks and gluons are the main constituents of baryonic matter, and they carry a (color) charge, giving rise to a (color) Coulomb interaction. Even though the color charge is non-abelian (the electrical charge is abelian) this analogy inspired early investigations in the field and even led to 
the term quark-gluon plasma (QGP) that was introduced by E.V.~Shuryak in 1978 \cite{SHURYAK_1978}. While quarks and gluons are an integral part of the standard model (Quantum chromodynamics, QCD), under ``normal'' conditions no free quarks and gluons exist. This is similar to plasmas which also do not exist under ``normal'' conditions, i.e., at ambient pressure and room temperature where electrons and ions are typically bound into atoms forming solids, liquids and gases. While creation of a normal (electrodynamic) plasma by heating or compression the solid, the liquid or the gas is routine and at the heart of plasma physics and technology, as discussed in this article, the situation is much more complex for the QGP. 

\begin{figure}[h]
    \centering
    \includegraphics[width=0.95\linewidth]{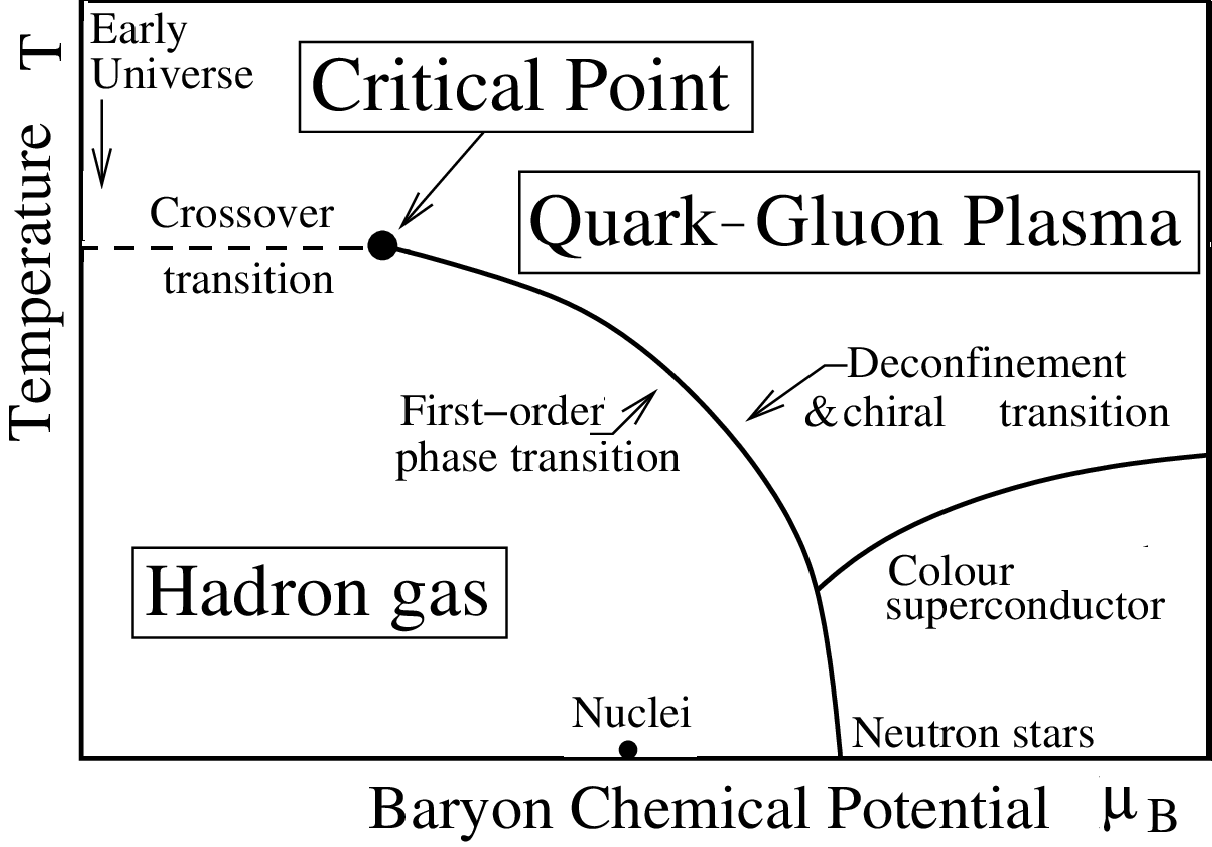}
    \caption{Conjectured phase diagram of QCD as a function of quark chemical potential $\mu$ and temperature T. Quark–gluon plasma is in the high-density, high-temperature part of the diagram. Note the similarities with the phase diagram of a hydrogen plasma, cf. Figs.~\ref{fig:0} and \ref{fig:h-phase-diagram}, for details, see text.
    Modified from Rajeev S. Bhalerao (Tata Inst.) - 1st Asia-Europe-Pacific School of High-Energy Physics (AEPSHEP 2012), pp. 219-239 Relativistic heavy-ion collisions DOI: 10.5170/CERN-2014-001.219.
    }
    \label{fig:qgp-phase-diag}
\end{figure}

Formation of the QGP is also expected to occur at sufficiently high temperature or at very high density (translating into a high baryonic chemical potential $\mu_B$). And indeed, the phase diagram shown in Fig.~\ref{fig:qgp-phase-diag} is reminiscent of the one of a dense plasma: the region of bound states is confined to low temperatures and low densities, cf. region below the red dashed line in Fig.~\ref{fig:0} (the line of zero degree of ionization is expected to proceed parallel, slightly below), and this line is very similar to the boundary between the hadron gas and the QGP, cf. full black line in Fig.~\ref{fig:qgp-phase-diag}.  Ionization of a hydrogen plasma via heating (at $T \gtrsim E_b$) is very similar to thermal breakup of hadrons (at $T \gtrsim 156$ MeV \cite{qcd_50}). In fact, this is the path the early universe is thought to have traversed -- in opposite direction -- during the first $10^{-10}\dots 10^{-6}$ s after the Big Bang, cf. Fig.~\ref{fig:qgp-phase-diag}. As in the case of hydrogen, this thermal breakup of bound states is not a phase transition but a gradual process (crossover). On the other hand, atoms can break up at low temperature, upon strong compression [see our discussion in Sec.~\ref{ss:fpimc-saha}], and a similar breakup of hadrons is indicated in Fig.~\ref{fig:qgp-phase-diag} as well. While in hydrogen, pressure ionization (in the limit of $T=0$ K) occurs at densities around $r_s \sim 1.2$, for the QGP this point translates to a value of $\mu_B \sim 1000$ MeV \cite{qcd_50}. As indicated in Fig.~\ref{fig:qgp-phase-diag}, this is expected to be a first order phase transition that might occur in the core of some neutron stars \cite{nstar_2023}. Interestingly, as shown in Fig.~\ref{fig:h-phase-diagram}, pressure ionization in cold dense hydrogen also proceeds as a phase transition: first from molecular hydrogen to atomic hydrogen and then to a proton liquid that is embedded into a Fermi sea of electrons, see our discussion in Sec.~\ref{ss:heavy-particles} and also Ref.~\cite{bonitz_pop_24}.

The quark-gluon plasma, for many years, remained a theoretical hypothesis, until it could be observed in laboratory experiments with heavy ion collisions at the Large Hadron Collider at CERN, in 2000, and, more conclusively, in 2005 at the Relativistic Heavy Ion Collider at Brookhaven National Lab \cite{rhic_2005}.
The analysis of the data revealed a surprising property: the QGP behaves as nearly perfect fluid characterized by low viscosity \cite{qgp-fluid_2006} and strong correlations.

Thus, there are indeed remarkable similarities between a conventional plasma and a quark-gluon plasma \cite{bonitz_rpp_10} with the important advantage that the properties of the former are much easier accessible in  laboratory experiments. There have also been many attempts to apply theoretical methods and simulations from conventional plasmas to the QGP. Examples are classical MD simulations \cite{SHURYAK_09} and color path integral Monte Carlo simulations to study the thermodynamic properties and conductivity, e.g.~\cite{filinov_cpp09,filinov_cpp11, filinov_prc_13, filinov_ppcf15}. Also, similar density response properties are expected which include collective excitations, plasmons and wake effects, in the case of a streaming plasma \cite{moldabekov_cpp15}. At the same time, to obtain reliable theoretical predictions of the properties of the QGP, requires nonperturbative lattice QCD simulations.
For a recent review on QCD and the QGP, see Ref.~\cite{qcd_50}.

%------------
\section{Methods for dense quantum plasmas and warm dense matter}\label{s:methods}
In recent years experimental diagnostics of warm dense matter have seen an unprecedented development. Both, the range of photon energies, the time resolution and the accuracy have improved dramatically. Of particular importance is currently  
X-Ray Thomson Scattering (XRTS) on which we focus in Sec.~\ref{ss:xrts} before discussing theoretical approaches.

\subsection{X-Ray Thomson Scattering Experiments}
\label{ss:xrts}
XRTS is sensitive to the quantum properties of WDM as well as to correlation effects.
XRTS probes the electronic dynamic structure factor
\begin{eqnarray}\label{eq:DSF}
    S_{ee}(\mathbf{q},\omega) = \sum_{l,m} P_m \left|\bra{l}\hat{n}(\mathbf{q})\ket{m}\right|^2\delta\left(\omega-\frac{E_l-E_m}{\hbar}\right)\ ,
\end{eqnarray}
which is given by the sum over transitions between the many-body eigenstates $m$ and $l$ with $P_m$ being the occupation probability of the initial state $m$.
From Eq.~(\ref{eq:DSF}), it becomes intuitively clear that XRTS gives one 
valuable insights into the nanophysics of the probed sample, including its density, temperature, ionization state, and ionization potential depression~\cite{redmer_glenzer_2009,Kraus_2018,Gregori_PRE_2003,Doeppner2023,Dornheim_NatComm_2025}, as well as dedicated physics effects such as miscibility~\cite{Frydrych_NatComm_2020}, bound--bound transitions~\cite{baczewski2021predictionsboundboundtransitionsignatures}, and potentially the roton feature that has been predicted to occur in solid density hydrogen~\cite{hamann_prr_23}.
A particularly useful property of the dynamic structure factor is given by the \emph{detailed balance} between negative and positive frequencies~\cite{giuliani2005quantum}
\begin{eqnarray}\label{eq:detailed_balance}
    S_{ee}(\mathbf{q},-\omega) = S_{ee}(\mathbf{q},\omega)\ e^{-\beta\hbar\omega}\ ,
\end{eqnarray}
which are related by a simple Boltzmann factor that is uniquely determined by the inverse temperature $\beta=1/k_\textnormal{B}T$. 
In stark contrast, it holds $S_\textnormal{cl}(\mathbf{q},-\omega)=S_\textnormal{cl}(\mathbf{q},\omega)$
for classical systems. 
\begin{figure}
    \centering
    \includegraphics[width=0.93\linewidth]{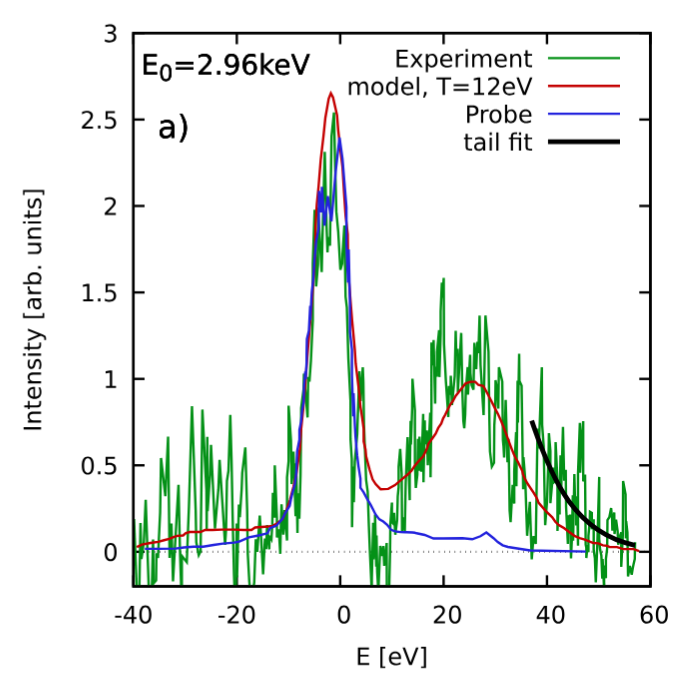}
    \caption{XRTS measurement of isochorically heated beryllium taken at the Omega laser facility with a beam energy of $E_0=2.96\,$keV collected at a scattering angle of $\theta=40^\circ$ by Glenzer \textit{et al.}~\cite{Glenzer_PRL_2007}, plotted as a function of photon energy loss $E$. Green: XRTS measurement; blue: source-and-instrument function $R(\omega)$; red: forward model reported in Ref.~\cite{Glenzer_PRL_2007}; black: empirical exponential fit to the high-energy tail of the XRTS spectrum.
    Adapted from Ref.~\cite{Dornheim_T_2022} with the permission of the authors.
    }
    \label{fig:XRTS}
\end{figure}
In Fig.~\ref{fig:XRTS}, we show an XRTS measurement of isochorically heated beryllium by Glenzer \textit{et al.}~\cite{Glenzer_PRL_2007} as a function of the photon energy loss. The central feature corresponds to the elastic peak, that is often attributed to the form factor of bound electrons and a screening cloud~\cite{Vorberger_Gericke_PRE_2015}. In addition, there is a distinct plasmon peak at $E=30\,$eV, which is damped on the up-shifted (photon energy gain) side at $-30\,$eV by the detailed balance.

Unfortunately, the measured XRTS intensity is (to a first, but often very good approximation~\cite{Gawne_JAP_2024,Gawne_CompPhysComm_2026,gawne2025spectraldeconvolutiondeconvolutionextracting})
given by a convolution of the dynamic structure factor with the combined source and instrument function $R(\omega)$ (blue curve in Fig.~\ref{fig:XRTS}),
\begin{eqnarray}\label{eq:convolution}
    I(\mathbf{q},\omega) = S_{ee}(\mathbf{q},\omega) \circledast R(\omega)\ ,
\end{eqnarray}
which destroys physical symmetry properties such as Eq.~(\ref{eq:detailed_balance}).
Moreover, a proper deconvolution of Eq.~(\ref{eq:convolution}) is usually rendered too unstable, e.g., by the experimental noise.
In that situation, the canonical approach for the interpretation of XRTS measurements has been the \emph{forward modeling approach}, where a theoretical model for $S_{ee}(\mathbf{q},\omega)$, usually computed from a chemical Chihara model~\cite{Chihara_1987,Gregori_PRE_2003,bohme2023evidencefreeboundtransitionswarm} or different flavors of time-dependent DFT~\cite{baczewski_prl_16,Mo_PRL_2018,Schoerner_PRE_2023,Gawne_ElectronicStructure_2025}, is convolved with $R(\omega)$ and then matched with the experimental signal by adapting a-priori unknown parameters such as the density, temperature, or ionization state~\cite{Kasim_POP_2019}; see the red curve in Fig.~\ref{fig:XRTS} computed from a Born-Mermin ansatz in Ref.~\cite{Glenzer_PRL_2007}.
The forward fitting approach has been successfully used for the interpretation of a great variety of XRTS experiments at both x-ray backlighter laser facilities such as Omega~\cite{Glenzer_PRL_2007,Saunders_PRE_2018,Falk_PRL_2018} and the NIF~\cite{Kraus_PRE_2016,Doeppner2023} in the US, the VULCAN laser in the UK~\cite{GarciaSaiz2008} and Shenguang-II in China~\cite{Lv_POP_2019}, as well as modern hard x-ray XFELs such as the European XFEL in Germany~\cite{bespalov2026experimentalevidencebreakdownuniformelectrongas} and LCLS in the US~\cite{Fletcher2015}.

An alternative approach has recently been proposed by Dornheim \textit{et al.}~\cite{Dornheim_T_2022,Dornheim_T2_2022,Dornheim_MRE_2023}, who have suggested to switch from the usual frequency domain to the imaginary time $\tau$.
Specifically, it has long been known that $S_{ee}(\mathbf{q},\omega)$ is related to the imaginary time density--density correlation function $F(\mathbf{q},\tau)$---corresponding to the usual intermediate scattering function $F_{ee}(\mathbf{q},t)$ evaluated for $t=-i\hbar\tau$ with $\tau\in[0,\beta]$---by a two-sided Laplace transform, see Eq.~(\ref{eq:Laplace}) below.
The key ingredient is then the deconvolution theorem of the Laplace transform, which allows for a stable deconvolution to obtain $F(\mathbf{q},\tau)$.
While the final inverse Laplace transform to recover the deconvolved dynamic structure factor $S_{ee}(\mathbf{q},\omega)$ is precluded by the ill-posed nature of this inverse problem~\cite{Jarrell_Review_1996}, it has been argued that this is not strictly necessary as, by definition, $F(\mathbf{q},\tau)$ and $S(\mathbf{q},\omega)$ contain exactly the same information, only in different representations~\cite{Dornheim_MRE_2023}.
For example, the \emph{detailed balance} Eq.~(\ref{eq:detailed_balance}) is directly transformed into the symmetry relation 
$F(\mathbf{q},\tau)=F(\mathbf{q},\beta-\tau)$, which allows for a model-free extraction of the temperature by locating the minimum of $F(\mathbf{q},\tau)$ at $\tau=\beta/2$~\cite{Dornheim_T_2022,Dornheim_T2_2022}.
The application of this idea to the XRTS measurement shown in Fig.~\ref{fig:XRTS} gives a model-free temperature estimate of $T=14.8\pm2\,$eV, which is reasonably close with the model-dependent forward fitting results of $T=12\,$eV that has been reported in the original Ref.~\cite{Glenzer_PRL_2007}.

Interestingly, the fact that $F(\mathbf{q},\tau)$ attains a minimum in the first place is a direct consequence of the quantum delocalization of the electrons and, thus, constitutes an important quantum effect in warm dense matter that is measured routinely in experiments.
Subsequently, this Laplace method has been generalized to the model-free extraction of a gamut of other properties such as the normalization of the XRTS spectrum from the f-sum rule~\cite{Dornheim_SciRep_2024,Dornheim_NatComm_2025}, the degree of non-equilibrium in the system~\cite{Vorberger_PLA_2024,Bellenbaum_APL_2025}, the Rayleigh weight characterizing the electronic localization around the ions~\cite{Dornheim_POP_2025}, and the static linear density response function~\cite{schwalbe2025staticlineardensityresponse}, see also Eq.~(\ref{eq:static_chi}) below.

The considerable success of XRTS measurements for the diagnostics of extreme states of matter has sparked some promising new developments, including the implementation of diced crystal analyzer (DCA) set-ups~\cite{mcbride2018,Wollenweber_RSI_2021} at XFEL facilities.
Very recently, XRTS measurements with an unprecedented meV resolution have been reported, which allow one to resolve ionic modes~\cite{Descamps_SciReports_2020,White_PRR_2024,White2025}, or to collect high-resolution measurements of electronic spectra over a spectral range of tens of eV~\cite{Gawne_PRB_2024,Gawne_ElectronicStructure_2025,gawne2025orientationaleffectslowpair}.

\subsection{Progress in theory and simulations}\label{ss:theory-progress}
Warm dense matter is a very complex state of matter because it is situated on the boundary between condensed matter and plasmas and is influenced by ground state and excited state behavior. Moreover, Coulomb correlation effects play an important role. However, the most challenging part is the importance of quantum effects, that were described in general terms in Sec.~\ref{s:q-effects}. There exist a variety of methods that were discussed frequently before, in reviews and text books, e.g. \cite{bonitz_pop_20,bonitz_pop_24}. Here we concentrate on a subgroup that are important for accurately treating quantum effects. We will discuss two first principles methods -- fermionic path integral Monte Carlo and density functional theory simulations. After this we consider a semiclassical method -- semiclassical molecular dynamics with quantum potentials. Finally, we briefly discuss hydrodynamic approaches, focusing on the accurate treatment of correlation effects.

\subsubsection{Fermionic path integral Monte Carlo (FPIMC)}\label{ss:fpimc-progress}

\begin{figure}
    \centering
    \includegraphics[width=0.93\linewidth]{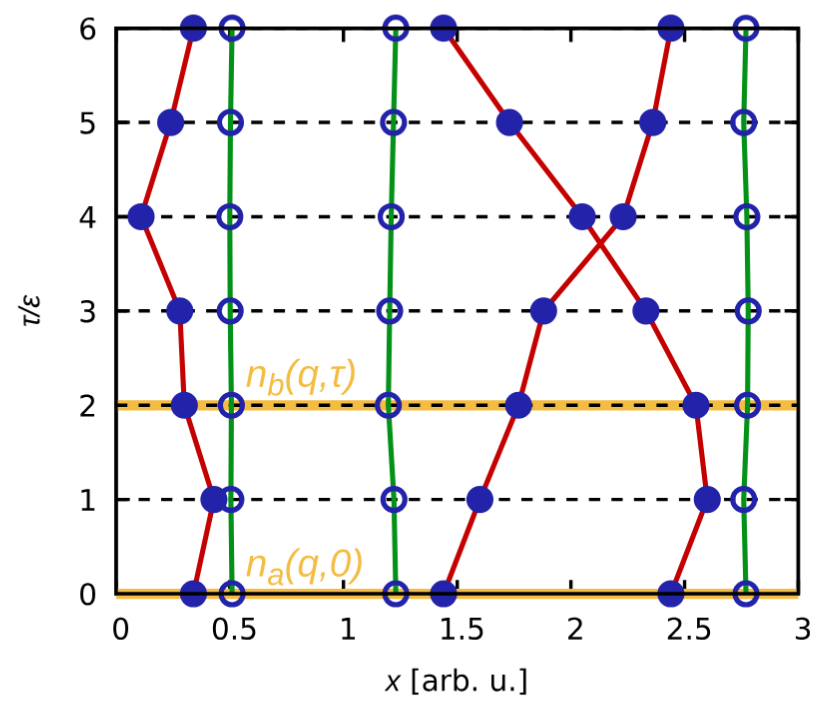}
    \caption{Schematic illustration of a PIMC configuration consisting of three electrons (red paths) and three protons (green paths) shown in the $\tau$-$x$-plane. The electrons exhibit substantially larger quantum delocalization (represented by the imaginary-time diffusion) due to their lower mass compared to the nuclei. Fermionic antisymmetry is taken into account
by sampling all permutation cycles~\cite{dornheim_jcp_19}, see the two electronic
paths on the right. The yellow horizontal lines indicate the
convenient possibility to evaluate imaginary-time correlation
functions within PIMC. Taken from Ref.~\cite{Dornheim_MRE_2024} with the permission of the authors.
    }
    \label{fig:paths}
\end{figure}

Having originally been introduced for the simulation of ultracold $^4$He in the 1960s~\cite{Fosdick_PhysRev_1966,Jordan_PhysRev_1968}, the \emph{ab initio} PIMC method~\cite{Ceperley_RevModPhys_1995} has emerged as one of the most successful methods for the simulation of correlated quantum many-body systems in thermal equilibrium.
PIMC is based on the celebrated \emph{classical isomorphism}~\cite{Chandler_JCP_1981}, where the quantum system of interest is mapped onto an effectively classical system of interacting ring polymers.
The latter are often interpreted as the eponymous \emph{paths}, which is illustrated in Fig.~\ref{fig:paths}.
Specifically, we show a configuration $\mathbf{X}$ of three hydrogen atoms, where the electrons and protons are represented by the red and green paths, respectively.
In essence, each particle is represented by a closed path throughout the imaginary time $t=-i\hbar\tau$ (with $\tau\in[0,\beta]$ and $\beta=1/k_\textnormal{B}T$ the inverse temperature), evaluated on $P$ discrete imaginary time slices of length $\epsilon=\beta/P$; this discretization becomes exact in the limit of $P\to\infty$, and more advanced factorization schemes have been explored in the literature~\cite{Sakkos_JCP_2009,Zillich_JCP_2010,Chin_PRE_2015}.
In the WDM regime, the protonic paths resemble straight lines, which correspond to classical point particles as nuclear quantum effects are negligible at medium to high temperatures.
In stark contrast, the electronic paths wriggle along their imaginary-time diffusion, and the extension of these paths is proportional to the thermal wavelength [see Eq.~(\ref{eq:lambda-def})].
The basic idea of the PIMC method is to use the Metropolis algorithm~\cite{metropolis} to randomly sample all possible paths according to the appropriate probability distribution $P(\mathbf{X})=W(\mathbf{X})/Z$, with $W(\mathbf{X})$ being the known configuration weight and
\begin{eqnarray}\label{eq:Z}
    Z = \sumint\textnormal{d}\mathbf{X}\ W(\mathbf{X})\,,
\end{eqnarray}
being the unknown partition function.
We note that explicit knowledge of the normalization $Z$ is not needed for Metropolis sampling.
An additional major complication is due to the fermionic nature of the electrons, as the required antisymmetrization under the exchange of particle coordinates makes it necessary to also sample over all $N!$ possible permutations $\sigma$ of the $N$-body permutation group $S_N$.
In the path integral picture, such permutations manifest as \emph{permutation cycles}~\cite{Dornheim_JCP_2019}, which are closed trajectories with more than a single particle in them. 
An example is given by the two electrons on the right side of Fig.~\ref{fig:paths}.
For bosons, all contributions to the partition function $Z$ [Eq.~(\ref{eq:Z})] remain strictly positive, and modern permutation sampling schemes such as the \emph{worm algorithm} by Boninsegni \textit{et al.}~\cite{Boninsegni_PRE_2006,Boninsegni_PRL_2006} (and potentially also the very recent exchange Monte Carlo approach presented in Ref.~\cite{zhao2026exchangemontecarlocontinuousspace}) allow for highly efficient simulations of $N\sim10^{4}$ particles.
For fermions, on the other hand, $W(\mathbf{X})$ can be both positive and negative, precluding the straightforward interpretation of $P(\mathbf{X})$ as a true probability suitable for Monte Carlo sampling.
As a workaround, we may define a modified probability $P'(X)=|W(\mathbf{X})|/Z'$ with the modified normalization 
\begin{eqnarray}
    Z' = \sumint\textnormal{d}\mathbf{X}\ |W(\mathbf{X})| = Z_\textnormal{Bose}\ ,
\end{eqnarray}
which, interestingly, corresponds to the bosonic partition function in the case of direct PIMC.
The exact fermionic expectation value of operator $\hat A$ is the computed as
\begin{eqnarray}\label{eq:ratio}
    \braket{\hat A} = \frac{\braket{\hat A \hat S}'}{\braket{S}'}\ ,
\end{eqnarray}
with $S(\mathbf{X})=W(\mathbf{X})/|W(\mathbf{X})|$ being the sign of configuration $\mathbf{X}$.
The denominator in Eq.~(\ref{eq:ratio}) is known simply as the \emph{average sign} $S$ in the literature,
\begin{eqnarray}
    S = \frac{1}{Z'} \sumint\textnormal{d}\mathbf{X}\ |W(\mathbf{X})| S(\mathbf{X}) = \frac{Z_\textnormal{Fermi}}{Z_\textnormal{Bose}}\ ,
\end{eqnarray}
and constitutes a straightforward measure for the amount of cancellation between positive and negative terms in a fermionic PIMC simulation~\cite{dornheim_pre_2019}.
In practice, the sign vanishes exponentially upon decreasing the temperature $T$ or increasing the system size $N$, leading to an exponential increase in the required compute time~\cite{troyer,dornheim_pre_2019}.
This is the notorious \emph{fermion sign problem}, which constitutes one of the most fundamental computational bottlenecks in many fields of physics, quantum chemistry and material science.

\begin{figure}
    \centering
\includegraphics[width=0.995\linewidth]{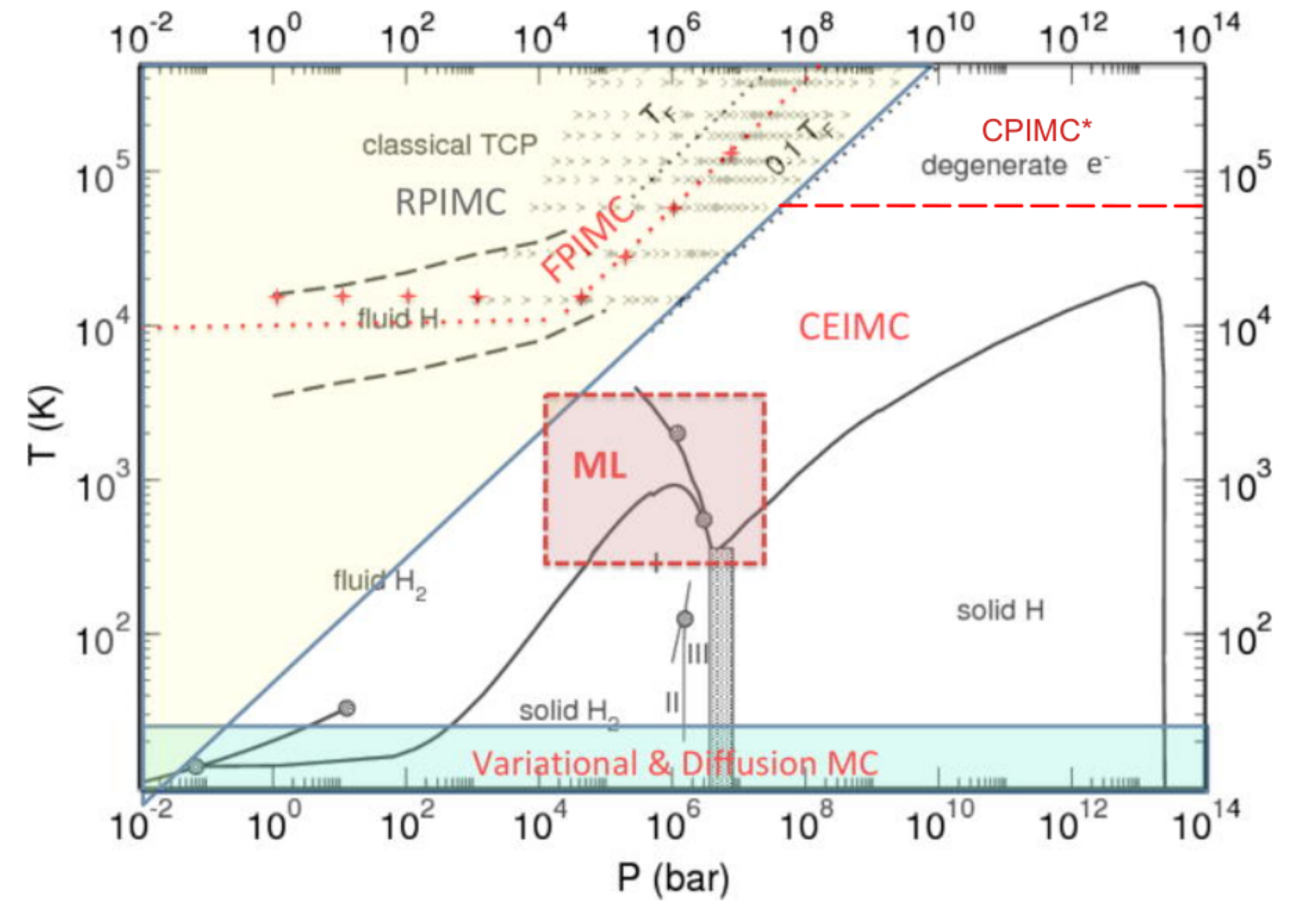}
    \caption{Capabilities of different quantum Monte Carlo methods for the example of partially ionized hydrogen. Restricted PIMC (RPIMC) is applicable inside of the yellow triangle and light black crosses indicate regions covered by the RPIMC database \cite{PhysRevB.84.224109}. The applicability range of Fermionic PIMC in coordinate space (FPIMC) is bounded by the red dotted lines, i.e. $T \gtrsim 0.5 T_F$, and was applied to $T\gtrsim 10,000$K \cite{filinov_pre_23}, whereas CPIMC extends from $r_s \sim 1$ to high densities, as was demonstrated for jellium \cite{schoof_prl15}. CEIMC denotes coupled electron-ion Monte Carlo \cite{pierleoni2016liquid}, and ML a region where machine-learned force fields were used in Ref.~\cite{PhysRevLett.130.076102}. Figure modified from Ref.~\cite{bonitz_pop_24}.
    }
    \label{fig:pimc-range}
\end{figure}

Over the years, a number of methods to deal with the sign problem have been considered in the literature~\cite{Ceperley1991,Ceperley_PRL_1992,Vorontsov_PRA_1993,Filinov_2001,Pierleoni_CPC_2005,Morales_PRE_2010,Pierleoni_PNAS_2016,Rillo_JCP_2017,Lyubartsev_PRE_2009,Chin_PRE_2015,dornheim_njp15,dornheim_cpp_19,schoof_cpp11,schoof_cpp15,schoof_prl15,Brown_2014,dubois,Hirshberg_JCP_2020,Dornheim_JCP_2020,yilmaz_jcp_20,Blunt_PRB_2014,Malone_JCP_2015,Malone_PRL_2016,Lee_JCP_2021,Militzer_PRL_2015,Xiong_JCP_2022,Xiong_PRE_2024,Xiong_JCP_2025,Dornheim_JCP_2023,Dornheim_MRE_2024,Dornheim_JCP_2024,Dornheim_NatComm_2025,dornheim2025taylorseriesperspectiveab}, including:
\begin{itemize}
    \item Restricted PIMC (RPIMC) denotes the application of the fixed-node approximation~\cite{Anderson01031995} to finite-$T$ PIMC~\cite{Ceperley1991,Ceperley_PRL_1992}. On the one hand, RPIMC formally avoids the sign problem, which means that simulations are available over a broad range of parameters~\cite{Brown_2014}, and even for second-row elements~\cite{Militzer_PRL_2015} and composite materials~\cite{Militzer_PRE_2021}.
    On the other hand, RPIMC constitutes a de-facto uncontrolled approximation as the true nodal surfaces are currently unknown. Moreover, the nodal restrictions preclude access to spectral information as it is encoded into various imaginary-time correlation functions.
    \item A number of authors~\cite{filinov_cpp05,Chin_PRE_2015,dornheim_njp15,Takahashi_Imada_PIMC_1984,Lyubartsev_2005,Filinov_CPP_2021,Xiong_JCP_2025} have suggested to directly use anti-symmetrized imaginary-time propagators, i.e., determinants for the Monte Carlo sampling. This attenuates but, crucially, does not remove the sign problem for $P>1$; indeed, the original sign problem is recovered for $P\to\infty$ as the ideal density matrices within the determinants become increasingly sparse. Nevertheless, the combination of the determinants with higher-order factorization schemes~\cite{dornheim_njp15,dornheim_cpp_19,Chin_PRE_2015,Filinov_CPP_2021,Filinov_PRE_2023}---coined \emph{permutation blocking PIMC} (PB-PIMC) by Dornheim \textit{et al.}~\cite{dornheim_njp15,dornheim_jcp15,dornheim_physrep_18,dornheim_cpp_19}---often allows to extend the domain of applicability of direct PIMC to higher densities and lower temperatures.
    \item Very recently, Xiong and Xiong~\cite{Xiong_JCP_2022} have suggested to avoid the sign problem based on direct PIMC (or path integral MD) simulations with fictitious identical particle statistics guided by a continuous parameter $\xi$; the physically meaningful cases of Fermi-, Boltzmann-, and Bose-statistics are recovered for $\xi=-1,0,1$, respectively. At weak to moderate levels of quantum degeneracy, it is often possible to infer the true fermionic limit of $\xi=-1$ exclusively from simulations in the sign-problem free domain of $\xi\geq0$~\cite{Xiong_JCP_2022,Dornheim_JCP_2023,Dornheim_JCP_2024,Dornheim_MRE_2024,Dornheim_NatComm_2025,Dornheim_JCP_2025,Morresi_PRB_2025,dornheim2025taylorseriesperspectiveab}, thereby effectively removing the exponential bottleneck with respect to the system size~\cite{Dornheim_JPCL_2024} while retaining access to spectral information~\cite{Dornheim_JCP_2023,Dornheim_MRE_2024,Dornheim_JCP_2024,Dornheim_NatComm_2025}.
    \item Bonitz, Schoof and others~\cite{schoof_cpp11,schoof_cpp15,schoof_prl15,dornheim_pop17,dornheim_physrep_18,yilmaz_jcp_20,groth_jcp17,groth_prb16} have suggested to switch from real space to second quantization in momentum space (with other representations being, in principle, also possible~\cite{schoof_cpp11}). The resulting configuration PIMC (CPIMC) method can be interpreted as a Metropolis Monte Carlo evaluation of the exact, infinite perturbation expansion around the ideal Fermi gas. Consequently, CPIMC is particularly efficient at high densities, which nicely complements real-space methods~\cite{schoof_prl15,dornheim_prb16,groth_prb16,dornheim_physrep_18} such as direct PIMC, PB-PIMC and RPIMC. A conceptually related approach with a similar range of applicability is given by density matrix QMC developed by Foulkes and others~\cite{Blunt_PRB_2014,Malone_JCP_2015,Malone_PRL_2016}.
    \item As the final method, we explicitly mention coupled electron ion Monte Carlo (CEIMC)~\cite{pierleoni_cpp19,morales2010equation,morales2010evidence,pierleoni2016liquid} developed by Pierleoni, Morales, Ceperley and others. CEIMC is based on the Born-Oppenheimer approximation, and combines a ground-state QMC treatment of the electrons with a classical or path integral MC sampling of the much heavier nuclei. Consequently, CEIMC is efficient at low temperatures and has been used extensively for the description of the infamous liquid--liquid phase transition in hydrogen at high pressure~\cite{pierleoni2016liquid}.
\end{itemize}
The often complementary parameter ranges where these methods are applicable are illustrated in Fig.~\ref{fig:pimc-range}, see also the recent review article by Bonitz \textit{et al.}~\cite{bonitz_pop_24} for a more extensive discussion specific to the simulation of dense hydrogen.

\begin{figure}
    \centering    \includegraphics[width=0.995\linewidth]{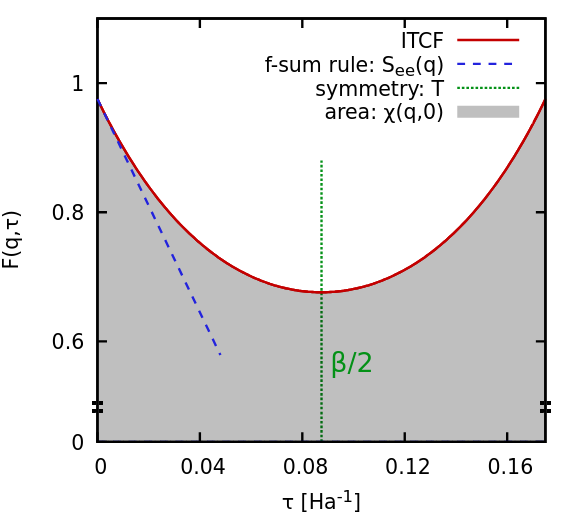}
    \caption{PIMC results for the electronic ITCF $F_{ee}(\mathbf{q},\tau)$ for warm dense beryllium at $T=155.5\,$eV, $\rho=7.5\,$g/cc, and $q=7.68\,$\AA$^{-1}$. The first derivative at $\tau=0$ is specified by the f-sum rule (dashed blue)~\cite{Dornheim_SciRep_2024}. Note the symmetry of the ITCF around $\tau=\beta/2$ (dotted green) due to detailed balance in thermal equilibrium [Eq.~(\ref{eq:detailed_balance}]~\cite{Dornheim_T_2022}. The area under the ITCF (grey) gives one direct access to the static linear density response [Eq.~(\ref{eq:static_chi})]~\cite{Dornheim_MRE_2023}.
    Adapted from Ref.~\cite{schwalbe2025staticlineardensityresponse} with the permission of the authors.
    }
    \label{fig:ITCF}
\end{figure}

Let us conclude this brief overview of QMC methods with a short discussion of a class of observables that depend particularly intimately on the quantum nature of the simulated systems: imaginary-time correlation functions (ITCF)~\cite{Dornheim_JCP_2021}.
A particularly important example is given by the density--density ITCF 
\begin{align}
F(\mathbf{q},\tau)=\braket{\hat{n}(\mathbf{q},0)\hat{n}(-\mathbf{q},\tau)}\,,   \label{eq:itcf}
\end{align}
where we correlate single-particle density operators at an imaginary-time difference $\tau$, see also the horizontal yellow lines in Fig.~\ref{fig:paths} above.
In Fig.~\ref{fig:ITCF}, we show PIMC results for the ITCF of beryllium at $T=155.5\,$eV, $\rho=7.5\,$g/cc and at a wavenumber of $q=7.68\,$\AA$^{-1}$.
We first note the symmetry of the ITCF around $\tau=\beta/2$, which directly follows from the detailed balance, Eq.~(\ref{eq:detailed_balance}). This is a pure quantum effect. In contrast, for classical systems, $S_\textnormal{cl}(\mathbf{q},-\omega)=S_\textnormal{cl}(\mathbf{q},\omega)$.
Similarly, the monotonic decay of the ITCF for $\tau<\beta/2$ is directly due to the imaginary-time diffusion illustrated in Fig.~\ref{fig:paths}, which, in turn, is caused by quantum delocalization.
The first derivative of $F(\mathbf{q},\tau)$ around $\tau=0$ follows from the f-sum rule~\cite{Dornheim_SciRep_2024,Dornheim_moments_2023,Dornheim_MRE_2023,Dornheim_PTRS_2023} and is proportional to $\sim q^2$. Consequently, the $\tau$-decay of the ITCF becomes steeper with increasing $q$, as we probe density correlations on smaller length scales, $\lambda=2\pi/q$, where quantum effects become increasingly important.
The area under the ITCF determines the static linear density response via the imaginary-time version of the fluctuation--dissipation theorem~\cite{Dornheim_MRE_2023},
\begin{eqnarray}\label{eq:static_chi}
    \chi(\mathbf{q},0) = -n\int_0^\beta\textnormal{d}\tau\ F(\mathbf{q},\tau)\ ,
\end{eqnarray}
which would revert to the classical relation $\chi_\textnormal{cl}(\mathbf{q},0)=-n\beta S_\textnormal{cl}(\mathbf{q})$ if there was no quantum delocalization such that $F(\mathbf{q},\tau)\equiv F(\mathbf{q},0) = S(\mathbf{q})$.
The quantum case, instead, demands a more elaborate Fourier Matsubara series, see Refs.~\cite{Tolias_JCP_2024,Dornheim_PRB_2024}.

Lastly, we mention the relation of $F(\mathbf{q},\tau)$ to the dynamic structure factor 
\begin{eqnarray}\label{eq:Laplace}
    F(\mathbf{q},\tau) = \mathcal{L}\left[ S(\mathbf{q},\omega) \right] = \int_{-\infty}^\infty \textnormal{d}\omega\ S(\mathbf{q},\omega)\ e^{-\hbar\omega\tau}\ ,
\end{eqnarray}
which can be used as the starting point for an analytic continuation, i.e., the numerical inversion of Eq.~(\ref{eq:Laplace}) to solve for $S(\mathbf{q},\omega)$~\cite{Jarrell_Review_1996,dornheim_prl_18,Filinov_PRB_2023,Chuna_JPA_2025,Chuna_JCP_2025,chuna2025noiselesslimitimprovedpriorlimit}.
We note that Eq.~(\ref{eq:Laplace}) also has become important for the model-free interpretation of XRTS experiments with WDM, see Sec.~\ref{ss:xrts}.
Other relevant ITCF that can be computed from PIMC simulations include the Matsubara Green function [see Sec.~\ref{ss:fpimc-mgf} below], the velocity--velocity ITCF, and also higher-order density ITCFs (i.e., three- and four-body correlators~\cite{Dornheim_JCP_2021}), which are related to different non-linear density response properties~\cite{Dornheim_JCP_2021,Dornheim_JPSJ_2021,Dornheim_CPP_2022,Vorberger2025,dornheim_pop_23}.

\subsubsection{Density functional theory simulations}\label{ss:dft-progress}

The fully-quantum treatment of non-ideal plasmas by FPIMC, discussed in the above section, is essentially a complete $3N$-dimensional problem. Its use is still limited to model systems like homogeneous electron gas and simple hydrogen system of certain size. Extending its applicability to partially ionized and mid-/high-Z plasmas needs overcoming many challenges, both computationally and conceptually. However, the FPIMC method, even applying to simple systems, can provide critical data of quantum electron exchange-correlation interaction for downfolding methods such as finite-temperature density-functional theory (DFT).

The density-functional theory (DFT) method was mainly developed in 1960s by Kohn, Sham, and Hohenberg \cite{hohenberg-kohn, kohn_sham}, which was extended to finite temperatures by Mermin \cite{mermin_1965}. As a downfolding method from FPIMC [cf. Sec.~\ref{s:downfolding}], finite-temperature DFT reduces the $3N$-dimensional quantum many-body problem to $N$-coupled $3$-dimensional problems. Namely, instead of having $N$-interacting electrons in a real system, one considers each individual electron in an effective potential of ions and other electrons \cite{Hu_OUP_Book_2026}. This dimensional reduction significantly speeds up simulations of quantum plasmas, while the price to pay is that the Hamiltonian needs to be approximated in the Kohn-Sham equation. To be specific, the exchange-correlation (XC) potential ($V_{\rm xc}$) is largely unknown in the following Kohn-Sham equation (in atomic units where $\hbar=m=e=1$) for many realistic systems:
\begin{equation}
    \left \{  -\frac{1}{2} \nabla ^2 + V_{\rm ie}[n({\bf r})] + V_{\rm ee}[n({\bf r})] + V_{\rm xc}[n({\bf r})] \right \} \phi_i({\bf r}) = \epsilon_i \phi_i({\bf r})\,,
    \label{eq:ks-equations}
\end{equation}
with $V_{\rm ie}$ and $V_{\rm ee}$ representing the electron-ion attraction and the electron-electron (Hartree) repulsion, respectively. The electron density is composed by using the Kohn-Sham orbitals:
\begin{equation}
    n({\bf r}) = \sum_i f_{\rm FD}(\epsilon_i) \left | \phi_i({\bf r})\right |^2
\end{equation}
with $f_{\rm FD}(\epsilon_i)=2/[1+{\rm exp}(\epsilon_i-\mu)/kT)]$ being the Fermi-Dirac distribution of electrons on the $i$-th eigenstate of a spin-unpolarized system (factor 2 for the two spin projections). Solving the Kohn-Sham equations (\ref{eq:ks-equations}), usually by iterative methods \cite{Hu_OUP_Book_2026},  gives the thermal ground-state electron density of a quantum plasma. 

The predictive power of DFT, however, depends critically on the exchange–correlation approximation employed. The search for predictive XC-functionals immediately started after the establishment of the above Kohn-Sham equation in the 1960s. It took about fifteen years to have the first reliable zero-temperature XC-functional which is the so-called local density approximation (LDA) \cite{perdew_zunger, PerdewWang_PRB_1992}. It was built from the parametrization of quantum Monte Carlo (QMC) simulations of the uniform electron gas \cite{ceperley_alder} (see also Sec.~\ref{ss:ueg-benchmarks}). In the following two to three decades after the LDA XC-functional, more and more accurate zero-temperature XC-functionals were systematically developed in different levels of approximations, such as generalized gradient approximation (GGA) \cite{PBE}, meta-GGA \cite{SCAN}, and hybrid functionals \cite{HSE}, mostly for condensed matter physics and materials science applications.  

Since the mid-1990s, the high-energy-density (HED) physics community has applied the quantum molecular dynamics (QMD or Kohn-Sham-MD) framework for simulating quantum plasmas, which couples the DFT calculation of electronic structure with molecular dynamics for classical ions \cite{collins_pre_95}. Over the past three decades, the DFT-based QMD  method has achieved remarkable success even using ground-state (zero-temperature) XC-functionals. An overview of different finite-$T$ functionals and their benchmarks against FPIMC is given in Sec.~\ref{ss:fpimc-dft}. Here, we present a few recent examples of successful DFT-based simulations of quantum plasmas (even using zero-temperature XC-functionals) for better understanding HED experiments. In some of these examples, we intend to highlight the importance of XC-functionals used in these {\it first-principles} DFT simulations, to close the gap between QMD simulations and experimental observations.

\begin{figure}[h]
    \centering
    \includegraphics[width=\linewidth]{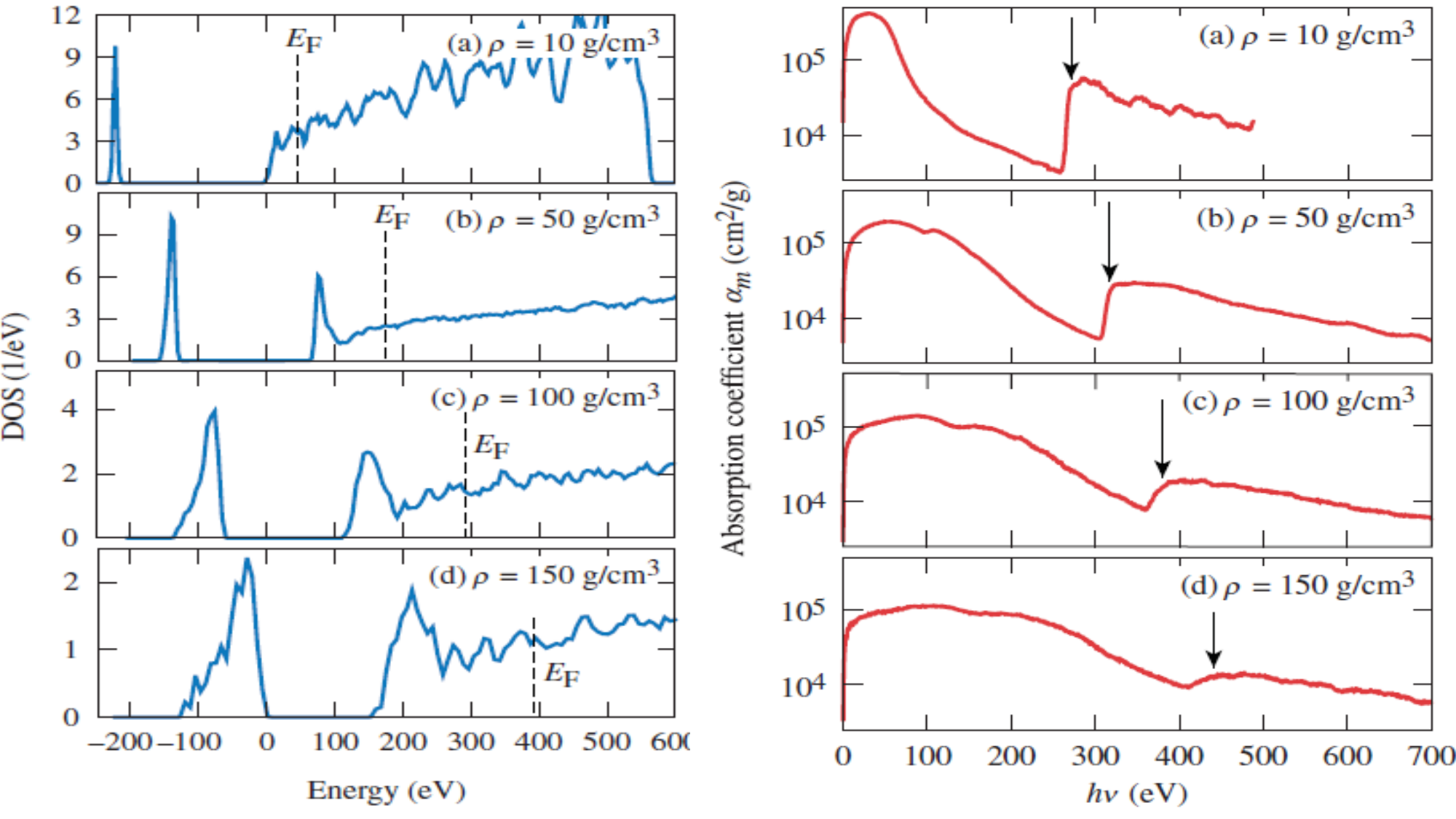}
    \caption{The DFT-based QMD (KS-MD) calculations of warm dense carbon at the same temperature of $T=15\,625$ K but different mass densities. Left panels show the density of states (DOS) for different density cases, while  right panels indicate the DFT-predicted X-ray absorption spectra for the corresponding cases. The quantum degeneracy effect of electrons is manifested by the up-shifting K-edge of carbon, due to the dense-plasma phenomenon so-called ``Fermi-surface raising" \cite{Hu_PRL_2017} (also known as ``Fermi barrier" \cite{bonitz_cpp_25} or ``Pauli blocking"). Adapted from Ref.~\cite{Hu_PRL_2017}.
    }
    \label{fig:Kedge}
\end{figure}

The first example we give here is the DFT-based QMD simulations of warm-dense carbon plasma \cite{Hu_PRL_2017}, which is indicated in Fig.~\ref{fig:Kedge} for different mass densities varying from $\rho=10$ g/cm$^3$ to $\rho=150$ g/cm$^3$ but at the same temperature of $T=15\,625$ K. DFT can properly describe the quantum behavior of electrons that are ``packed" within a smaller volume (high densities), while molecular dynamics of ions can sample ionic configurations of the warm-dense carbon plasma. To be specific, the quantum degeneracy of electrons and the Pauli blocking (i.e., maximum occupation of two electron for each state that is required by Fermions) are fully incorporated into the quantum treatment of many-electron system through the Kohn-Sham DFT scheme. The QMD simulations have employed $64-128$ atoms in a cubic box with periodic boundary condition, a bare Coulomb potential being used for electron-ion interactions with an energy cutoff of $E_{cut}=60-100$ keV, and the Perdew-Burke-Ernzerhof (PBE) exchange-correlation functional \cite{PBE}. These simulations were done with VASP (Vienna Ab-Initio Simulation Package). 

Figure~\ref{fig:Kedge}(a) shows the density-of-state (DOS) from these QMD calculations, in which the discrete $1s$ state moves up in energy as the carbon mass density increases. The up-shifting of $E_{1s}$ energy is just a consequence of the increased free-electron screening to the nuclear charge, which is essentially the phenomenon often called ``continuum lowering" or ``ionization potential depression" in classical plasma physics. DFT treatment of dense plasmas properly catches that behavior without the need of any ad hoc continuum lowering models. The other quantum effect of such a many-electron system is the so-called ``Fermi-surface rising" \cite{Hu_PRL_2017}, which is marked by the vertical dashed line in Fig.~\ref{fig:Kedge}(a). Namely, the electrons being ``ionized" are still fully occupying some of the positive-energy states, even though they are ``free". Consequently, any bound electrons cannot transition to these ``free" but still ``occupied" states, which leads to the X-ray absorption K-edge being moved to higher energies as shown in Fig.~\ref{fig:Kedge}(b). This quantum feature of ``Pauli blocking" has been properly attributed to the ``Fermi barrier" effect in a recent FPIMC treatment of dense plasmas \cite{bonitz_cpp_25}. It is noted that any treatment of dense plasmas based on Kohn-Sham DFT, including average-atom models \cite{PhysRevLett.120.119501, PhysRevResearch.3.023026} or the single-atom-in-a-periodic-box model \cite{Hu_PRL_2017, PhysRevLett.120.119502}, can essentially capture this quantum many-electron effect.            

\begin{figure}[h]
    \centering
    \includegraphics[width=1.05\linewidth]{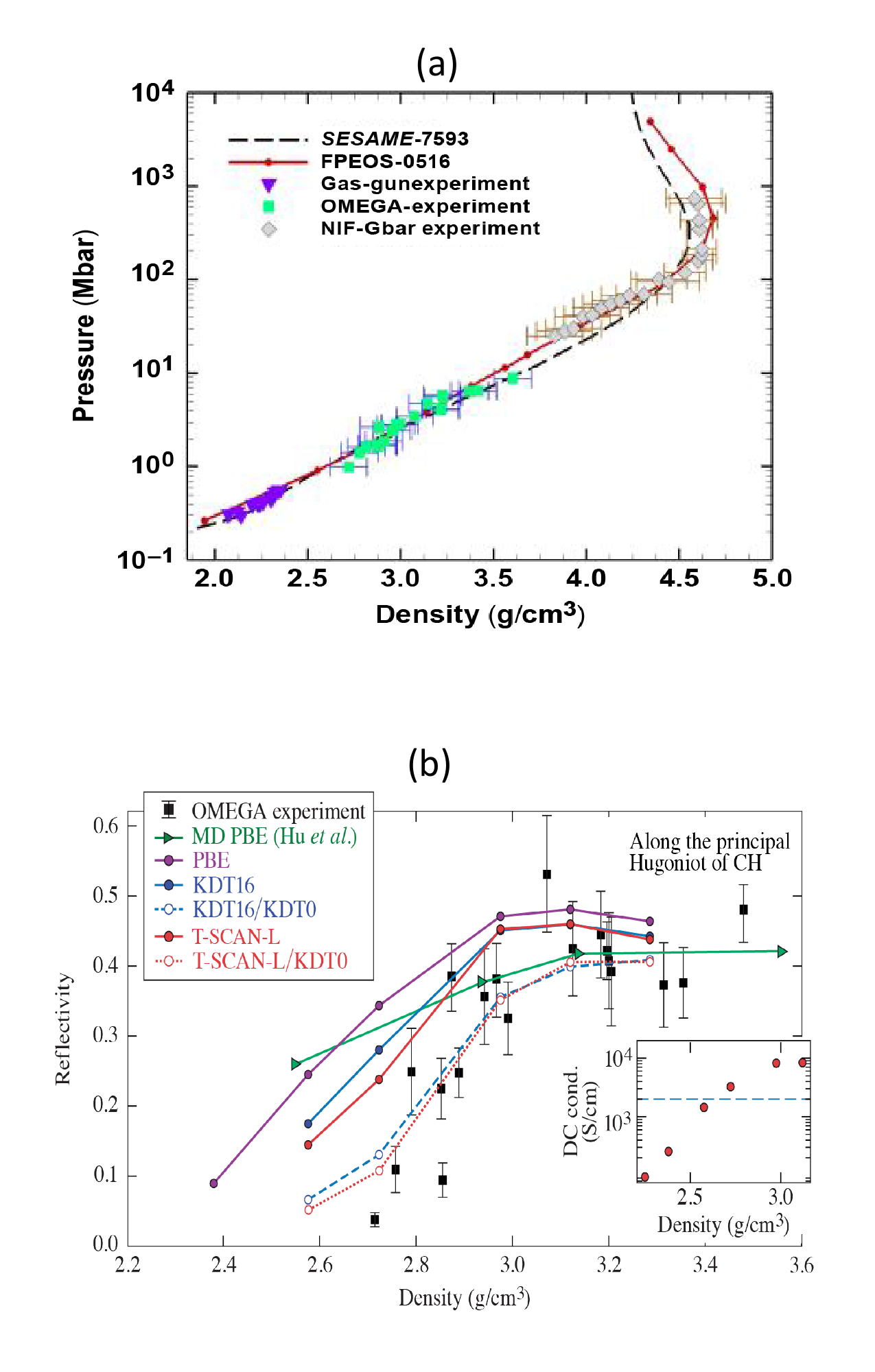}
    \caption{Comparisons of static and optical properties of shock compressed polystyrene (CH) between DFT-based QMD simulations and experimental data: (a) The shock  pressure {\it versus} density along the principal Hugoniot; and (b) the reflectivity of CH-shock up to $\sim 10$ Mbar pressures. Adapted partially from Ref.~\cite{goshadze2023shock}.
    }
    \label{fig:CHshock}
\end{figure}

The second example of DFT simulations for HED experiments is the shock compression of polystyrene (CH). As a cheap and easy to make material, CH is often used as a proper ablator material for manufacturing ICF capsules. Its static, dynamic, and optical properties under extreme HED conditions have been extensively investigated in the past, both experimentally using gas guns and high-energy lasers \cite{Gasgun_1983, cauble1997absolute, koenig2003optical, ozaki2009shock,  barrios2010high, Doeppner2018, Kritcher2020, mccoy2020measurement} and theoretically with PIMC and DFT-based QMD simulations \cite{wang2011thermophysical, hamel2012lazicki, hu2014properties, hu2015first, huser2015experimental, colin2016dissociation, hu2016first, hu2017optical, zhang2017first, hu2018review}. In shock experiments, the principal Hugoniot of CH was often measured to obtain the shock pressure as a function of shock density. Such experimental Hugoniot results are shown by symbols in Fig.~\ref{fig:CHshock}(a), for experiments from gas-gun shocks \cite{Gasgun_1983} at low pressures of $P<1$ Mbar as well as from energetic laser drives at high pressures up to $P\sim 1$ Gbar \cite{barrios2010high, Doeppner2018, Kritcher2020}. It can be seen that the DFT-based first-principles equation-of-state (FPEOS) table (FPEOS-0516) \cite{hu2015first} gives overall good agreement with experimental data in the whole range of pressures, while the SESAME-7593 model shows noticeable difference in the high-pressure region at $P>10$ Mbar.

In addition to shock pressure-density measurement, some laser-driven experiments \cite{barrios2010high} have also measured the reflectivity of VISAR light ($\lambda=532$ nm) from the CH-shock front, along the principal shock Hugoniot. These reflectivity measurements provide necessary data for testing XC-functional approximations used in DFT simulations. The optical-property comparison of CH shock between experiments and DFT-based Kubo-Greenwood (KG) calculations is made in Fig.~\ref{fig:CHshock}(b). It indicates that the Kubo-Greenwood calculations using both GGA and meta-GGA XC functionals, such as PBE, KDT16~\cite{Karasiev_PRL_2018}, T-SCAN-L~\cite{PhysRevB.105.L081109}, do not reproduce the experimentally observed reflectivity of CH shock, as they overestimate the CH band-gap closing induced by shocks \cite{wang2011thermophysical,hu2014properties}. Only the thermal hybrid KDT0~\cite{PhysRevB.101.245141} XC functional used in KG-calculations can properly catch the shock-induced band-gap closing of CH,  in which the temperature-dependent XC functionals of KDT16 and T-SCAN-L are necessary for predicting the molecular structure of CH shock in QMD simulations \cite{goshadze2023shock} (see Sec. \ref{ss:fpimc-dft} for a description of the thermal XC functionals development).         

\begin{figure}[h]
    \centering
    \includegraphics[width=\linewidth]{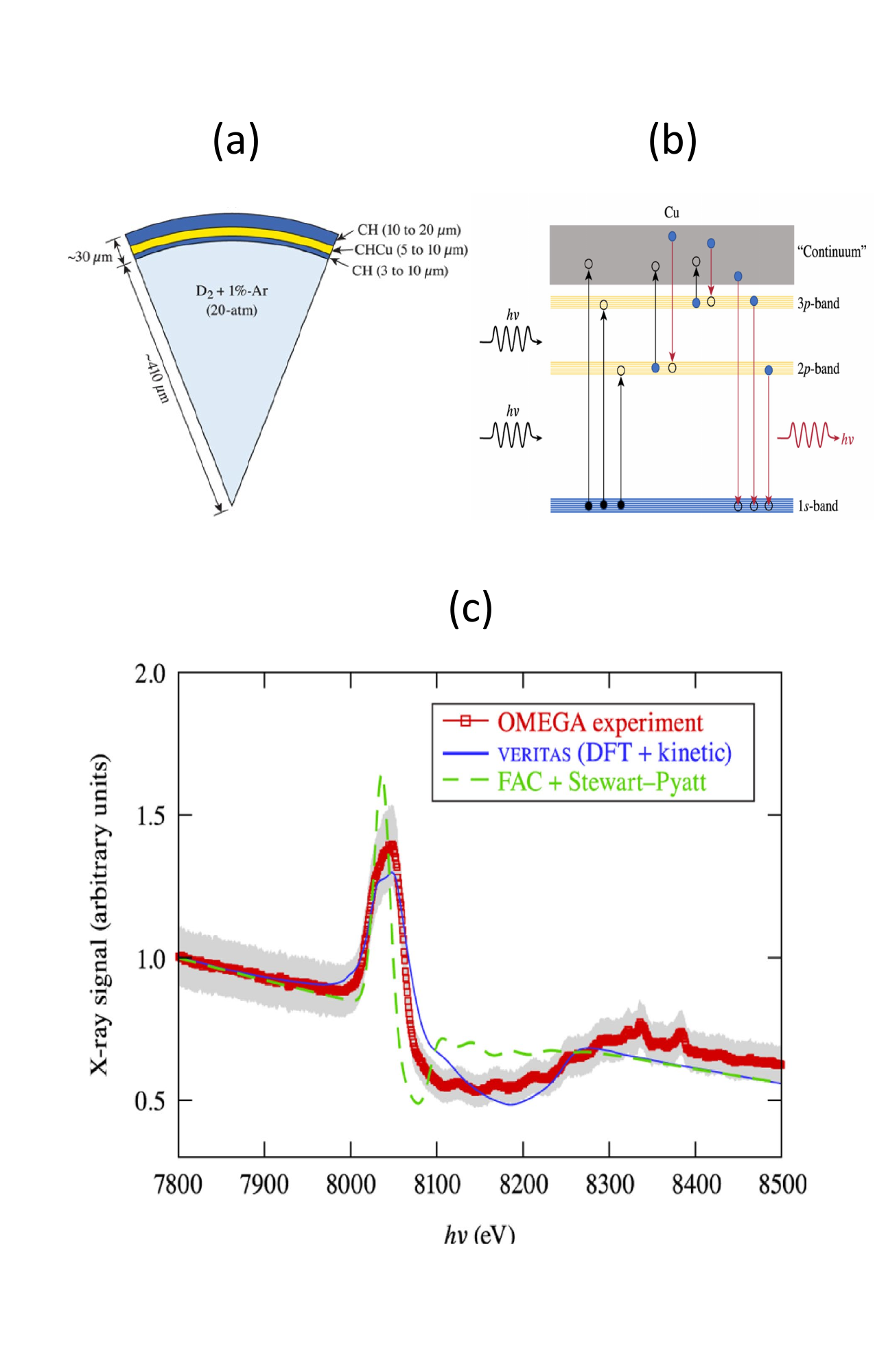}
    \vspace{-1.2cm}
    \caption{(a) Schematic diagram of targets used for implosion X-ray spectroscopy experiments on the Omega Laser Facility \cite{Hu_Nature_Communications_2022}; (b) The DFT-based kinetic model --- {\it VERITAS} \cite{Hu_PoP_2025}---invokes the electronic bands relevant to quantum radiative transition of electrons in the sample element copper (Cu); and (c) The comparison of the time-integrated X-ray emission/absorption spectrum among experiment, DFT-based {\it VERITAS} model, and a traditional  collision-radiative equilibrium (CRE) model. The quantum treatment of the dense plasma environment is key to mostly reproduce the detailed  experimental features. Adapted  from Ref.~\cite{Hu_PoP_2025}.
    }
    \label{fig:Xspec}
\end{figure}

The last example of DFT simulation of HED experiments is shown in Fig.~\ref{fig:Xspec}, in which DFT-based kinetic model is used to properly interpret X-ray spectroscopy experiment with laser-driven implosion \cite{Hu_Nature_Communications_2022}. X-ray emission and absorption in plasmas are essentially related to energy transitions of quantum electrons within radiative atoms. X-ray spectroscopy is often used to infer the  plasma density and temperature conditions \cite{FloridoPhysRevE2011} and to reveal novel phenomena such as inter-species radiative transitions occurred in super-dense plasmas \cite{hu2020interspecies}. We recall that modern quantum mechanics was originated from Bohr's hydrogen model being proposed to explain the discrete spectral lines of hydrogen. Thus, quantum treatment of electrons is necessary for interpreting X-ray spectroscopy experiments, in particular for dense plasmas. How a dense plasma environment affects the quantum electron behavior inside a radiative atom is truly an important aspect of quantum plasmas. 

X-ray spectroscopy experiments {\it via} laser-driven implosions have recently been demonstrated on the Omega Laser Facility \cite{Hu_Nature_Communications_2022, Hu_PoP_2025}. For such experiments on OMEGA, the  schematic diagram of a typical target is shown in Fig.~\ref{fig:Xspec}(a) in which some portion of the CH capsule was doped with 2\%-4\% copper (Cu) in atomic fraction. The capsule was filled by D$_2$ gas with 1\%-Ar doping. The target is imploded by 60 OMEGA laser beams with a total energy of $\sim 25$ kJ. Once the imploding shell stagnates, a hot spot is formed which has a temperature of $k_BT=1-2$ keV providing X-rays to ``backlight" the stagnated Cu-doped CH shell of high densities and temperatures ($\rho=10-20$ g/cm$^3$ and $k_B T=200-550$ eV). The intense radiation from Ar-doped hot spot can pump $K_{\alpha}$ emission at around $h\nu \simeq 8050$ eV and $1s\rightarrow 2p$ absorption at relatively high photon energies of $h\nu \simeq 8100-8300$ eV, shown by red squares in Fig.~\ref{fig:Xspec}(c). The time-integrated x-ray spectral data was obtained using x-ray spectrometer. To interpret these experimental data, a DFT-based kinetic model ({\it VERITAS}) was developed \cite{Hu_PoP_2025}, in which the relevant energy bands of Cu were invoked by the kinetic rate equation. These energy bands of Cu in the dense CH plasma, depicted in Fig.~\ref{fig:Xspec}(b), were calculated from DFT-based QMD simulations. With these DFT-determined energy bands, the {\it VERITAS} model uses rate equations to track the time evolution of electron populations on these DFT energy bands, through considering the possible radiative transition pathways among them. Such a kinetic treatment of band populations was further coupled with radiation transport to obtain the simulated spectra. The {\it VERITAS} results was plotted as the blue solid line in Fig.~\ref{fig:Xspec}(c), which overall reproduced both emission and absorption features. The traditional collisional-radiative models using ad-hoc continuum lowering models with isolated atomic data, such as the one marked as ``FAC+Stewart-Pyatt" with green dashed line in Fig.~\ref{fig:Xspec}(c),  misses the detailed absorption feature seen in experiments. This demonstrated the necessary DFT treatment of quantum dense plasma environment.                      

\subsubsection{Semi-classical Molecular Dynamics simulations}\label{ss:md-progress}
A broadly applied method is semiclassical MD simulations with effective interactions (force fields) that are derived from zero temperature many-body simulations. MD simulations with force fields have been significantly extended by including neural networks, see Ref.~\cite{vorberger_wdm_26} for a review. This concept has been extended to charged particles and finite temperatures in quantum plasma simulations. Here effective quantum pair potentials were derived within many-body perturbation theory by Kelbg and others \cite{kelbg_ap_63,kelbg_ap_63_2, kelbg_ap_64,bonitz_cpp_23}. These potentials take into account quantum delocalization and yield a finite value at zero particle separation, in line with the sketch in Fig.~\ref{fig:q-effects}.
This was extended to strong coupling, where perturbation theory fails, by Filinov, Bonitz and others who solved the two-particle density matrix exactly, resulting in improved Kelbg potentials between electrons and ions in hydrogen~\cite{filinov_jpa03, filinov_pre04}. The results for the thermodynamic properties of dense quantum plasmas above approximately $T \sim 60\,000\,$K are very accurate as was demonstrated in Ref.~\cite{filinov_pre04, bonitz_pop_24}.

While the original Kelbg potential reproduces the correct value of the derivative at zero particle separation, at $r=0$, the value of the potential at $r=0$ is not correct. Therefore, one derives an \textit{improved Kelbg potential}~\cite{filinov_pre04} that contains an additional parameter, which leaves the first derivative at $r=0$ unaffected but can be adjusted such that the exact value of the potential at $r=0$ is matched as well. Accurate fits for its temperature dependence, both for electron-electron and electron-proton interactions, have been provided~\cite{filinov_pre04}. Electron spin effects can be incorporated via additional temperature-dependent potentials. Applications include, e.g., electron-ion temperature relaxation~\cite{benedict2012pre,lavrinenko2024cpp}. Semi-classical MD simulations have also been applied recently to compute the dynamic structure factor of dense hydrogen~\cite{kaehlert2026pop}. 

\begin{figure}
    \centering
    \includegraphics[width=\linewidth]{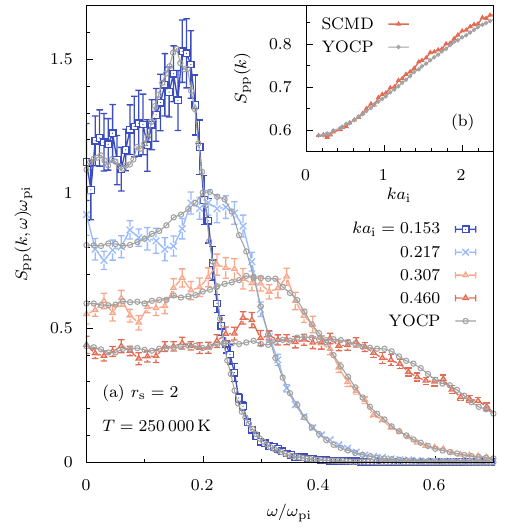}
    \caption{(a) Proton-proton dynamic structure factor at various wave numbers $k$ from semi-classical MD simulations (SCMD) of a dense hydrogen plasma at $r_s=2$ and a temperature of $T=250\,000\,\text{K}$ using the improved Kelbg potential of Filinov \textit{et al.}~\cite{filinov_pre04}. Results from YOCP simulations with $\kappa=1.55$ and $\Gamma=0.63$ are shown for comparison. (b) Static structure factors. Adapted from Ref.~\cite{kaehlert2026pop}.
    }
    \label{fig:dsfpp}
\end{figure}

Results for the proton-proton dynamic structure factor at $r_s=2$ and $T=250\,000\,\text{K}$ are shown in Fig.~\ref{fig:dsfpp}. At the smallest wave number $k$, a pronounced peak can be observed, corresponding to an ion-acoustic mode. Although the protons behave as classical particles at these conditions, quantum effects enter through the electrons, which effectively screen the proton-proton interaction. Simulations for a one-component Yukawa plasma (YOCP, see Sec.~\ref{ss:hydro-progress}) are shown for comparison, where the screening parameter is $\kappa=a_i/\lambda_s$, where $\lambda_s$ is the screening length of the Yukawa potential, that was adjusted to match the proton-proton static structure factor of the semi-classical MD simulations at small $k$. They reproduce the dynamic structure factor very accurately, at a fraction of the computational cost. With the recent availability of meV resolution in x-ray scattering experiments~\cite{mcbride2018}, the ionic contribution to the total electron dynamic structure factor can be resolved, providing access to ion transport properties such as viscosity or thermal conductivity. Recently, this has been successfully demonstrated on the example of warm dense methane, where the sound speed could be determined~\cite{white2024pre}.

\subsubsection{Hydrodynamics for correlated quantum plasmas}\label{ss:hydro-progress}

Hydrodynamic models offer a practical alternative to first principles simulations for analyzing dynamical processes in macroscopic plasmas on large length and time scales.
However, in the regime where $\Gamma_i \geq 1$ and ions are strongly coupled, the standard hydrodynamic equations must be modified to properly account for interaction and correlation effects.
Such modifications have been developed for the effective model of a Yukawa one-component plasma.
In this model, only the classical ions are dynamical and interact via the screened Coulomb (Yukawa) potential, $\phi(r)=Q_i^2e^{-r/\lambda_s}/r$, where $r$ is the distance between particles and $\lambda_s = \lambda_s(n_e,T)$ is the fixed screening length due to quantum electrons, which are assumed to form a neutralizing background.
The validity of this model is justified by the large mass ratio between ions and electrons, which leads to a separation of time scales.
\par 

The YOCP is conveniently characterized by the dimensionless parameters $\Gamma_i$  [Eq.~\eqref{eqn:gammai}] and $\kappa = a_i/\lambda_s$, where quantum effects of the electrons are effectively included through an appropriate choice of the screening parameter $\kappa$.
Various approaches have been proposed to determine their values for a given two-component plasma.
These include direct calculations based on a specific physical model for $\lambda_s$~\cite{murillo2008, haxhimali2015pre, stanton_pre_15, moldabekov2018pre} using known physical parameters (electron density and temperature), e.g., employing interpolations between the Thomas-Fermi screening length, at $T=0$, and the classical Debye length, at high $T$~\cite{murillo2008}. 
For the ICF compression path shown in Fig.~\ref{fig:overview}, at an intermediate point with $T = 67\,600\textrm{ K}$  and $r_s = 1.6$,
%$T = 316\,000\textrm{ K}$ and $r_s = 1$
and estimating the degree of ionization [Eq.~\ref{eq:alpha-ion}] using FPIMC data in Fig.~\ref{fig:ipd_ieff} as $\alpha^{\rm ion} = 0.92$, this interpolation yields $\Gamma_i = 2.85$ and $\kappa = 2.92$.
Other approaches include fitting $\Gamma_i$ and/or $\kappa$ to various data, such as the pair distribution function~\cite{ott_pre_11, ott_pop_14, ott_cpp15, clerouin2016prl, wang2020pre, clerouin2022pre, issanova2025aps}, dynamic structure factor~\cite{moldabekov2019pre}, and velocity autocorrelation function~\cite{ott2011pop}. 
As an example, a recently proposed method that uses both short-range (through the pair distribution function $g(r)$ near $r \sim a_i$) and long-range (through the static structure factor $S(k)$ in the limit $k \to 0^+$) static properties of ions~\cite{krimans2025cpp} can be applied to the two-component hydrogen plasma discussed in the previous subsection.
For parameters $T = 250\,000$\textrm{ K} and $r_s = 2$, this method results in $\Gamma_i = 0.67$ and $\kappa = 1.56$.

Several approaches have been proposed to construct hydrodynamic models for classical strongly coupled plasmas such as the YOCP.
These include generalized hydrodynamics~\cite{kaw1998plasmas}, 
and the viscoelastic-density functional model~\cite{diaw2015pre}. 
However, these models require additional external input, such as relaxation time or viscosity.
Simpler approaches include standard Euler hydrodynamic equations supplemented by a YOCP equation of state and a mean-field interaction term~\cite{khrapak2016physcontrolfusion, salin2007plasmas, rao1990planetspacesci, khrapak2015pre}.

Here we focus on a recently proposed approach that derives the hydrodynamic equations from a variational principle (variational hydrodynamics, VH)~\cite{krimans2024fluids, krimans2025pre}.
This approach uses an averaged, macroscopic Lagrangian
\begin{equation}
    \label{eq:hydro_lagrangian}
    \begin{gathered}
        L = \int\left( \frac{1}{2}m\vec{v}^2-\frac{3}{2}k_BT \right)n(\vec{x})\,\mathrm{d}\vec{x}
        \\-\frac{1}{2}\iint\phi(|\vec{x}-\vec{x}'|)n(\vec{x})n(\vec{x}')g(n_0,T,|\vec{x}_0-\vec{x}_0'|)\mathrm{d}\vec{x}\mathrm{d}\vec{x}'.
    \end{gathered}
\end{equation}
\par 
The first integral arises from averaging the kinetic energy using the one-particle number density $n$.
It is separated into a macroscopic contribution, written in terms of the macroscopic velocity field $\vec{v}$, and a microscopic contribution, approximated using the local temperature field $T$. 
The second integral represents the interaction energy, where the potential $\phi$ is averaged over the two-particle number density $n_2(\vec{x},\vec{x}')=n(\vec{x})n(\vec{x}')g(\vec{x},\vec{x}')$, which is expressed in terms of the one-particle densities and the pair distribution function $g$, where the latter is approximated by assuming local thermodynamic equilibrium.
In this case, $g$ depends on the local temperature $T$ and, additionally, on the reference (initial or equilibrium) positions $\vec{x}_0$ and $\vec{x}_0'$, where the number density is given by $n_0$.
Within this framework, the effects of strong ion coupling and correlations are incorporated explicitly through the pair distribution function in the Lagrangian, whose dependence on the physical parameters can be taken from experimental data, theoretical models, or MD simulations.

The resulting equations of motion are given by
\begin{equation}
    \begin{gathered}
        mn\left(\frac{\partial \vec{v}}{\partial t} + \left( \vec{v}\cdot\vec{\nabla} \right)\vec{v} \right) = 
        \\
        - \vec{\nabla}p - \int \vec{\nabla}\phi\Big|_{\vec{x}'}n(\vec{x})n(\vec{x}')\left( \frac{g+g^T}{2} \right)\mathrm{d}\vec{x}',
    \end{gathered}
\end{equation}
which have a form similar to the standard Euler hydrodynamic equations.
The key difference is that the integral over $\vec{\nabla}\phi$ generalizes the usual mean-field term by including additional weighting through the pair distribution function $g$ (where $g^T$ denotes its transpose).
At the same time, the pressure term $p$ generalizes the thermodynamic pressure to
\begin{equation}
    \begin{gathered}
        p = -\frac{\partial T}{\partial (1/n)}\Bigg|_s \left( \frac{3}{2}k_B + \frac{1}{2} \int n(\vec{x}')\phi(|\vec{x}-\vec{x}'|) \frac{\partial g}{\partial T}\mathrm{d}\vec{x}' \right),
    \end{gathered}
\end{equation}
where the derivative of the temperature corresponds to the adiabatic derivative at constant entropy $s$.
The second term, given by the integral over the potential $\phi$, represents the contribution from correlation effects, in addition to the first term that corresponds to the ideal gas pressure.

\begin{figure}[t!]
    \centering
\includegraphics[width=0.49\textwidth]{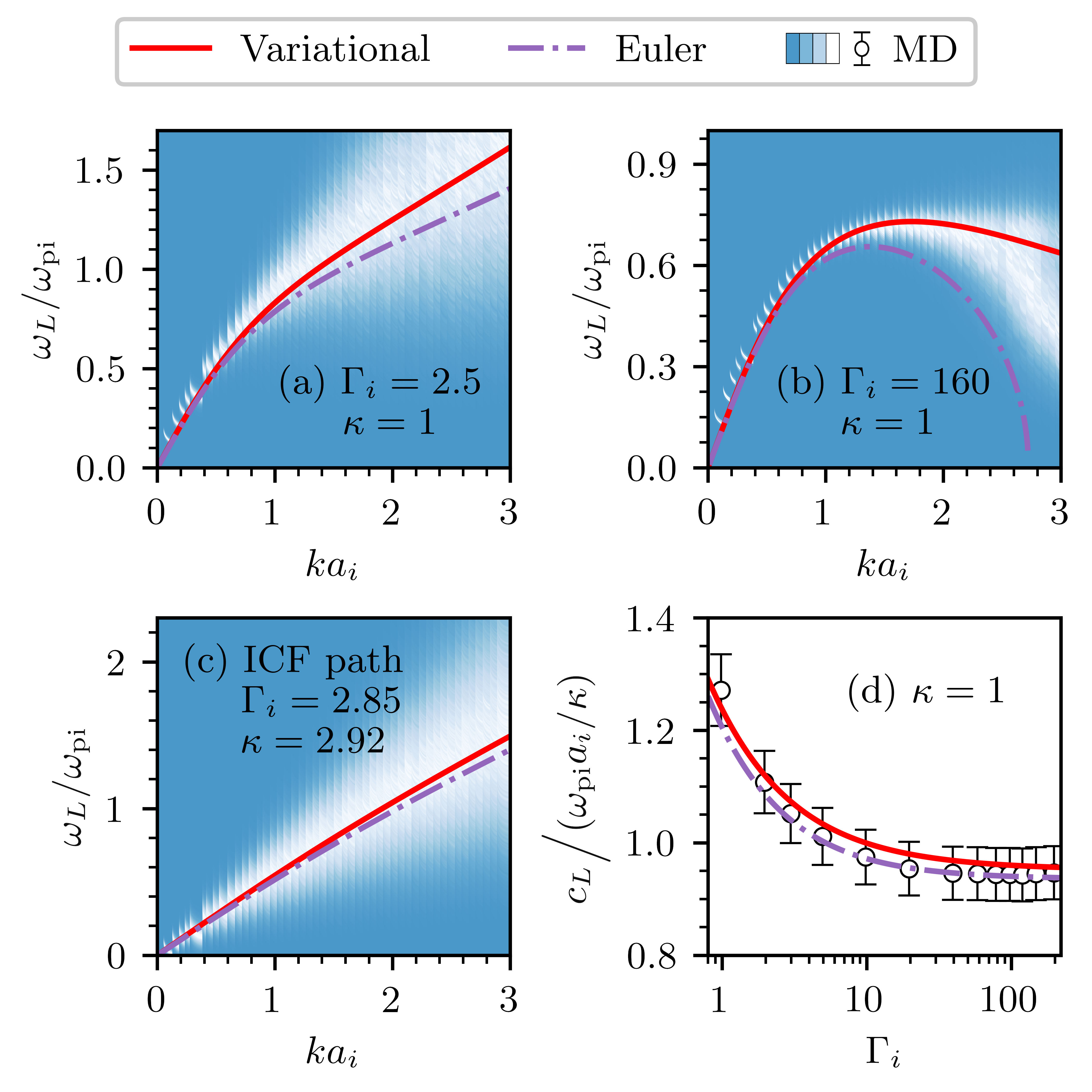}
    \caption{
    (a) and (b): Longitudinal ion-acoustic dispersion for a  screening parameter $\kappa = 1$, with moderate ($\Gamma_i = 2.5$) and strong ($\Gamma_i = 160$) ion coupling, respectively.
    Solid (red) line: VH; dash-dotted line: Euler hydrodynamics. The longitudinal current fluctuation spectrum obtained from MD is shown as blue-white background, with peak positions shown in white.
    (c): Longitudinal ion-acoustic dispersion for parameters corresponding to an intermediate point along the ICF compression path where $r_s=1.6$, cf. Fig.~\ref{fig:overview}. (d): Longitudinal sound speed  over a wide range of ion coupling strengths for both hydrodynamic models and MD data~\cite{silvestri2019physreve} (empty circles) when $\kappa = 1$.
    }
    \label{fig:hydro_comparison}
\end{figure}

The main challenge in solving these equations arises from their nonlocal character due to the integrals over the pair distribution function, as well as its dependence on the reference variables, as shown in Eq.~\eqref{eq:hydro_lagrangian}.
Nevertheless, in the linear regime VH yields analytic expressions for the longitudinal and transverse collective modes that are completely determined by equilibrium properties.
The longitudinal dispersion laws for the typical screening parameter $\kappa = 1$ and for both moderate and strong coupling strengths, $\Gamma_i = 2.5$ and $160$, are shown in Fig.~\ref{fig:hydro_comparison}(a) and Fig.~\ref{fig:hydro_comparison}(b).
The figure compares predictions from the Euler model and VH with MD simulations that computed the longitudinal current fluctuation spectrum~\cite{krimans2025pre} and serve as benchmarks. 
Both hydrodynamic approaches accurately reproduce the MD results in the large-wavelength limit $k \to 0^+$.
In addition, VH shows particularly good agreement at shorter length scales comparable to the inter-ionic distance $a_i$ and strong coupling with $\Gamma_i \gg 1$.

\par 
Figure~\ref{fig:hydro_comparison}(c) shows the longitudinal dispersion relations for an intermediate point along the ICF compression path discussed above, corresponding to $T = 67\,600\textrm{ K}$  and $r_s = 1.6$.
Both hydrodynamic approaches accurately reproduce the MD results in the long-wavelength limit as well as at finite wavelengths, since strong coupling effects are suppressed by the large value of the estimated screening parameter.
The advantage of the VH model becomes more pronounced for other compression paths where the ionic coupling is stronger and the screening weaker.
On the other hand, the success of VH in describing the linear regime, even when the ion coupling is strong and when length scales are comparable to $a_i$, motivates their further application to nonlinear problems such as shock waves or free expansion of the plasma.

Finally, Fig.~\ref{fig:hydro_comparison}(d) 
shows the long-wavelength ($k \to 0^+$) behavior by examining the longitudinal speed of sound for two typical values of screening parameters.
For $\kappa = 1$, the agreement is excellent for both hydrodynamic models across the entire ion coupling range $\Gamma_i > 1$.
For larger screening parameters, $\kappa \geq 2$, the Euler hydrodynamic equations continue to show excellent agreement with the MD results for the predicted speed of sound.
In contrast, VH exhibits increasing deviations, with peak errors $9\%$ at $\kappa = 2$ and $17\%$ at $\kappa = 3$.

In conclusion, we note that VH also provides a physically clear starting point for formulating hydrodynamics through a single Lagrangian, which enables rigorous generalizations to quantum hydrodynamics (QHD) and dynamically evolving electrons, going systematically beyond previous formulations of QHD \cite{manfredi_prb_01,manfredi_fields_05,zhandos_pop18,bonitz_pop_19}.
For example, 
it is possible to formulate an analogous variational principle for quantum particles corresponding to a quantum one-component plasma.
In that case, the Lagrangian in Eq.~\eqref{eq:hydro_lagrangian} would be modified by replacing the microscopic kinetic energy $(3/2)k_BT$ with the corresponding quantum generalization $\varepsilon= \varepsilon_{\textrm{id}}+\varepsilon_{\textrm{xc}}$.
Here, $\varepsilon_{\textrm{id}}$ denotes the ideal Fermi kinetic energy, while $\varepsilon_{\textrm{xc}}$ accounts for exchange and correlation contributions.
In addition, the pair distribution function should be taken from the quantum regime, for example, using data obtained from first-principles PIMC simulations, e.g.~\cite{filinov_pre_23,bonitz_pop_24}.

\section{Toward predictive simulations of dense quantum plasmas}\label{s:predictive}
In Sec.~\ref{s:methods} we have discussed a variety of theoretical methods to simulate quantum plasmas. Among them are two first principles methods -- fermionic path integral Monte Carlo and Kohn Sham density functional theory simulations. On the other hand, we also briefly discussed two mesoscopic methods -- semiclassical MD with quantum potentials and hydrodynamics -- that are of lower accuracy. The high accuracy of FPIMC and DFT-MD comes at a high price: the computational effort is very large and simulations are typically restricted to small simulation cells of a few hundred particles. In contrast, the mesoscopic methods are significantly less expensive and are capable to describe much larger systems and also longer time scales. The ultimate goal is, of course, to make large simulation scales possible and achieve high accuracy, at the same time, i.e., to achieve large simulations that have predictive capability.

\subsection{Simulations with predictive capability}\label{ss:predictive}
One strategy to achieve the goal of predictive capability are multiscale simulations. For example, 
Linke \textit{et al.} have presented a multiscale simulation framework that combines molecular dynamics simulations with finite elements \cite{linke_pre_25}. 

Here we proceed differently. Instead of developing a single multiscale framework we propose to develop different kinds of simulations in parallel which include semiclassical MD, hydrodynamics, DFT-MD and FPIMC
 simulations and possibly additional methods. An important first step in this procedure is to benchmark the lower level methods against exact data that are provided by FPIMC. In Sec.~\ref{ss:ueg-benchmarks} we start by discussing benchmarks for the uniform electron gas. This is extended to hydrogen and other quantum plasmas in Sec.~\ref{s:benchmarks-h}. The final step of combining simulations is discussed in Sec.~\ref{s:downfolding}.

\subsection{Benchmarks for the uniform electron gas}\label{ss:ueg-benchmarks}

The UEG constitutes the archetypical system of interacting electrons~\cite{loos_gill_2016,giuliani2005quantum} that was originally developed for electrons in metals, cf. Sec.~\ref{ss:jellium}. For example, the availability of accurate parametrizations of ground-state QMC results~\cite{ceperley_alder,Ortiz_PRB_1994,Ortiz_PRL_1999,Moroni_PRL_1995,Spink_PRB_2013,Holzmann_PRL_2011} for the exchange-correlation energy $E_{\rm xc}$ and a variety of other UEG properties~\cite{Corradini_PRB_1998,vosko_wilk_nusair,perdew_zunger,PerdewWang_PRB_1992,GoriGiorgi_PRB_2002,GoriGiorgi_PRB_2000} has facilitated the arguably unrivaled success of DFT for the description of real materials at ambient Conditions~\cite{Jones_RevModPhys_2015}. For dense quantum plasmas and warm dense matter, however, ground state results become unreliable and the effects of finite temperature have to be explicitly taken into account. 

\subsubsection{Benchmarks for the thermodynamic functions of the UEG}\label{ss:ueg-td}
First parametrizations of the exchange--correlation free energy $F_\textnormal{xc}(r_s,\Theta;\xi)$ that explicitly take into account temperature have been presented based on quantum-to-classical mappings~\cite{Perrot_Dharma_2000,Liu_JCP_2014} and dielectric theories~\cite{Perrot_Dharma_1984,ICHIMARU198791,Tanaka_1985,doi:10.1143/JPSJ.55.2278,sjostrom_dufty_2013}, see also Refs.~\cite{Tanaka_2016,Tanaka_2017,Tolias_JCP_2021,Lucco_Castello_2022,Tolias_JCP_2023,Tolias_PRB_2024,Tolias_CPP_2025,kalkavouras2026dielectricformalism2duniform} for novel, significantly improved dielectric methodologies.

\begin{figure}
    \centering
    \includegraphics[width=0.95\linewidth]{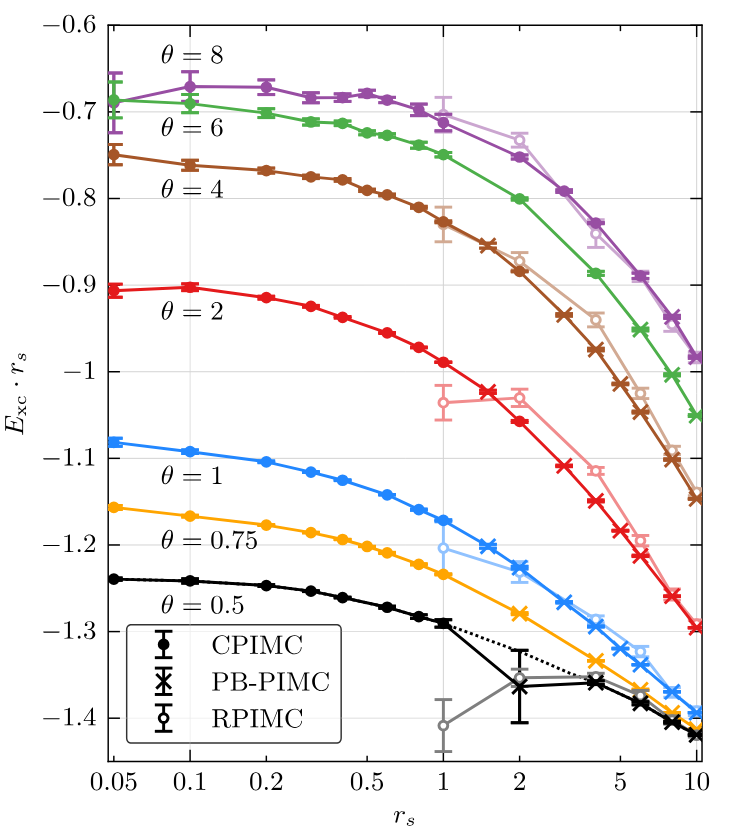}
    \caption{Isochores of the exchange--correlation energy $E_\textnormal{xc}$  for the spin-polarized UEG with $N=33$ electrons. A combination of CPIMC ~\cite{schoof_prl15,groth_prb16} and PB-PIMC \cite{dornheim_njp15,dornheim_jcp15} yields exact results for the entire density range, for temperatures $\Theta \gtrsim 0.5$. The results confirm the limited accuracy of  RPIMC~\cite{Brown_2014} around $r_s=1$. Taken from Groth \textit{et al.}~\cite{groth_prb16} with the permission of the authors.
    }
    \label{fig:UEG_overlap}
\end{figure}

The first extensive PIMC results for the exchange--correlation energy $E_\textnormal{xc}$ of the warm dense UEG have been presented by Brown \textit{et al.}~\cite{Brown_2014} using the fixed-node approximation with ideal surfaces.
Although the accuracy remained unknown, these results have been used to build corresponding parametrizations of $F_\textnormal{xc}$ and $E_\textnormal{xc}$~\cite{sjostrom_dufty_2013,ksdt,Brown_PRB_2013}.
Subsequently, Schoof \textit{et al.}~\cite{schoof_prl15} have presented approximation-free CPIMC results [cf. Sec.~\ref{ss:fpimc-progress}] for the spin-polarized UEG, which allowed for the first time to gauge the accuracy of RPIMC, finding systematic errors exceeding $10\%$ in the latter, in particular at high density and low temperature.
These inaccuracies have subsequently been fully confirmed by independent DMQMC~\cite{Malone_PRL_2016} and auxiliary-field QMC~\cite{Lee_JCP_2021} calculations.
At the same time, Dornheim \textit{et al.}~\cite{dornheim_jcp15,groth_prb16,dornheim_prb16}
have presented highly accurate PB-PIMC results for the UEG, which nicely complement CPIMC by being available at lower densities and, hence, stronger coupling; see Fig.~\ref{fig:UEG_overlap} for a direct comparison of CPIMC, RPIMC and PB-PIMC.
Taken together, the combination of CPIMC and PB-PIMC has allowed for a very accurate description of the UEG over a substantial part of the WDM regime. 
The subsequent introduction of an improved finite-size correction to extrapolate from the finite simulation cell to the thermodynamic limit~\cite{dornheim_prl16,dornheim_physrep_18,DornheimVorberger_JCP_2021}, combined with a thermal correction computed within the Singwi-Tosi-Land-Sj\"olander (STLS)~\cite{doi:10.1143/JPSJ.55.2278,sjostrom_dufty_2013} scheme, then allowed for the construction of the current state-of-the-art thermal UEG parametrizations, namely GDSMFB~\cite{groth_prl17,dornheim_physrep_18} and corrKSDT~\cite{Karasiev_PRL_2018,Karasiev_PRB_2019}, which boast an overall accuracy of $\sim0.3\%$.
On the one hand, the quality of GDSMFB and improvedKSDT is expected to be high enough for practical applications such as thermal DFT simulations on the level of the local density approximation~\cite{karasiev_2016,ramakrishna2020influence,Moldabekov_JCTC_2023,Moldabekov_JCTC_2024,bonitz_pop_24}. On the other hand, Karasiev \textit{et al.}~\cite{Karasiev_PRB_2019} have shown that some unphysical oscillations are observed in derived properties, such as the heat capacity, and overcoming such inconsistencies remains an important task for future work.

The existence of accurate FPIMC results for the model of the warm dense UEG is also highly valuable for the test and  improvement of other simulations, including DFT. Detailed benchmarks have been presented in Refs.~\cite{dornheim_pop17,dornheim_physrep_18}.

\subsubsection{Benchmarks for the dynamic properties of the UEG}\label{ss:ueg-dynamics}

The achievements for the thermodynamic functions of the UEG have been complemented by the QMC based study of a plethora of other UEG properties.
Dornheim \textit{et al.}~\cite{dornheim_prl_18,groth_prb_19,Dornheim_PRE_2020,hamann_prb_20,hamann_cpp_20} have presented first accurate results for the dynamic structure factor $S(\mathbf{q},\omega)$ of the UEG [and related properties such as the dynamic linear density response $\chi(\mathbf{q},\omega)$]
 via the analytic continuation of the ITCF $F(\mathbf{q},\tau)$, i.e., by numerically inverting Eq.~(\ref{eq:Laplace}).
This was done by the stochastic sampling of the dynamic local field correction $G(\mathbf{q},\omega)$, which is formally equivalent to the dynamic XC kernel $K(\mathbf{q},\omega)=-4\pi/q^2 G(\mathbf{q},\omega)$ from the linear-response TDDFT formalism~\cite{Marques_TDDFT_review_2004,Moldabekov_JCTC_2023}.
A particularly interesting result of these studies has been the analysis of a \emph{roton} like non-monotonic dispersion relation [by which we here simply mean the position of the maximum of $S(\mathbf{q},\omega)$, $\omega(q)$], which has attracted considerable interest both at ambient conditions~\cite{Takada_PRB_2016,Koskelo_PRL_2025} and at finite temperatures~\cite{dornheim_comphys_22,Dornheim_JCP_2022,filinov_prb_23,Chuna_JCP_2025,Chuna_PRB_2025}, and which has subsequently also been predicted to be, in principle, observable in XRTS measurements on optically pumped hydrogen jets~\cite{hamann_prr_23}.
Other recent results for the dynamic properties of the UEG include QMC results for the dynamic XC kernel at low temperatures~\cite{LeBlanc_PRL_2022}, and
PIMC results for the single-particle spectral function $A(\mathbf{p},\omega)$ [see Sec~\ref{ss:fpimc-mgf}] by analytically continuing the Matsubara Green function $G_\textnormal{M}(\mathbf{p},\tau)$~\cite{hamann_26} (see also earlier results for the related momentum distribution $n(\mathbf{p})$ in Refs.~\cite{Militzer_PRL_2002,hunger_pre_21,Dornheim_PRB_nk_2021,Dornheim_PRE_2021,Militzer_PRL_2002}).

\begin{figure}
    \centering
    \includegraphics[width=0.995\linewidth]{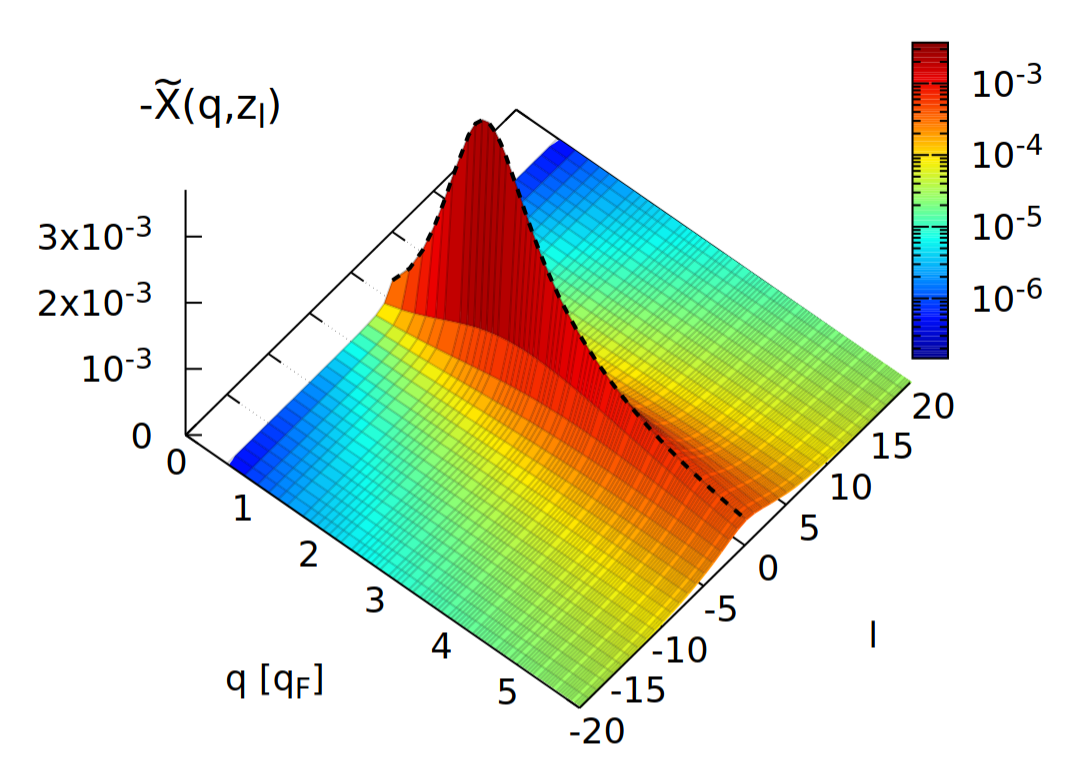}\\\vspace*{-0.2cm}
    \includegraphics[width=0.995\linewidth]{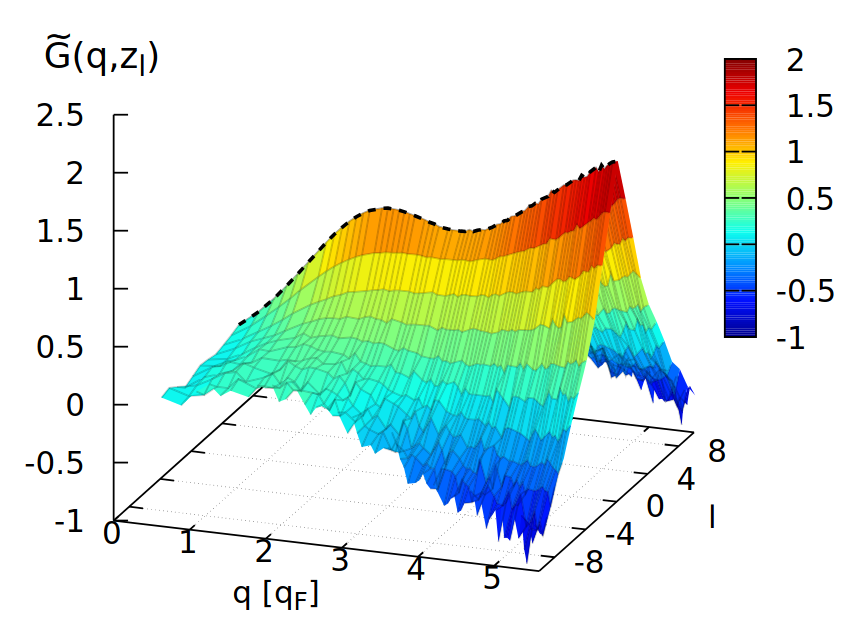}
    \caption{PIMC results for the dynamic Matsubara density response function $\widetilde{\chi}(\mathbf{q},z_l)$ [top] and Matsubara local field correction $\widetilde{G}(\mathbf{q},z_l)$ [bottom] of the strongly coupled unpolarized UEG at $r_s=20$ and $\Theta=1$. Taken from Dornheim \textit{et al.}~\cite{Dornheim_PRB_2024} with the permission of the authors.
    }
    \label{fig:Matsubara_LFC}
\end{figure}

% dornheim_pre17,groth_jcp17,

A second rich contemporary domain of research is given by the investigation of linear response properties.
Dornheim and co-workers~\cite{dornheim_jcp_19-nn,Dornheim_PPCF_2020,Dornheim_PRB_2020,Dornheim_PRL_2020_ESA,Dornheim_PRB_2021,Tolias_JCP_2021,Dornheim_PRR_2022,Dornheim_HEDP_2022} have presented extensive results for static linear response of the UEG covering a broad range of parameters via Eq.~(\ref{eq:static_chi}).
These studies have to both a neural network fit~\cite{dornheim_jcp_19-nn} and an analytical representation~\cite{Dornheim_PRB_2021}, which are freely available and can be directly used for a great variety of practical applications, e.g., Refs.~\cite{dornheim_pop_23,moldabekov_pre_20,Ramakrishna_PRB_2021,Zan_PRE_2021,Poole_PRR_2024}, and have been verified independently by novel diagrammatic QMC results by Hou \textit{et al.}~\cite{Hou_PRB_2022}.

A logical and potentially highly impactful extension of these developments is the investigation of the dynamic Matsubara density response function $\widetilde{\chi}(\mathbf{q},z_l)$, where $z_l=i2\pi l/\beta$ are the bosonic imaginary Matsubara frequencies~\cite{ICHIMARU198791}.
Tolias and co-workers~\cite{Tolias_JCP_2024} have introduced a corresponding Fourier--Matsubara series formalism, which allows to seemingly switch between the ITCF $F(\mathbf{q},\tau)$ and $\widetilde{\chi}(\mathbf{q},z_l)$ without the need for any analytic continuation or numerically tedious integration over poles at small wavenumbers $q$.
Subsequently, Dornheim \textit{et al.}~\cite{Dornheim_PRB_2024,Dornheim_EPL_2024,MOLDABEKOV2025104144,Dornheim_CPP_2025}
have presented extensive results for both $\widetilde{\chi}(\mathbf{q},z_l)$ and also the corresponding dynamic Matsubara local field correction $\widetilde{G}(\mathbf{q})$ for a broad range of system parameters.
The top panel of Fig.~\ref{fig:Matsubara_LFC}
shows $\widetilde{\chi}(\mathbf{q},z_l)$ for the unpolarized UEG at $r_s=20$ and $\Theta=1$, i.e., at the margin of the strongly coupled electron liquid regime.
The dashed black line shows the static density response $\chi(\mathbf{q},0)=\widetilde{\chi}(\mathbf{q},0)$, and we observe a symmetric decay of the Matsubara density response with increasing frequency index $|\pm l|$. 
Interestingly, all $|l|\neq 0$ contributions to $\widetilde{\chi}(\mathbf{q},z_l)$ are zero for a classical system, for which we have the aforementioned direct relation to the classical static structure factor $\chi_\textnormal{cl}(\mathbf{q},0)=-S_\textnormal{cl}(\mathbf{q})/n\beta$.
Since it nevertheless holds $\lim_{q\gg q_\textnormal{F}} S_\textnormal{cl}(\mathbf{q}) = \lim_{q\gg q_\textnormal{F}} S(\mathbf{q}) = 1$, the decay of $\chi(\mathbf{q},0)$ for $q\gg q_\textnormal{F}$ due to quantum delocalization effects must be compensated by increasing contributions to $\widetilde{\chi}(\mathbf{q},z_l)$ in the quantum case for increasing $q$, i.e., for smaller length scales where quantum delocalization effects are particularly important.
The observed combination of a shark fin like feature in the static limit combined with a wake like form for finite $l$ can thus be unambiguously be interpreted as a manifestation of quantum delocalization in the warm dense UEG, see also the more extensive discussions in Refs.~\cite{Dornheim_PRB_2024,Dornheim_EPL_2024}.
The bottom panel of Fig.~\ref{fig:Matsubara_LFC} shows the corresponding dynamic Matsubara local field correction $\widetilde{G}(\mathbf{q},z_l)$, which has been discussed extensively in the recent Ref.~\cite{Dornheim_CPP_2025}.
We note that $\widetilde{G}(\mathbf{q},z_l)$ exhibits a smooth behavior with a lot of known limits, which opens up the intriguing opportunity for a future four-point parametrization of $\widetilde{G}(\mathbf{q},z_l;r_s,\Theta)$.
The latter would be directly useful for a gamut of practical application, including the construction of advanced, non-local XC functionals for thermal DFT simulations~\cite{Pribram_Jones_PRL_2016}, see also Sec.~\ref{ss:dft-progress}, as well as the intriguing possibility to estimate electron--electron correlation functions such as $S_{ee}(\mathbf{q})$ directly from DFT simulations, see the very recent work by Moldabekov \textit{et al.}~\cite{Moldabekov_MRE_2025} for first results on warm dense hydrogen.
Finally, we mention recent QMC results for the
Matsubara spin XC kernel of the low temperature UEG by Li \textit{et al.}~\cite{Li_PRB_2025}.

\subsubsection{Nonlinear response properties of the UEG}\label{ss:nonlinear-response}
The final aspect of the density response of the warm dense UEG that we choose to mention here is the quantification of \emph{nonlinear effects}, which are known to play an important role, e.g., in stopping power calculations~\cite{Echenique_PRA_1986,Nagy_PRA_1989,Nagy_PRA_1989}. The first PIMC results for the non-linear density response of the warm dense electron gas were presented in Ref.~\cite{dornheim_prl_20} based on the direct perturbation approach, i.e., by applying an external harmonic perturbation and then measuring the response of the system~\cite{Moroni_PRL_1992,Bowen_PRB_1992,Moroni_PRL_1995,dornheim_pre17}.
This investigation has sparked a number of further developments, including the development of semi-analytical framework that is capable of accurately describing also the excitation of higher harmonics in terms of the linear local field correction~\cite{dornheim_prr_21}, the investigation of mode-coupling effects~\cite{dornheim_cpp22}, the generalization of the imaginary-time fluctuation--dissipation theorem Eq.~(\ref{eq:static_chi}) to the calculation of non-linear response from higher-order ITCFs~\cite{Dornheim_JCP_2021}, and the investigation of the non-linear density response of the spin-polarized UEG~\cite{Dornheim_CPP_2021}.
These simulation results have, in turn, sparked interesting conceptual developments, including the exploration of the connection between non-linear effects and higher-order correlators~\cite{Dornheim_JPSJ_2021,Vorberger_JStatPhys_2025}, the generalization of the recursion relation for the ideal density response at the first three harmonics by Mikhailov~\cite{Mikhailov_PRB_2016} to arbitrary orders by Tolias \textit{et al.}~\cite{Tolias_EPL_2023}, as well as the consideration of non-linear effects in other observables such as different energies~\cite{Dornheim_JCP_spatial_2023}.

In this regard, the archetypical nature of the UEG is further illustrated by the importance of these developments related to the linear and non-linear density response of the electron gas to density functional theory, see Refs.~\cite{MOLDABEKOV2025104144,Moldabekov_JCTC_2023,Moldabekov_JCTC_2022,Moldabekov_PRB_2023,Moldabekov_JCP_2023,Moldabekov_JPCL_2023,moldabekov2026generalizeddensityfunctionaltheory,Moldabekov_MRE_2025, Moldabekov_jcp_2021, Moldabekov_prb_2022, Moldabekov_electronic_structure_2025}, and by the utilization of UEG results as a benchmark for novel methods~\cite{Morresi_PRB_2025,li2026twoelectroncorrelationsmetallicelectron,Lee_JCP_2021,Dornheim_JCP_2023,Dornheim_JPCL_2024,dornheim2025taylorseriesperspectiveab,Chuna_JPA_2025,chuna2025noiselesslimitimprovedpriorlimit,Robles_CPC_2026,Hou_PRB_2022,LeBlanc_PRL_2022,Chen_NatComm_2019,yilmaz_jcp_20}.
Finally, we mention recent direct free energy~\cite{Dornheim_PRB_2025,Dornheim_PRR_2025,Dornheim_JCTC_2025,Svensson_JPCL_2025} and chemical potential~\cite{dornheim_prb_25}
 PIMC calculations, which are capable of giving highly accurate results for up to $N\sim10^3$ electrons and consistently confirm the high quality of the aforementioned UEG parametrizations.

\subsection{Benchmarks for dense hydrogen and other quantum plasmas}\label{s:benchmarks-h}

\begin{figure}
    \centering
    \includegraphics[width=0.995\linewidth]{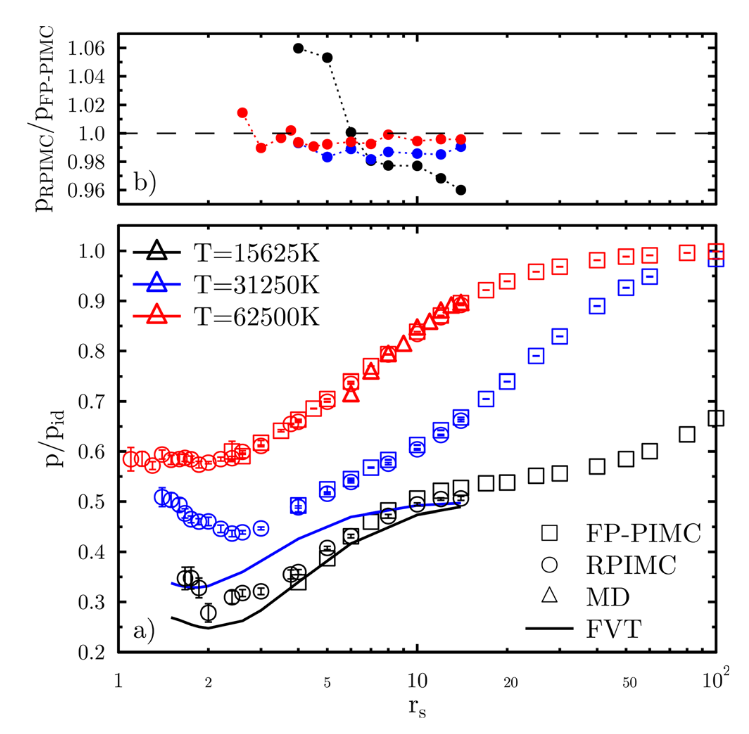}
    \caption{(a): Pressure of partially ionized hydrogen along three isotherms, in units of the ideal pressure (classical ions plus Fermi gas of electrons). Squares: FPIMC simulations (due the fermion sign problem no simulations are possible for $r_s\lesssim 3$), circles: RPIMC data of Ref.~\cite{Militzer_PRE_2021}, triangles: semiclassical MD simulations with improved Kelbg potential [cf. Sec.~\ref{ss:md-progress}], lines: chemical model (FVT) of atomic and molecular hydrogen \cite{juranek_jcp_02}. (b): Ratio of RPIMC and FPIMC data for three isotherms and densities where both simulations are possible.
    Taken from Bonitz \textit{et al.}~\cite{bonitz_pop_24} with the permission of the authors.
    }
    \label{fig:FPIMC_vs_RPIMC}
\end{figure}

Despite the high value of the model of the uniform electron gas as a limiting case and as a testbed for models and approximations, the UEG neglects important features of real quantum plasmas. In a real plasma the ions do not form a uniform background but have a spatial structure and can attract electrons to form atoms and molecules. The simplest test case to take bound states and partial ionization into account is hydrogen. While there have been numerous PIMC simulations for hydrogen before, including RPIMC, e.g., \cite{militzer_path_2000,militzer_pre_2001}
and fermionic PIMC, e.g., \cite{filinov_jetpl_00,filinov_jetpl_01,filinov_ppcf_01}, their accuracy has remained unclear. Here, recently a breakthrough was achieved by A.~Filinov and M.~Bonitz \cite{filinov_pre_23} who performed an extensive convergence analysis and computed the thermodynamic properties of dense hydrogen over a broad density range, for $T\gtrsim 15\,000$K. These results, for the first time, have allowed to rigorously benchmark RPIMC, DFT and other simulation methods in Refs.~\cite{filinov_pre_23,bonitz_pop_24}. 

\subsubsection{FPIMC benchmarks for the hydrogen equation of state}\label{ss:h-eos}
An example is presented in Fig.~\ref{fig:FPIMC_vs_RPIMC} where three pressure isotherms are depicted over a broad density range spanning six orders of magnitude. The FPIMC simulations (squares in the bottom panel) nicely reflect the expected behavior: at low densities, $r_s \gtrsim 100$ the plasma is a nearly ideal classical gas mixture of electrons and protons. With increasing density the pressure decreases due to Coulomb attraction and formation of hydrogen bound states. The pressure reduction (in units of the ideal pressure of classical ions and a Fermi gas of electrons) is particularly strong for the lowest temperature, $T=15\,625$K, due to the formation of molecules. Further density increase to $r_s \lesssim 3$ yields again a pressure increase that is related to a lowering of the ionization potential and break up of bound states due to Coulomb and quantum effects, see also Sec.~\ref{ss:fpimc-saha}. The figures contains benchmarks of three other methods against FPIMC results: restricted PIMC simulations of Militzer \textit{et al.} \cite{Militzer_PRE_2021}, semiclassical  molecular dynamics with improved Kelbg potential, cf. Sec.~\ref{ss:md-progress} and fluid variational theory of Juranek \textit{et al.}~\cite{juranek_jcp_02}. Without going into details, the benchmarks are extremely valuable to establish the range of validity of various approximate methods as well as to compare the quality of different approximations, for more details, see Ref.~\cite{bonitz_pop_24}. We will return to the high potential of such benchmarks for future method developments below, in Sec.~\ref{sss:downfolding}.

\subsubsection{Further FPIMC benchmarks}\label{ss:h-more-fpimc}
In this context, we also mention the recent PIMC set-up for the direct estimation of the free energy presented in Ref.~\cite{Dornheim_PRB_2025} (see also Refs.~\cite{Dornheim_PRR_2025,Dornheim_JCTC_2025,Dornheim_ChemPot_2025,Svensson_JPCL_2025}), which has allowed to assess the accuracy of thermal DFT results for a single proton snapshot comparing both zero-temperature and thermal LDA functionals; indeed, this analysis has further substantiated the importance of thermal XC-functionals in the WDM regime~\cite{karasiev_2016,ramakrishna2020influence,bonitz_pop_24,Moldabekov_JCTC_2024}.

\begin{figure}
    \centering
    \includegraphics[width=0.995\linewidth]{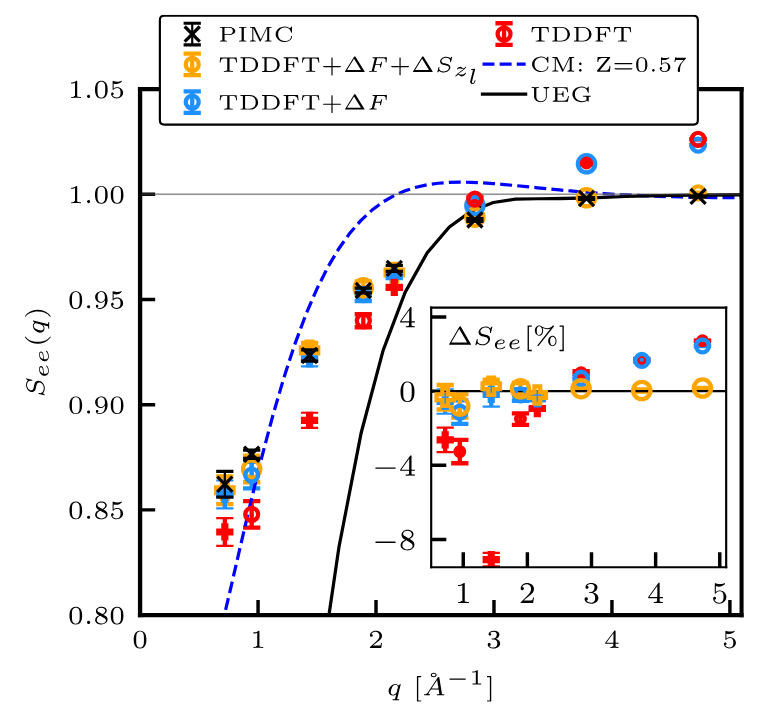}
    \caption{Electron--electron static structure factor $S_{ee}(\mathbf{q})$ for warm dense hydrogen at $r_s=3.23$ and $\Theta=1$ ($\rho=0.08\,$g/cc  and $T=4.8\,$eV. Shown are quasi-exact PIMC reference results (black crosses), raw TDDFT results (red), density-response corrected TDDFT (blue) and fully corrected TDDFT results (orange), as well as a simple chemical model (dashed dark blue). The inset shows relative deviations towards PIMC. Taken from Moldabekov \textit{et al.}~\cite{Moldabekov_MRE_2025} with the permission of the authors.
    }
    \label{fig:see-hydrogen}
\end{figure}

A particularly important class of PMIC results is given by the static and dynamic density of warm dense hydrogen, which has been used extensively to benchmark a broad range of properties computed from DFT.
B\"ohme \textit{et al.}~\cite{Bohme_PRL_2022,Bohme_PRE_2023,Dornheim_PRE_2023} have used PIMC results for the electron density and density response of hydrogen snapshots to assess the accuracy of KS-DFT.
These efforts have been explored further in Ref.~\cite{Moldabekov_JCTC_2024}, where a generalized reduced density gradient measure has been developed for the more systematic assessment of the accuracy of both thermal and non-thermal XC-functionals.
Moldabekov \textit{et al.}~\cite{Moldabekov_JPCL_2023,Moldabekov_JCP_2023,Moldabekov_JCTC_2023} have worked extensively on the computation of the static density response and the related static XC-kernel using DFT across Jacob's ladder of XC-functionals~\cite{PerdewSchmidt}.
The XC-kernel can then be used in linear-response TDDFT simulations~\cite{Moldabekov_MRE_Lanczos_2025,Moldabekov_MRE_2025, moldabekov2026_tddft}, which can be compared with quasi-exact reference data in the imaginary-time domain, cf.~Eq.~(\ref{eq:Laplace}).
A recent highlight is shown in Fig.~\ref{fig:see-hydrogen}~\cite{Moldabekov_MRE_2025}, depicting the electron--electron static structure factor $S_{ee}(\mathbf{q})$ of hydrogen at $\rho=0.08\,$g/cc ($r_s=3.23$) and $T=4.8\,$eV ($\Theta=1$).
The black crosses show PIMC results, which are exact within the given statistical error bars. Naively, DFT constitutes an effective single-electron theory and, thus, does not give one direct access to many-electron correlation functions.
However, linear-response TDDFT can be used to compute the dynamic density response function $\chi_{ee}(\mathbf{q},\omega)$, which in turn gives one the dynamic structure factor $S_{ee}(\mathbf{q},\omega)$---and thus the sought-after static structure factor $S_{ee}(\mathbf{q})$---via the fluctuation--dissipation theorem; direct results for this procedure are shown as the red pluses in Fig.~\ref{fig:see-hydrogen}, and systematically and substantially deviate from the PIMC baseline over the entire depicted range of wavenumbers $q$, see also the inset showing relative deviations from PIMC.
Moldabekov \textit{et al.}~\cite{Moldabekov_MRE_2025} have resolved the origins of these deviations, which are different for short and long wavelengths. 
Moreover, they have introduced proper corrections for both, which, importantly, do not require additional external input apart from the usual static XC-functional that is needed in all DFT calculations.
The thus corrected results have been included as the orange circles in Fig.~\ref{fig:see-hydrogen} and are in perfect agreement with PIMC everywhere.
This opens up the intriguing possibility to use DFT for the estimation of electron correlation functions for a broad range of parameters and materials, for which PIMC is currently not available due to the aforementioned fermion sign problem.

\begin{figure}
    \centering
    \includegraphics[width=0.995\linewidth]{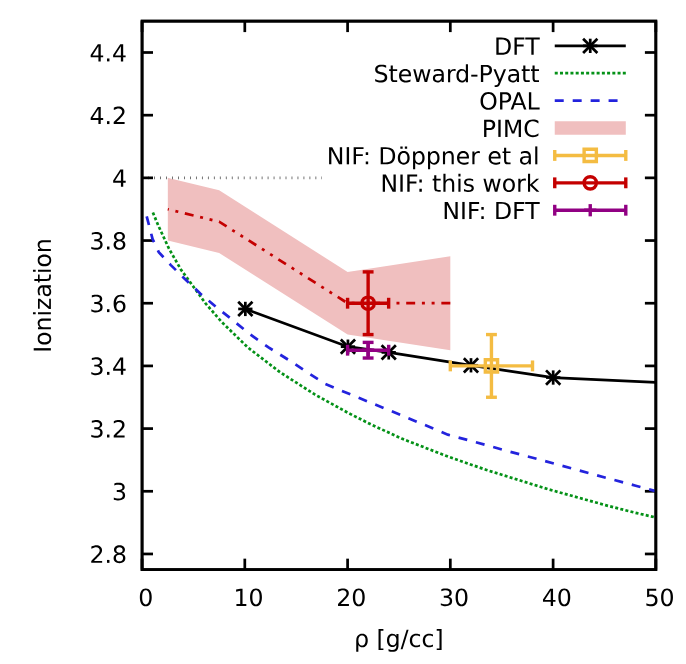}
    \caption{Ionization degree $\alpha^{\rm ion}$ of strongly compressed beryllium at $T\approx150\,$eV. Red curve: PIMC; black crosses: DFT~\cite{Doeppner2023}; dashed blue: OPAL~\cite{OPAL}; dotted green: Steward-Pyatt~\cite{SPyatt}. The red, purple and yellow crosses show interpretations of an XRTS measurement at the NIF by D\"oppner \textit{et al.}~\cite{Doeppner2023} using PIMC, DFT, and a chemical Chihara model, respectively. Taken from Dornheim \textit{et al.}~\cite{Dornheim_NatComm_2025} with the permission of the authors. 
    }
    \label{fig:fpimc-be-ionization}
\end{figure}

The advent of the $\xi$-extrapolation technique~\cite{Xiong_JCP_2022,Dornheim_JCP_2023,dornheim2025taylorseriesperspectiveab} has facilitated PIMC simulations beyond hydrogen, and relatively extensive results for warm dense beryllium have become available over the last few years~\cite{Dornheim_JCP_2024,Dornheim_NatComm_2025,Dornheim_POP_2025,schwalbe2025staticlineardensityresponse}.
This has allowed Dornheim \textit{et al.}~\cite{Dornheim_NatComm_2025} to use PIMC simulations for the interpretation of an XRTS measurement on strongly compressed beryllium that was taken at the NIF by D\"oppner \textit{et al.}~\cite{Doeppner2023}, and to assess the accuracy of previous models.
Some aspects of this investigation are shown in Fig.~\ref{fig:fpimc-be-ionization}, depicting the ionization degree $\alpha^\textnormal{ion}$ as a function of the mass density $\rho$ for the experimentally relevant temperature of $T\approx150\,$eV.
The dashed blue and dotted green curves correspond to the OPAL datatable~\cite{OPAL} and the Steward-Pyatt model~\cite{SPyatt}, respectively, which are often used for the modeling of IFE applications.
The black stars have been computed using DFT-MD, where the ionization is estimated using the Thomas-Reiche-Kuhn sum rule as proposed by Bethkenhagen~\textit{et al.}~\cite{Bethkenhagen_PRR_2020}.
Finally, the red curve has been obtained from PIMC, where the ionization degree is estimated by analyzing the static density response in comparison to a free electron gas~\cite{Bohme_PRL_2022,Dornheim_NatComm_2025}.

The red, purple and yellow symbols correspond to alternative interpretations of a single NIF dataset~\cite{Doeppner2023} using PIMC, DFT-MD and a traditional chemical model~\cite{redmer_glenzer_2009}, respectively. 
Evidently, using either of the two \textit{ab initio} methods leads to a substantially lower mass density compared to the Chihara model, whereas the inferred degrees of ionization are comparable.

Clearly, the presented results for hydrogen and beryllium are very interesting in their own right and have already given a number of important new insights into the physics of warm dense quantum plasmas, and into the performance of widely used models and approximations.
These investigations are expected to continue over the next years, and might potentially also lead to the investigation of light mixtures (e.g., lithium hydride) and somewhat heavier elements, with warm dense carbon being of particular importance for a gamut of practical applications.

%------------ 

\subsection{FPIMC downfolding. Systematic improvement of models.}
\label{s:downfolding}
First principles benchmark data provide important opportunities, even beyond accuracy tests for other, more approximate methods. If several approximations exist, first principles simulations can be used to select, among the approximate models or approximations, the most accurate one, for the given parameter range. 
\subsubsection{Idea of FPIMC downfolding}\label{sss:downfolding}

Moreover, we put forward a ``downfolding'' procedure that is
suitable for rigorously deriving model descriptions from first principles. 
Downfolding concepts have been used widely in electronic structure theory. There, the idea is   to map the exact many-body hamiltonian onto a reduced model, such as a lattice model, which is supplemented with accurate model parameters that are taken from exact simulations, density matrices \cite{wagner_fp_18} or from DFT, e.g. \cite{wehling_scipost_24, alvertis_prap_25}, for a recent overview see Ref.~\cite{aryasetiawan_downfolding}.  For quantum plasmas, however, a ground state description is not applicable. Therefore,  we follow a different approach that is based on three main concepts: 1. We take into account finite temperature effects, along with Coulomb interaction, bound states, quantum and spin effects, simultaneously. For this goal,  fermionic PIMC simulations are ideally suited as they provide numerically exact (up to statistical errors) results for many observables. 2. Instead of simplifying the hamiltonian we concentrate on a suitable set of observables. 3. Furthermore, the goal is not to consider a model hamiltonian but simplified - as compared to FPIMC - theoretical approaches. We consider only such approaches that can be strictly derived from the exact many-body hamiltonian whereas the level of approximation is characterized by a single input quantity.

Below we discuss this approach in application to Kohn-Sham density functional theory, Matsubara Green functions, quantum kinetic equations and to chemical models (Saha equation).

\subsubsection{Application to DFT}\label{ss:fpimc-dft}

\begin{figure}
    \centering
    \includegraphics[width=0.99\linewidth]{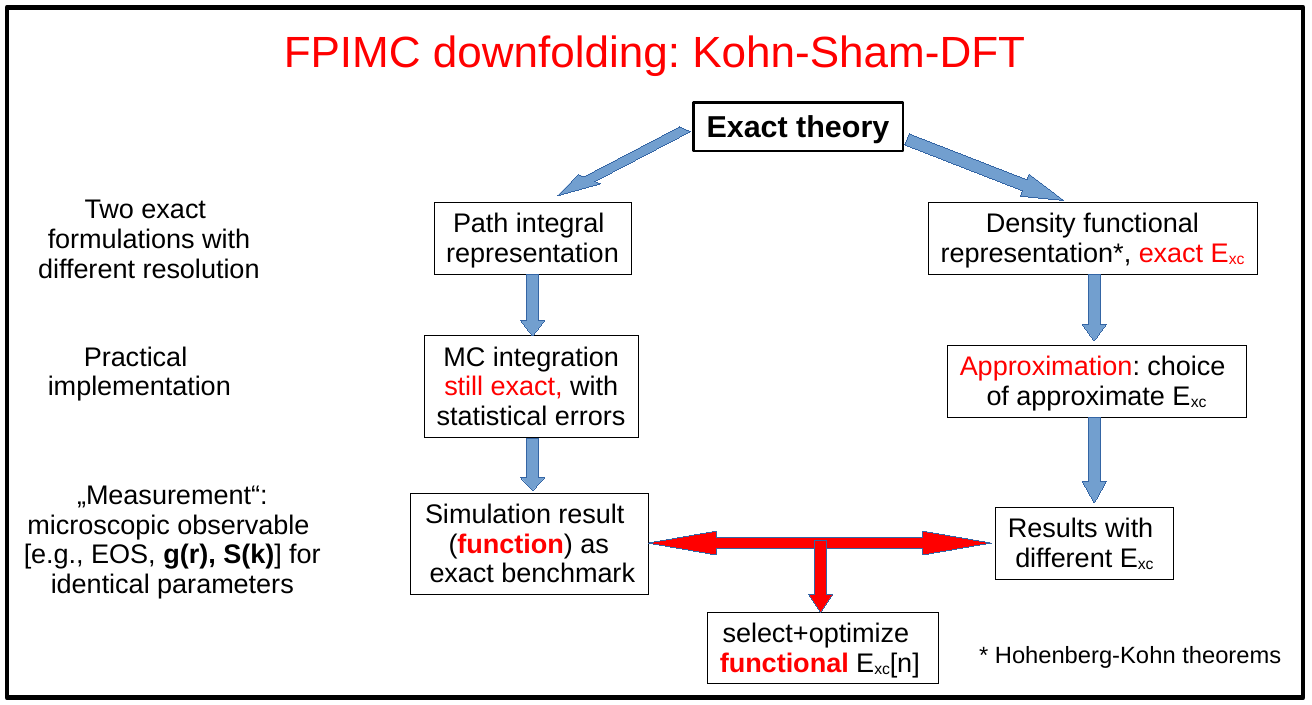}
    \caption{FPIMC downfolding for Kohn-Sham density functional theory. KS-DFT represents an exact representation of the many-body problem, provided the exact exchange-correlation energy would be used (Hohenberg Kohn theorems). FPIMC has the capability to discriminate between different approximations for $E_{\rm xc}$ [or ${\cal F}_{\rm xc}$] and to construct systematically improved approximations.}
    \label{fig:fpimc-dft}
\end{figure}

Let us demonstrate the FPIMC downfolding idea for the example of Kohn-Sham density functional theory. DFT contains a single input quantity -- the exchange-correlation energy $E_{\rm xc}$ or free energy ${\cal F}_{\rm xc}$. The theorems of Hohenber and Kohn \cite{hohenberg-kohn} establish that DFT is -- in principle -- exact, if the exact expression, $E^{\rm exact}_{\rm xc}$ is being used. In Fig.~\ref{fig:fpimc-dft} the downfolding procedure is sketched: We start from the exact many-body problem which is then mapped exactly onto two alternative formulations -- FPIMC and KS-DFT, respectively. While FPIMC simulations are performed exactly yielding accurate results for observables, such as equation of state, energy, pair distribution functions or static structure factor, the situation is different for KS-DFT (right branch). For practical computations an approximation for $E_{\rm xc}$ [or ${\cal F}_{\rm xc}$] has to be selected the accuracy of which is in general unknown. One way to proceed is to repeat the simulations with another choice for $E_{\rm xc}$ [or ${\cal F}_{\rm xc}$]. Comparing the results for observables to FPIMC data allows to benchmark various functionals and select the most accurate one, for a given set of system parameters, e.g., density and temperature, as was done  in Ref.~\cite{filinov_pre_23} and \cite{bonitz_pop_24}. Alternatively, the set of FPIMC simulations can be extended by invoking different parameter sets or additional observables, in order to better understand the behavior of various approximations for $E_{\rm xc}$ [or ${\cal F}_{\rm xc}$] and predict for what parameters which is expected to be more appropriate. Moreover, one can develop strategies to optimize $E_{\rm xc}$ [or ${\cal F}_{\rm xc}$], e.g. by making an improved ansatz for the density and/or temperature dependence.

As discussed in Sec.~\ref{ss:dft-progress}, QMD simulations based on DFT using zero-temperature XC-functionals have shown some success in simulating quantum plasmas under HED conditions. However, as experimental precision in HED physics improved and fermion path-integral Monte Carlo (FPIMC) calculations became computationally feasible, systematic discrepancies emerged between experiments, PIMC benchmarks, and QMD simulations based on zero-temperature XC functionals. The underlying issue was clear: ground-state XC functionals, developed primarily for condensed-matter physics and quantum chemistry, lack explicit temperature dependence in the exchange–correlation free energy evaluated as ${\cal F}_{\rm xc}[n,T]\approx E_{\rm xc}[n]$ completely missing the XC entropy contribution \cite{PhysRevE.93.063207}. In contrast, HED plasmas routinely span temperatures from tens of thousands to several million Kelvin—far beyond the regime for which traditional zero-temperature XC approximations were designed. This realization, crystallizing in 2010s, motivated the development of rigorous finite-temperature XC functionals. The basic chronology of the XC density functionals with explicit temperature dependence, ${\cal F}_{\rm xc}$, is shown in Table \ref{tab:table1}. The development followed the finite-$T$ analog of the Perdew–Schmidt ladder \cite{PerdewSchmidt}. Functionals can be arranged by the level of theoretical refinement as rungs of that ladder. 

The first rung, the local density approximation (LDA), is represented by the (corr)KSDT~\cite{ksdt,Karasiev_PRL_2018} and GDSMFB~\cite{PhysRevLett.119.135001} functionals which depend on the electron density and temperature. Accurate analytical parameterization for the exchange-correlation free energy of the homogeneous
electron gas based on the FPIMC reference data is the most representative example of the downfolding procedure applied to the development of two non-empirical LDA XC free-energy density functionals.

The second, GGA rung includes two levels of theoretical refinement: (i) a simple one with additive Perdew-Burke-Ernzerhof (addPBE)~\cite{Sjostrom_Gradient_2014} and multiplicative (ltPBE)~\cite{ltPBE} LDA-level thermal corrections; and (ii) the fully thermal GGA, with explicit temperature dependencies on reduced density exchange and correlation gradients, represented by the Karasiev-Dufty-Trickey (KDT16)~\cite{Karasiev_PRL_2018}.

The meta-GGA rung is also represented by two approaches: (i) a simple scheme which uses a universal additive GGA-level thermal correction applied to the deorbitalized versions of strongly constrained, and appropriately normed (SCAN-L) and to the regularized-restored r$^2$SCANL functional, termed T-SCAN-L and T-r$^2$SCAN-L~\cite{PhysRevB.105.L081109}; and (ii) the fully thermal $f$TSCAN meta-GGA~\cite{PhysRevMaterials.9.L050801}.

\begin{table}
\caption{\label{tab:table1}
Time line of modern XC  free-energy density functional development.
}
\begin{ruledtabular}
\begin{tabular}{lll}
Year & Functional & Reference \\
\hline\\
	2014 & KSDT LDA  & Karasiev {\it et al.}, Ref. \citenum{ksdt} \\
	     &                                              & (see Ref. \citenum{Karasiev_PRL_2018} for corrKSDT) \\
	2014 & additive addPBE  & Sjostrom {\it et al.}, Ref. \citenum{Sjostrom_Gradient_2014} \\
	2017 & GDSMFB LDA  & Groth {\it et al.}, Ref. \citenum{PhysRevLett.119.135001} \\
	2018 & KDT16 GGA   & Karasiev {\it et al.}, Ref. \citenum{Karasiev_PRL_2018} \\
    2020 & KDT0 hybrid & Mihaylov {\it et al.}, Ref. \citenum{PhysRevB.101.245141} \\
	2022 & additive T-(r$^2$)SCAN-L  & Karasiev {\it et al.}, Ref. \citenum{PhysRevB.105.L081109} \\
    2023 & multiplicative ltPBE & Kozlowski {\it et al.}, Ref. \citenum{ltPBE} \\
    2025 & $f$TSCAN meta-GGA & Hilleke {\it et al.}, Ref. \citenum{PhysRevMaterials.9.L050801}\\
    2025 & RS-KDT0 hybrid & Ellaboudy {\it et al.}, Ref. \citenum{RS-KDT0-PRB-0} \\
\end{tabular}
\end{ruledtabular}
\end{table}

\begin{figure}
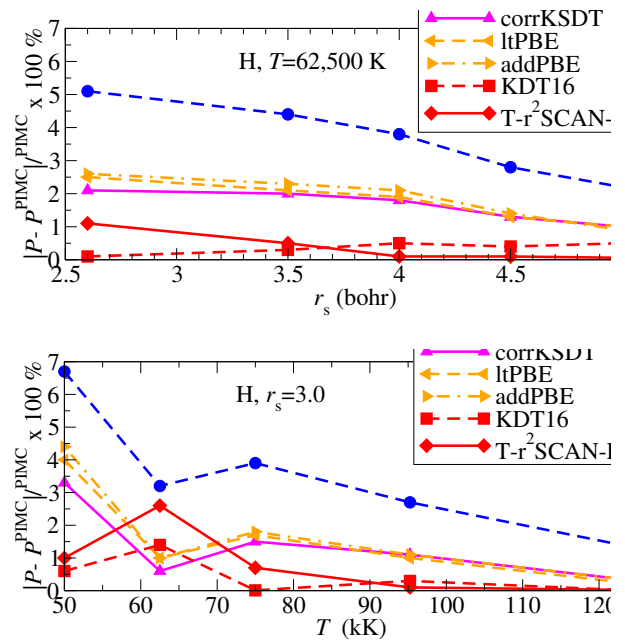
%[H]
\includegraphics*[width=8.4cm,angle =00]{P-Ppimc-rel-vs-rs.PBE-corrKSDT-ltPBE-addPBE-KDT16-TR2SCANL.H.T62.5kK.v1b.eps}
\includegraphics*[width=8.4cm,angle =00]{P-Ppimc-rel-vs-rs.PBE-corrKSDT-ltPBE-addPBE-KDT16-TR2SCANL.H.rs3.0.v1b.eps}
\caption{
The relative error of total pressure from the KS-MD simulations of warm dense H using PBE (ground-state), ltPBE and addPBE (GGAs with LDA thermal corrections), KDT16 (thermal GGA), and T-r$^2$SCANL (meta-GGA with GGA thermal corrections) XC functionals, calculated with respect to the reference PIMC data \cite{Filinov_PRE_2023}, and shown as a function of $r_{\rm s}$ along the $T=62,500$ K isotherm (upper panel), and as a function of temperature along the $r_{\rm s}$=3.0 isochore (bottom panel).
}
\label{Ptot-H}
\end{figure}
%%

%% Figure
\begin{figure}
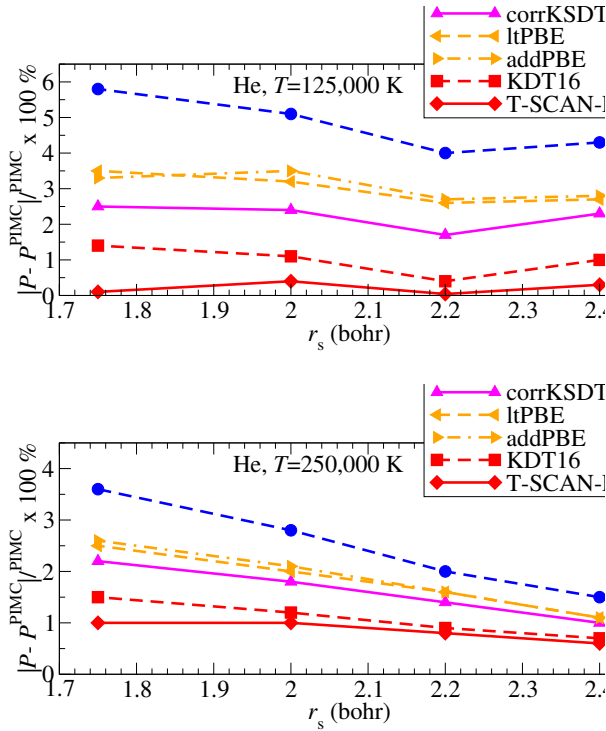
%[H]
\includegraphics*[width=8.4cm,angle =00]{P-Ppimc-rel-vs-rs.PBE-corrKSDT-ltPBE-addPBE-KDT16-TSCANL.He.T125kK.v1b.eps}
\includegraphics*[width=8.4cm,angle =00]{P-Ppimc-rel-vs-rs.PBE-corrKSDT-ltPBE-addPBE-KDT16-TSCANL.He.T250kK.v1b.eps}
\caption{
The relative error of total pressure from molecular dynamics simulations of warm dense He using PBE (ground-state), ltPBE and addPBE (GGAs with LDA thermal corrections), KDT16 (thermal GGA), and T-r$^2$SCANL (meta-GGA with GGA thermal corrections) XC functionals calculated with respect to the reference restricted PIMC (RPIMC, cf. Sec.~\ref{ss:fpimc-progress}) data \cite{PhysRevB.79.155105}, and shown as a function of $r_{\rm s}$ along the $T=125,000$ K and $250,000$ K isotherms.
}
\label{Ptot-He}
\end{figure}

To illustrate how the FPIMC downfolding procedure works for Mermin-Kohn-Sham DFT for assessment of the accuracy of XC functionals, we consider the ground-state PBE and finite-$T$ corrKSDT, ltPBE, addPBE, KDT16 and T(r$^2$)SCANL. Figure \ref{Ptot-H} shows the relative error of total pressure from the KS-MD simulations of warm-dense hydrogen, for all six XC functionals, with respect to the PIMC reference. All data are taken from Ref. \citenum{10.1063/5.0315749}. The relative pressure error of the ground-state PBE, which can also be considered as the  magnitude of explicit thermal XC effects, ranges between 1.3\% and 6.7\% with a mean absolute relative error (MARE) of 3.6\%. Thermal corrKSDT LDA reduces that error roughly by a factor of two with MARE of 1.5\%.

The two LDA thermal-corrected GGAs, ltPBE, and addPBE
also improve on the ground-state PBE, but these two GGAs are less accurate compared to the corrKSDT LDA, with MARE values of 1.7\% and 1.8\% for the ltPBE and addPBE functionals, respectively. Both the ltPBE and addPBE functionals include only the LDA level thermal corrections, while spatial inhomogeneity effects are
described by the ground-state PBE gradient-dependent terms without any explicit $T$-dependence. Thus, as explained in Ref. \citenum{10.1063/5.0315749}, these two functionals do not reduce to the second-order finite-$T$ gradient expansion in the weakly varying density limit. The fully thermal KDT16 GGA, and the GGA-thermal corrected Tr$^2$SCANL meta-GGA provide a very similar performance. The relative error with respect to the PIMC reference does not exceed 1\% (with exception at one thermodynamic condition, which may need an additional investigation) and MARE values of 0.4\% and 0.6\% respectively. Both of these functionals include reduced density X and C gradients with explicit, correct temperature dependence derived from the second-order finite-$T$ gradient expansion. The situation for warm dense He, shown in Fig. \ref{Ptot-He} is very similar. The corrKSDT LDA systematically outperforms the two LDA-thermal corrected GGAs, with MARE values of 1.8\% (corrKSDT), 2.4\% (ltPBE) and 2.5\% (addPBE). The PIMC downfolding procedure confirms that the high-level theoretical refinement thermal XC starting from the fully thermal KDT16 GGA rung and above are the most reliable choice for KS-DFT-based simulations of matter at warm-dense conditions.

\begin{figure}
    \centering
    \includegraphics[width=0.995\linewidth]{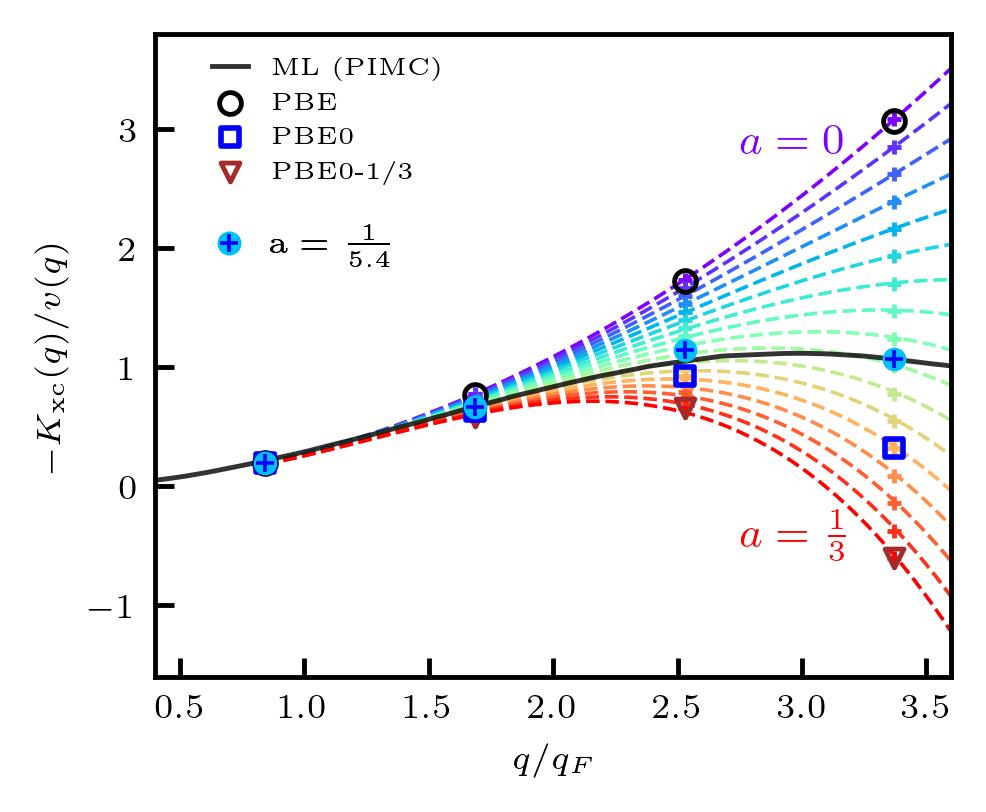}
    \caption{Static XC-kernel of the UEG at $\theta=1$ for $r_s=2$. The solid line is the machine learning (ML) representation of the exact PIMC data by Dornheim \textit{et al.}~\cite{dornheim_ML}.
    The mixing parameter $a$ in the PBE-based hybrid XC functional is varied in the range $0\leq a \leq 1/3$. The KS-DFT data corresponding to different $a$ values are presented by dashed lines. In addition, we show the data points computed using $a=1/5.4$. Adapted from Moldabekov \textit{et al.} \cite{Moldabekov_JPCL_2023} with the permission of the authors. }
    \label{fig:Kxc_hybrid}
\end{figure}

\subsubsection{From FPIMC density response to XC functionals}\label{ss:LFC_in_DFT}

As discussed in Sec.~\ref{ss:fpimc-dft}, the UEG model provides key ingredients for the construction of non-empirical XC functionals with high transferability for KSDFT calculations of extended systems, such as solids and warm dense matter.
In addition to thermodynamic properties of the UEG, such as the free energy density, static density response properties provide further constraints for the construction of XC functionals. For the UEG, the static local field correction $G(\vec q)$ of the UEG is related to the second-order functional derivative of the XC functional (i.e., the XC-kernel)~\cite{MOLDABEKOV2025104144},
\begin{eqnarray}\label{eq:Kxc}
    K_\textnormal{xc}(\mathbf{q}) = \mathcal{F}\left[ \frac{\delta^2 F_\textnormal{xc}(n)}{\delta n(\mathbf{r}) \delta n(\mathbf{r}')} \right]\ ,
\end{eqnarray}
via $G(\vec q) = -K_{\rm xc}(\vec q)/v(\vec q)$, where $v(\vec q)$ denotes the Coulomb potential. In the ground state, the XC-free energy in Eq.~(\ref{eq:Kxc}) is replaced by the XC-energy $E_\textnormal{xc}$.

For example, $G(\vec q)$ of the UEG has been used as an ingredient in the construction of the highly successful ground-state meta-GGA XC functional SCAN \cite{SCAN}.  
For warm dense matter, the exact PIMC results for $G(\vec q)$ can be used to identify the mixing coefficient in hybrid XC functionals. Hybrid XC functionals are typically constructed by mixing a fraction of Hartree--Fock (HF) exchange with the exchange energy from an XC approximation \cite{doi:10.1063/1.472933}. The commonly cited rationale for the choice of the mixing coefficient in the hybrid XC functionals was given by Perdew \textit{et al.} \cite{doi:10.1063/1.472933}, who analyzed atomization errors of typical molecules using the M{\o}ller--Plesset perturbation expansion and proposed the mixing parameter $a = 1/4$.
Clearly, there is considerable room for variation in the mixing parameter for systems beyond molecules under ambient conditions. For instance, in WDM, atomic properties such as atomization energies can become ill-defined due to the blurring of the distinction between bound and free states at high pressure or temperature \cite{Moldabekov_JCTC_2024}, see also Fig.~\ref{fig:bound-states-pimc}. 

Exact PIMC results for the local field correction enable non-empirical determination of hybrid functional parameters across different densities and temperatures \cite{Moldabekov_JPCL_2023}.
One such approach using $G(\vec q)$ is illustrated in Fig.~\ref{fig:Kxc_hybrid}, where the PIMC results for the local field correction  of the UEG at $r_s = 2$ and $\theta = 1$ are compared with data obtained from hybrid XC functionals based on PBE~\cite{PBE}. The implicit dependence on electronic temperature is accounted for in the HF contribution through occupation-number smearing. The choice $a = 1/4$ corresponds to PBE0~\cite{Adamo1999}, while $a = 1/3$ yields PBE0-1/3~\cite{Cortona2012}. In the limiting case $a = 0$, the HF exchange is omitted, and the standard PBE functional is recovered.  From Fig.~\ref{fig:Kxc_hybrid}, one can see that, choosing $a = 1/5.4$, leads to close agreement with PIMC data for the local field correction at wavenumbers $q \leq 3q_F$. Starting from any LDA or GGA-level XC functional, a similar analysis can be carried out across a range of densities and temperatures relevant for warm dense matter. Therefore, exact PIMC results for the local field correction of the UEG provide valuable input for both, benchmarking and downfolding -- the development of XC functionals. In this context, we note that the large-$q$ behavior of the static local field correction, although computed exactly, can lead to spurious results in the electronic structure factor. This issue can be mitigated by employing the effective static approximation developed by Dornheim \textit{et al.}~\cite{Dornheim_PRL_2020_ESA, Dornheim_PRB_2021}.

\subsubsection{Application to Green functions}\label{ss:fpimc-mgf}
Let us now turn to the FPIMC downfolding for the Matsubara (thermodynamic) Green function approach. 
The general one-particle Green's function $G(\vec{r},t,\vec{r}',t')$ describes the propagation of a particle (respectively of the created vacancy) when adding/removing a particle to/from the system. For uniform systems in thermodynamic equilibrium, this quantity depends on the relative arguments $|t-t'|$ and $|\vec{r}-\vec{r}'|$ only, and it is beneficial to switch to momentum space. Without interaction, an added particle remains in its initial state, $G_0(\vec{p},t-t') \sim e^{i\epsilon_\vec{p}(t-t')/\hbar}$, corresponding to an excitation spectrum with a single peak $A(\vec{p},\omega) = 2\pi \delta(\hbar\omega - \epsilon_\vec{p})$, where $\epsilon_\vec{p} = \vec{p}^2/2m -\mu$ is the energy of a free particle. This dispersion is modified when including many-body effects, which are usually described in terms of the self-energy $\Sigma(\vec{p},\omega) = G_0(\vec{p},\omega)^{-1} - G(\vec{p},\omega)^{-1}$. 
The selfenergy is the only input quantity in Green functions theory for the ground state, thermodynamic equilibrium or nonequilibrium. It was shown by Martin and Schwinger \cite{martin-etal.59}, Baym and Kadanoff \cite{kadanoff-baym}, Keldysh \cite{keldysh65,bonitz_pss_19_keldysh} and others that, if the exact functional, $\Sigma^{\rm{exact}}[G]$, would be used, 
Green functions simulations would be exact. However, in practice, approximations have to be used which leads to approximate results for the Green function and the spectral function. 
For example, when the frequency dependence of $\Sigma$ is weak, the spectral function approaches a Lorentzian, 
\begin{align}
A(\vec{p},\omega) = \frac{2 \Im \Sigma(\vec{p})}{[\hbar\omega - \epsilon_\vec{p} - \Re\Sigma(\vec{p})]^2 + [\Im \Sigma(\vec{p})]^2 }\,,    
\label{eq:lorentzian}
\end{align}
 centered around $\hbar\omega = \epsilon_\vec{p} + \Re\Sigma(\vec{p})$, corresponding to a quasi-particle with a finite lifetime governed by $\Im \Sigma(\vec{p})$.

\begin{figure}[h]
    \centering
    \includegraphics[width=1.02\linewidth]{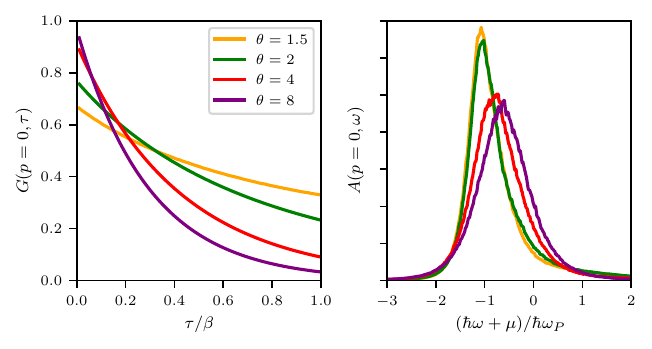}
    \caption{First principles PIMC results for the Matsubara Green's function (left) and reconstructed one-particle spectral function (right) of the uniform electron gas for electrons with zero momentum, at $r_s=4$ and multiple temperatures, obtained from a grand canonical simulation with an average of $\langle N\rangle \approx 20$ electrons.}
    \label{fig:mgf-spectral-fct}
\end{figure}

\begin{figure}[h]
    \centering
    \includegraphics[width=0.99\linewidth]{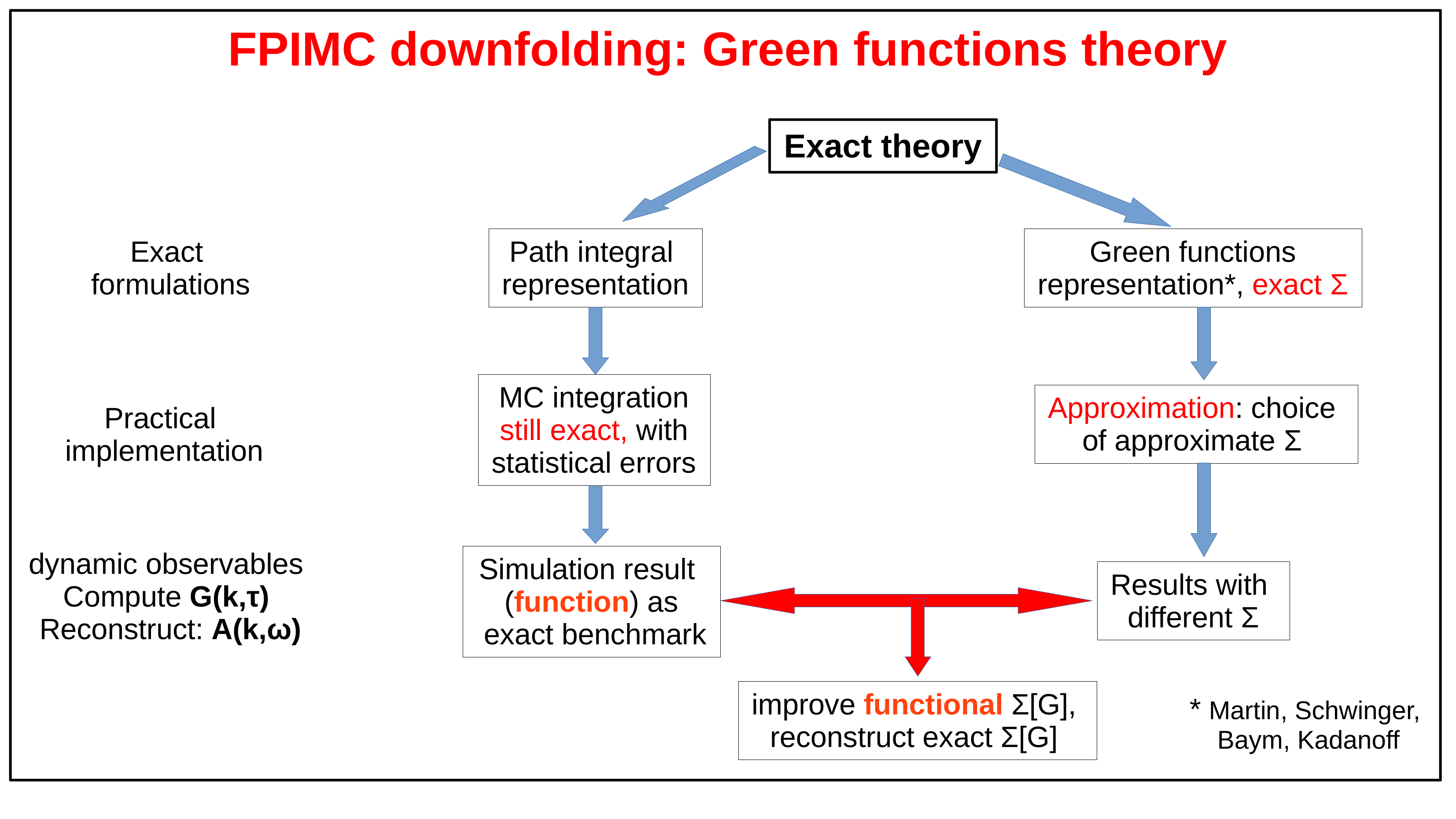}
    \caption{FPIMC downfolding for equilibrium (Matsubara) Green functions (MGF) theory. FPIMC simulations allow for the computation of the exact imaginary time GF, $G(\textbf{p},\tau)$ and of the spectral function $A(\textbf{p},\omega)$ \cite{hamann_26}. Furthermore, a  discrimination between various selfenergies $\Sigma$, of MGF theory is possible and, ultimately, FPIMC data can be used to derive an improved and possibly exact selfenergy. }
    \label{fig:fpimc-negf}
\end{figure}

Thus the conditions are fulfilled to apply the FPIMC downfolding concept of Sec.~\ref{sss:downfolding} to Green functions theory. The idea is similar to Kohn-Sham DFT and is sketched in Fig.~\ref{fig:fpimc-negf}. The key is that fermionic PIMC simulations are capable to deliver exact Matsubara Green functions. To this end, PIMC simulations involving trajectories with open ends (worm algorithm) are performed and yield
 an exact result, $G(\vec{p},\tau)$, in the imaginary-time domain (Matsubara Green's function, MGF). A few illustrative examples for the uniform electron gas at WDM conditions are shown in Fig.~\ref{fig:mgf-spectral-fct}.
 
 As a next step, the FPIMC-produced imaginary time MGF allows one to reconstruct the exact spectral function, thereby going beyond the simple quasiparticle approximation \eqref{eq:lorentzian}. Indeed, using the Kadanoff-Baym relation, the spectral function follows from an inverse Laplace transform:
\begin{equation}\label{eq:inversion}
  G(\vec{p},\tau)= \int_{-\infty}^\infty\frac{\textnormal{d}\omega}{2\pi} e^{-\tau\hbar\omega} A(\vec{p},\omega) [ 1 - f(\omega)]\ ,
\end{equation}
where $f(\omega) = 1/[e^{\beta\hbar\omega}+ 1]$ is the Fermi function. This procedure was successfully demonstrated in Refs. 
Refs.~\cite{hamann_msc_2025,hamann_26}, and a few results are depicted in  Fig.~\ref{fig:mgf-spectral-fct}.

Returning to the downfolding scheme of Fig.~\ref{fig:fpimc-negf}, the next step would be to benchmark different approximations for $\Sigma$ against the exact FPIMC results for the Matsubara Green function or the spectral function, for first results, see Ref.~\cite{hamann_26}. Ultimately, this should allow one to reconstruct an exact selfenergy, $\Sigma^{\rm{exact}}[G]$, from the FPIMC data.

\subsubsection{Application to Quantum Kinetic equations}\label{ss:fpimc-qkin}
An alternative approach to the many-particle dynamics are density operators. Their main difference to  Green functions theory is the restriction to a single time. The basic equations of motion are derived from the von Neumann equation for the N-particle density operator, $\rho_{1\dots N}$, and are given by the BBGKY-hierarchy for the reduced density operators $F_{1\dots s}$ \cite{bonitz_qkt},
\begin{align}\label{bbgkyf1}
i\hbar\frac{\partial}{\partial t} F_{1}-[H_{1},F_{1}] =&
\mbox{Tr}_{2}[V_{12},F_{12}],
\\\label{bbgkyf2}
i\hbar\frac{\partial}{\partial t} F_{12}-[H_{12},F_{12}] =&
\mbox{Tr}_{3}[V_{13}+V_{23},F_{123}],
\end{align}
to be complemented by an equation for $F_{123}$, and so on, and the reduced density operators follow from the partial trace $F_{1\dots s} = \frac{N!}{(N-s)!}\mbox{Tr}_{s+1 \dots N} \,\rho_{1\dots N}$. Here $H_1$ and $H_{12}$ are the single-particle and two-particle hamiltonian, respectively, and $V_{ij}$ is the pair interaction between particles i and j. The BBGKY-hierarchy is exact, and even its first equation is exact, if the exact functional form $F^{\rm exact}_{12}[F_1]$ is used. The latter is, of course, not known in general and is, in practice, replaced by approximations that are related to the choice of the collision integral in quantum kinetic equations \cite{bonitz_qkt}. Here the FPIMC downfolding approach is again applicable, for plasmas in thermodynamic equilibrium.
A related concept has been recently presented for a classical plasmas where a generalized collision operator was reconstructed from classical molecular dynamics simulations \cite{zhao_prl_25}.

\subsubsection{Application to chemical models and the Saha equation}\label{ss:fpimc-saha}
As a final example for the downfolding concept we consider a standard chemical model for a partially ionized plasma which was recently presented for the ionization equilibrium of warm dense hydrogen \cite{bonitz_cpp_25}. Chemical models map a spatially uniform equilibrium plasma onto a mix of free and bound particles (e.g. atoms), see for example Refs.~\cite{ebeling-richert_85, schlanges_cpp_95,juranek_jcp_02} and references therein. The chemical equilibrium of both follows from the equality of chemical potentials of free and bound particles, 
\begin{align}
    \mu_e + \mu_i = \mu_A\,,
    \label{eq:chem-eq}
\end{align}
where the chemical potentials depend on temperature $T$, total density $n$, as well as on the fractions, $\alpha$ and $x_A$ of free and bound particles.
Equation~\eqref{eq:chem-eq} can be transformed into an equation for $\alpha=\alpha^{\rm ion}$ as a function of $x_A, n, T$ which is the Saha equation \cite{saha_20} that has been broadly used in plasma physics. While at low densities, the chemical potentials of electrons and ions are those of an ideal gas whereas the atomic chemical potential involves the ionization potentials of atomic bound states, $I_{nl}$, with increasing density, Coulomb interaction, quantum and spin effects give rise to corrections, $\mu^{\rm id} \longrightarrow \mu^{\rm id} + \mu^{\rm int}$. As a consequence all bound states levels shift, and the effective ionization energies decrease with density, $I_{nl} \to I^{\rm eff}_{nl}$ and, at critical densities (Mott density), bound states vanish one by one and atoms break up. Would the values $I^{\rm eff}_{nl}$ be known exactly, then the nonideal Saha equation would yield exact results for the fractions of free and bound particles.
In practice, of course, the effective ionization potentials $I^{\rm eff}_{nl}$ or their reduction
[``ionization potential depression (IPD)''] have to be computed approximately, and there have been extensive efforts in plasma physics in this direction, based a variety of models and approximations, e.g. \cite{inglis-teller_39, ebeling-richert_85, ecker-kroell_63,stewart-pyatt_66, green-book, schlanges_cpp_95} and references therein.
\begin{figure}
    \centering
    \includegraphics[width=0.99\linewidth]{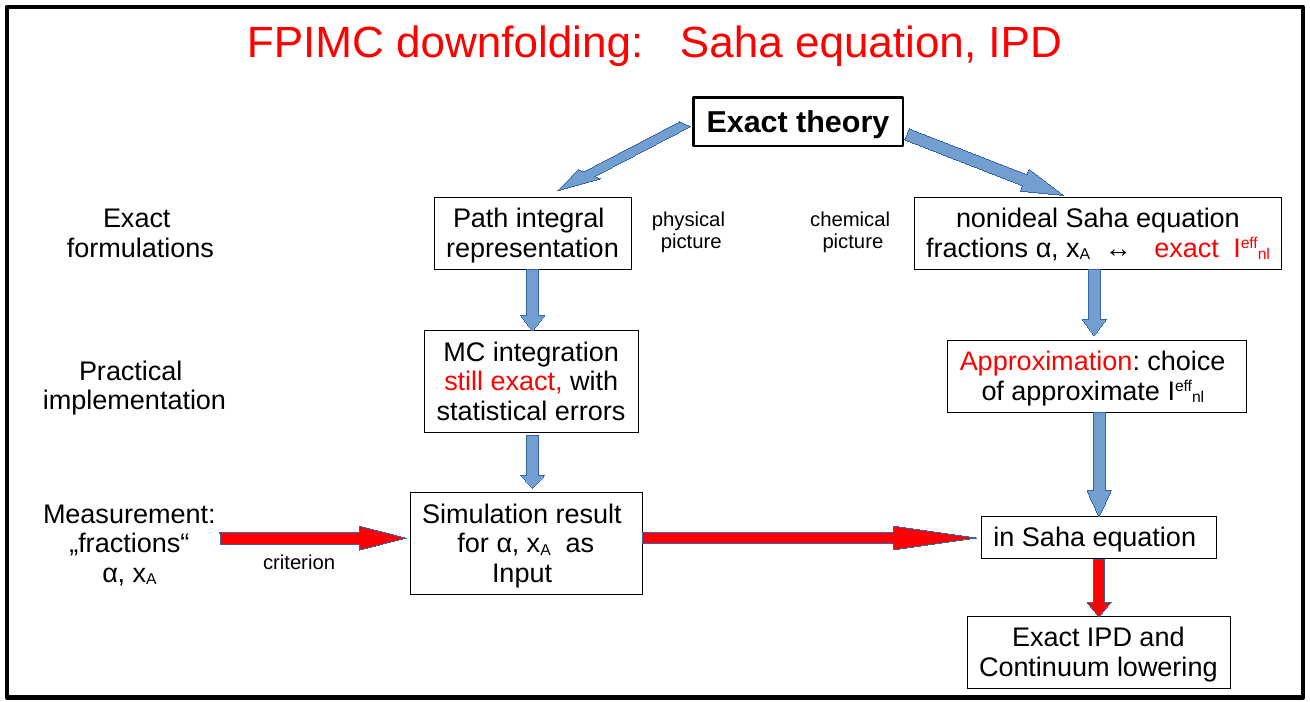}
    \caption{FPIMC downfolding for the Saha equation and the ionization potential depression (IPD). Right branch: the nonlinear Saha equation is typcially solved for the free particle and atomic fractions, $\alpha$ and $x_A$, taking approximations for the effective ionization potentials as input. Left branch: first principle FPIMC simulation allow to compute $\alpha$ and $x_A$. Feeding this into the Saha equation allows in principle to compute the exact ionization potential depression.} 
    \label{fig:fpimc-saha}
\end{figure}

The direct connection $\alpha, x_A \leftrightarrow \{I^{\rm eff}_{nl}\}$ allows for the application of  a downfolding procedure. In fact, performing an accurate simulation of $\alpha$ and $x_A$ would allow one to invert the problem and compute the IPD from the Saha equation, using $\alpha$ and $x_A$ as input. This concept was first put forward in Ref.~\cite{bonitz_pop_24} and presented in more detail in Ref.~\cite{bonitz_cpp_25}. The idea is sketched in Fig.~\ref{fig:fpimc-saha}. The FPIMC simulations for a partially ionized hydrogen plasma are depicted in the left branch, as in the applications to DFT and Green functions, and are again exact up to statistical errors. But compared to the previous cases of Figs.~\ref{fig:fpimc-dft} and \ref{fig:fpimc-negf}, here the observables that are being computed by FPIMC are different: the fractions of free and bound electrons, $\alpha$ and $x_A$. Even though FPIMC works in the ``physical picture'' where no strict discrimination between bound and free states is performed (as is done in chemical models), one can introduce criteria how to distinguish the two cases, based on the shape of the QMC trajectories in the vicinity of the nuclei, as was done e.g. in Refs.~\cite{militzer_pre_2001} and \cite{filinov_pre_23} and was illustrated in Fig.~\ref{fig:bound-states-pimc} above. Such a procedure carries, of course, some ambiguity. At the same time, the influence of the chosen criterion on the results for the IPD has been found to be small, on the order of one percent \cite{bonitz_cpp_25}. 
\begin{figure}
    \centering
    \includegraphics[width=0.49\textwidth]{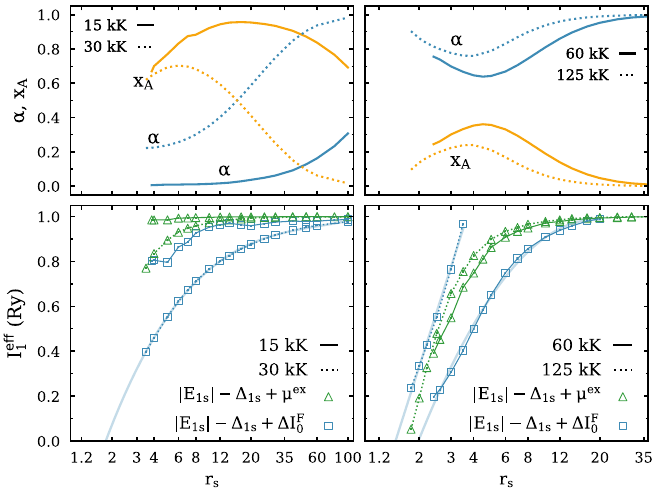}
    \caption{\textbf{Top:} Density dependence of free electron and atom fractions in hydrogen for four temperatures, calculated with FPIMC-simulations in Ref.~\cite{filinov_pre_23}. At the two lower temperatures the plasma also contains molecules,  therefore, $\alpha+x_A < 1$. \textbf{Bottom:} Effective ionization energy of the  ground state, $I^{\rm eff}_1$, with level shift $\Delta_{1s}$ included, for $31\,250$ K and $15\,625$ K (left), and  $62\,500$ K and $125\,000$ K (right). %Triangles (squares): FPIMC1 (FPIMC2) results with (without) renormalization of the bound state energies.
     Triangles: continuum shift based on Ebeling's Padé formula \cite{ebeling-richert_85}. Squares: continuum shift from FPIMC-data. Light blue lines: analytical fit and extrapolation to the Mott density. Figure reproduced from Ref.~\cite{bonitz_cpp_25} with the permission of the authors.
}
    \label{fig:ipd_ieff}
\end{figure}

The results for the free and bound electron  fractions in dense hydrogen in a very broad density range are shown, for four temperatures, in the top parts of Fig.~\ref{fig:ipd_ieff}. At the two lowest temperatures ($15\,000$ K and $30\, 000$ K) atoms and molecules dominate in a broad density range. At low densities they vanish, due to decreasing probability of encounter of an electron and a proton. On the other hand, at high density, $r_s \lesssim 4$, atoms break up due to lowering of the binding energies, $I^{\rm eff}_{nl}$. At higher temperature (top right figure) the fraction of atoms is lower and their finite density range of existence is even more pronounced.
The results for the ground state ionization potential, $I^{\rm eff}_{1}$, computed via the nonlinear Saha equation using FPIMC input in Ref.~\cite{bonitz_cpp_25} are shown in the bottom panels of Fig.~\ref{fig:ipd_ieff} by the blue lines with squares. As expected, at low density (large $r_s$), the ionization potential equals 1 Ry and, with increasing density, decreases monotonically. Since fermionic PIMC simulations are afflicted with the fermion sign problem, cf. Sec.~\ref{ss:fpimc-progress}, computations are only possible for $r_s \gtrsim 3$ \cite{filinov_pre_23}. Nevertheless, the smooth behavior of the curve allows one to extrapolate to the density where $I^{\rm eff}_{1} \to 0$, and all bound states vanish. This density is commonly called ``Mott density''. Thus FPIMC simulations, together with the present downfolding idea, allow one to obtain first principles results for the Mott density of hydrogen at moderately low temperatures, $T\gtrsim 30\, 000$K \cite{bonitz_cpp_25}.

Note that the present downfolding procedure works well only in case that atoms are predominantly in the ground state. Otherwise the Saha equation contains not just the ground state ionization potential but also those of the excited states. FPIMC, in the present implementations, does not resolve individual bound state contributions, and the procedure requires additional input for the density dependent lowering of the bound state levels which is typically available only in approximations \cite{bonitz_cpp_25}. 

For completeness we mention an 
alternative FPIMC approach to IPD. In fact, Bellenbaum \textit{et al.} \cite{bellenbaum_25} have used the FPIMC results for the XRTS spectra together with the Chihara decomposition \cite{Chihara_1987} to simultaneously reconstruct the degree of ionization and the IPD.

\subsection{Benchmarks for ICF modeling}\label{ss:icf-benchmarks}
A typical laser-driven ICF target capsule consists of a solid DT-ice layer enclosed by an ablator shell. As was discussed in Sec.~\ref{ss:icf-overview}, either direct laser ablation (laser-direct-drive: LDD) or laser-driven x-ray ablation (laser-indirect-drive: LID) can launch shocks into the ICF target, which compress the DT fuel up to tens of times of its solid density and heat the DT layer to tens of electro-volts temperature. The shock-compressed/heated shell then starts to implode by the continuous drive pressure, which leads to higher densities due to the spherical convergence. The density and temperature conditions of the in-flight DT-shell condition is illustrated in Fig.~\ref{fig:overview} as marked by ``LID ICF path'' \cite{abu-icf_prl_24}.

Simulations of inertial confinement fusion experiments are traditionally performed using  radiation-hydrodynamics codes. The NIF shots were accompanied by simulations using the three-dimensional HYDRA code \cite{hydra_1996,hydra_1998,hydra_2001}. Application of this code to the recent high gain shots was briefly described by Kritcher \textit{et al.} in Ref.~\cite{kritcher_pre_22}.
Radiation hydrodynamic codes succesfully resolve the implosion geometry and predict hydrodynamic instabilities. At the same time, they require extensive external input \cite{kritcher_pre_22} for thermodynamic (equation of state), transport and optical properties of the fusion fuel and the capsule (ablator) materials that are not easily available for the extreme density and temperature conditions of the implosion. A central input are updated versions of the SESAME tables \cite{sesame_2003}. As we point out below, it has to be expected that these simulations, as well as the used input, are not reliable for large parts of the fusion path.
\begin{figure}
    \centering
    \includegraphics[width=0.95\linewidth]{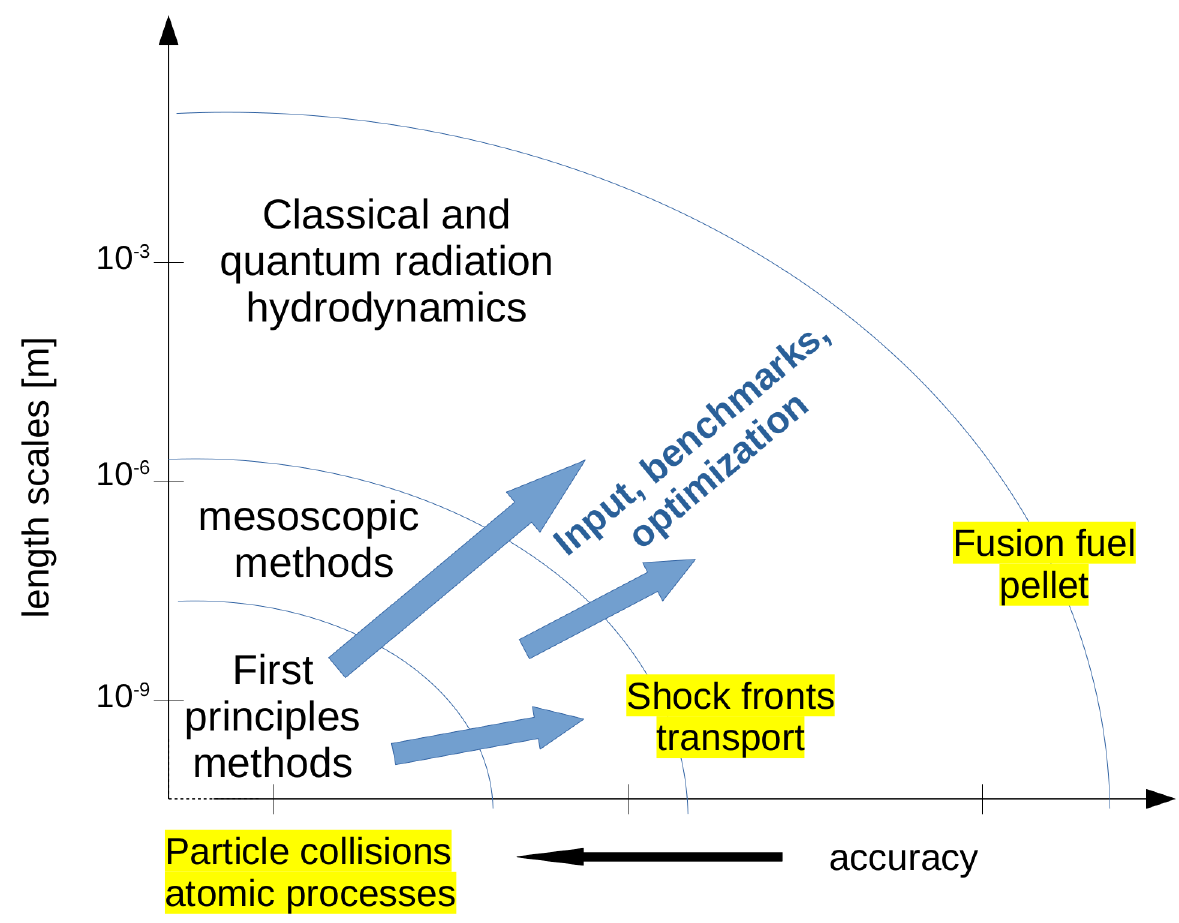}
    \caption{Illustration of the length scales and different groups of simulation methods for ICF modeling. First principles methods (such as FPIMC, DFT, NEGF) have the highest accuracy and resolve the elementary collision processes but are restricted to the smallest length and time scales. The macroscopic scales of the entire fusion pellet can be captured by radiation hydrodynamics, but its accuracy is limited. An intermediate ``mesoscopic'' level of description is provided by semiclassical MD or average atom models. A more detailed picture is presented in Fig.~\ref{fig:combination}.}
    \label{fig:methods-space}
\end{figure}

An alternative to radiation hydrodynamics are \textit{ab initio} simulations. Morales \textit{et al.} performed KS-DFT-MD simulations, as well as coupled electron-ion QMC simulations, for the early stage of the implosion \cite{morales_icf_12} and reported good agreement with the modified SESAME tables \cite{sesame_2003}.
However, these simulations are computationally costly and only possible for limited particle number and system sizes.
On the other hand, Hu \textit{et al.} \cite{hu_militzer_PhysRevLett.104.235003,PhysRevB.84.224109} 
performed restricted PIMC simulations and constructed a wide range equation of state. They underlined the importance of strong coupling and degeneracy effects \cite{PhysRevB.84.224109, hu2015impact}. In addition, Hu \textit{et al.} used QMD simulations, based on both Kohn-Sham DFT and orbital-free DFT, to have established  first-principle EOS tables for fusion fuel deuterium \cite{mihaylov2021improved} and common ablator materials such as  polystyrene \cite{hu2015first}, beryllium \cite{ding2017first}, as well as silicon \cite{hu2017first}. These tables provide EOS data in a wide range of density and temperatures to enable radiation-hydrodynamics simulation of ICF implosions. For a recent overview on equation of state models for ICF, see Ref.~\cite{gaffney_hedp_18}. 

The most accurate EOS data for dense hydrogen are the recent FPIMC results of Filinov and Bonitz that are available for $T\gtrsim 15\,000$ K \cite{filinov_pre_23} and that were used extensively for benchmarks in this paper, cf.  Secs.~\ref{s:benchmarks-h} and \ref{ss:fpimc-dft}.
\begin{figure}[h]
    \centering
    \includegraphics[width=0.995\linewidth]{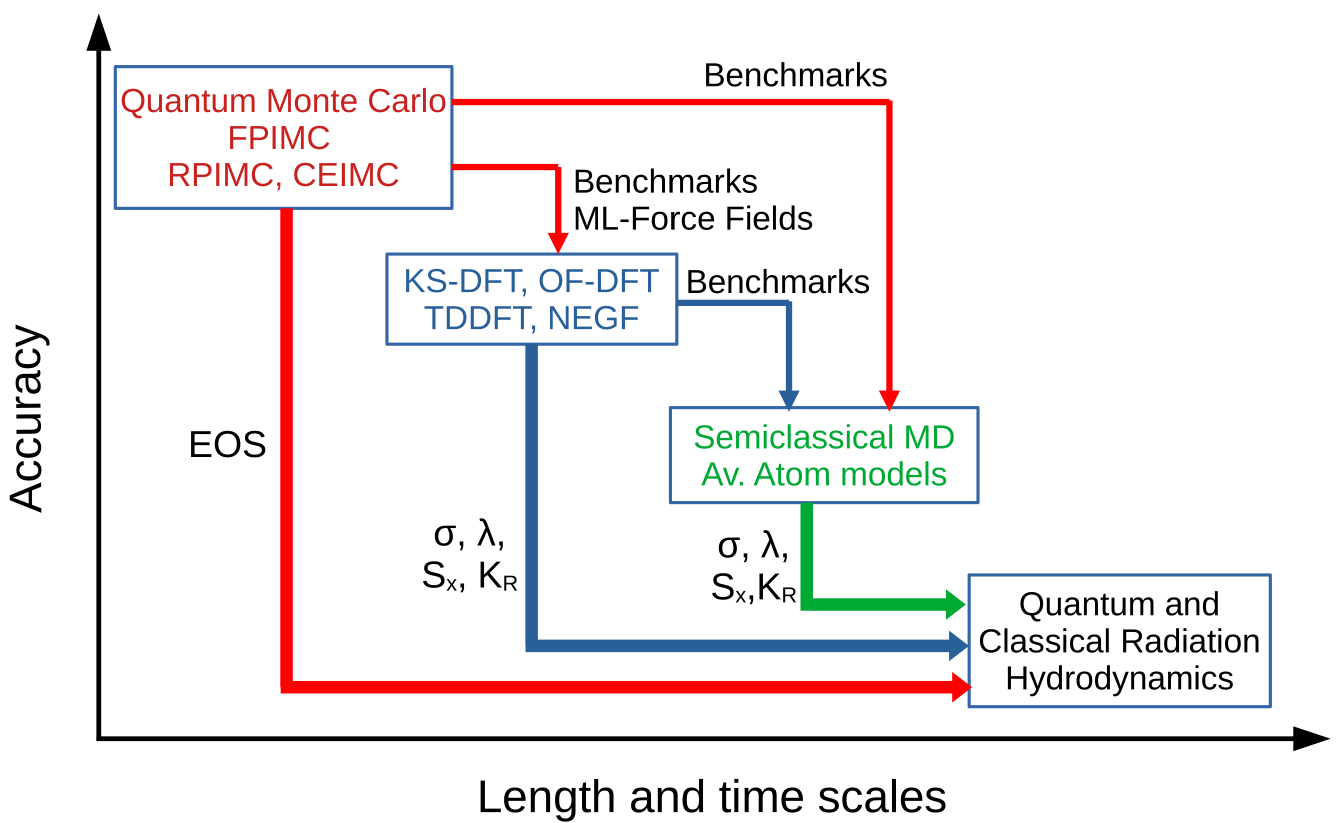}
    \caption{Combination of simulation methods that are relevant for ICF modeling, extending Fig.~\ref{fig:methods-space}. Large scale simulations are possible with radiation-hydrodynamic methods that require extensive input from more accurate methods that are indicated by the arrows. Examples for input are EOS: equation of state, $\sigma$: electrical conductivity, $\lambda$: thermal conductivity, $S_x$: stopping power, $K_{\rm R}$: opacity. Adapted from Ref.~\cite{bonitz_pop_24}.}
    \label{fig:combination}
\end{figure}
The importance of using accurate EOS and transport data for ICF modeling can be clearly seen by considering
 a typical ICF implosion path in the temperature-density plane depicted in Fig.~\ref{fig:overview}.
Both, in LDD and LID scenarios, this path indeed passes through the so-called warm-dense matter regime, in which the quantum electron degeneracy and the strong ion-ion coupling have a crucial influence on the EOS, thermal transport, and opacity of the in-flight DT shell \cite{hu2018review,abu-icf_prl_24}. Typically, the in-flight shell temperature is less than half of its Fermi temperature. Furthermore, the DT shell spends more than 80\% of the implosion time in this WDM regime, before it stagnates into a super-dense shell surrounding a hot spot for fusion burn. This is why accurate knowledge of both DT shell and ablator materials in this exactly WDM condition is so crucial for reliable modeling of ICF implosions using radiation-hydrodynamic simulations. It is noted that the in-flight DT shell condition essentially determines whether or not an ICF implosion can reach ignition and energy gain \cite{abu-icf_prl_24}.  We thus conclude that accurate input, in particular for the WDM part of the implosion process,  is crucial for a reliable modeling of ICF.

As we have demonstrated above, recent fermionic PIMC simulations for dense hydrogen \cite{filinov_pre_23} allow one to benchmark RPIMC simulations as well as KS-DFT simulations with ground state and finite temperature exchange correlation functionals which are of direct relevance for ICF conditions \cite{bonitz_pop_24}. The results of these benchmarks suggest that ICF simulations can achieve significantly increased accuracy if the appropriate DFT functionals are being used and the strengths of different approaches are being combined.
The idea of such a combination was briefly formulated in Ref.~\cite{bonitz_pop_24} and is sketched 
in Fig.~\ref{fig:methods-space}.
There we loosely group the available methods in First principles approaches that include quantum Monte Carlo and DFT simulations, mesoscopic methods, such as semiclassical MD or average atom models, 
and macroscopic approaches -- primarily radiation hydrodynamics. As the figures demonstrates, each of the groups has a different accuracy level and works on different length (and time) scales and captures different physical processes. The blue arrows indicate that the higher level methods can provide input and tests for the lower level methods and can be used to optimize the approximations being used in the latter. This scheme is presented in more detail 
in Fig.~\ref{fig:combination}.

\section{Conclusions and outlook} \label{s:conclusion}
In this paper we presented an overview on the relevance of quantum effects in modern plasma physics. Motivated by the international year of quantum science and technology 2025 we looked back at the first steps of quantum theory that were made by Max Planck with the discovery of the formula for the spectrum of thermal electromagnetic radiation and the hypothesis of elementary quanta of action. We also recalled pioneering experiments in low pressure plasmas showing spatially separated glowing striations that today can be seen as clear and beautiful demonstration of the quantum properties of atoms and that led to the famous Franck-Hertz experiment. 

After an elementary overview on quantum effects we considered various kinds of plasmas, including low-pressure and high temperature plasmas, and discussed the importance of quantum effects for them. The tremendous range of temperatures and densities covered by plasmas is astonishing, as is the universal role of Coulomb interaction effects that unite such extremely diverse systems. We also pointed out plasma-like behavior in other systems including condensed matter (electron gas in metals, electron-hole plasma in semiconductors) and in high-energy systems (e.g., relativistic laser plasmas and the quark-gluon plasma). These systems exhibit a number of qualitative similarities with traditional plasmas. At the same time, their state of the art description, however, requires detailed system-specific information and methods that are beyond standard plasma physics.

The main focus of the reminder of this article was devoted to high density plasmas, including warm dense matter and inertial fusion plasmas, where the electrons exhibit quantum effects. We discussed recent advances in experiments and simulations, including first principles simulations (fermionic PIMC and Kohn-Sham DFT),  mesoscopic simulations such as semi-classical MD, as well as macroscopic hydrodynamic simulations. These simulations have complementary advantages and disadvantages: hydrodynamics and mesoscopic simulations are conceptionally simple and computationally moderately expensive. On the other hand, first principles simulations achieve high accuracy, but they are computationally costly. To combine the ``best of all worlds'', we put forward a novel downfolding approach that is based on computationally exact fermionic PIMC simulations. This approach includes FPIMC benchmarks of other less accurate methods. But the downfolding idea goes beyond benchmarks. It includes systematic comparisons of FPIMC results for a large variety of quantum plasma observables including the equation of state, static structure factor, Matsubara Green function, dynamic structure factor and others. This extensive information can be used to systematically improve other methods, such as Kohn-Sham DFT of Green functions theory by optimizing the exchange correlation functional or the selfenergy, respectively. Possible strategies include mixing of existing approximations for the exchange-correlation functional, in DFT, or selfenergies, in Green functions theory.
%Future developments along this path include ...

%

The present discussion focused on quantum plasmas in thermodynamic equilibrium where exact data can be obtained from fermionic PIMC simulations. On the other hand, similar concepts are also of high interest for plasmas that are excited by short pulse lasers or by shock compression. In this case, nonequilibrium phenomena, including nonthermal distributions \cite{kosse-etal.97, bonitz_qkt, kremp_pop_00} and nonlinear response or wake effects \cite{moldabekov_cpp15,moldabekov_pre15}, are relevant. Here, instead of FPIMC, a suitable starting point are nonequilibrium Green functions simulations that are currently under active development and have already produced many benchmarks for model systems, e.g.~\cite{bonitz_qkt, balzer-book, schluenzen_jpcm_19, bonitz_pssb23, schroedter_26}.

  \section*{Acknowledgments}
   MB acknowledges valuable discussions with Arne Schirrmacher (HU Berlin) on Planck's derivation of the radiation law. 
   This work has been supported by the Deutsche Forschungsgemeinschaft via grants BO1366/13-2, BO1366/16-1 and DO2670/1-1.

 This work has received funding from the European Research Council (ERC) under the European Union’s Horizon 2022 research and innovation programme (Grant agreement No. 101076233, ``PREXTREME''). 
Views and opinions expressed are however those of the authors only and do not necessarily reflect those of the European Union or the European Research Council Executive Agency. Neither the European Union nor the granting authority can be held responsible for them.

Computations were performed on a Bull Cluster at the Center for Information Services and High-Performance Computing (ZIH) at Technische Universit\"at Dresden and at the Norddeutscher Verbund f\"ur Hoch- und H\"ochstleistungsrechnen (HLRN) under grant mvp00024.

SXH and VVK would like to acknowledge the  support by the U.S. Department of Energy
(National Nuclear Security Administration), University of
Rochester “National Inertial Confinement Fusion Program” under
Award No. DE-NA0004144, and U.S. National Science Foundation PHY Grant No. 2020249.

\bibliography{ref,mb-ref,mb-ref-1}

\end{document}